\documentclass[rmp,aps,eqsecnum]{revtex4-1}

\usepackage{amsmath}
\usepackage{amssymb}
\usepackage{color}
\usepackage{graphicx}
\usepackage{epsfig}
\usepackage{psfrag}
\usepackage{latexsym}
\usepackage{bm}
\usepackage[hyperfootnotes=false]{hyperref}
\usepackage{xspace}
\usepackage{textcomp}
\usepackage{bbold} 

\newcommand{\be}{\begin{equation}}
\newcommand{\ee}{\end{equation}}

\newcommand{\bea}{\begin{eqnarray}}
\newcommand{\eea}{\end{eqnarray}}

\newcommand{\ba} {\begin{eqnarray}}
\newcommand{\ea} {\end{eqnarray}}

\newcommand{\nn}{\nonumber}

\newcommand {\sd}[1]{\left(#1\right)}

\newcommand {\re}[1]{\text{Re}\sd{#1}}

\newcommand {\dm}{\Delta m^2}

\newcommand{\vev}[1]{\left\langle #1 \right\rangle}

\def\article{\@ifnextchar[{\earticle}{\oarticle}}

\def\beq{\begin{equation}}
\def\eeq{\end{equation}}
\def\bea{\begin{eqnarray}}
\def\eea{\end{eqnarray}}
\def\bet{\begin{tabular}}
\def\eet{\end{tabular}}
\def\bes{\begin{subequations}\bea}
\def\ees{\eea\end{subequations}}

\def\a{\alpha}
\def\b{\beta}

\def\c{\chi}
\def\g{\gamma}

\def\e{\epsilon}

\def\u2{\dd\frac{1}{\sqrt{2}}}

\def\be{\begin{equation}}
\def\ee{\end{equation}}
\def\bc{\begin{center}}
\def\ec{\end{center}}
\def\bea{\begin{eqnarray}}
\def\eea{\end{eqnarray}}
\def\dd{\displaystyle}
\def\nn{\nonumber}

\catcode`@=11
\def\marginnote#1{}
\newcount\hour
\newcount\minute
\newtoks\amorpm
\hour=\time\divide\hour by60
\minute=\time{\multiply\hour by60 \global\advance\minute by-\hour}
\edef\standardtime{{\ifnum\hour<12 \global\amorpm={am}%
        \else\global\amorpm={pm}\advance\hour by-12 \fi
        \ifnum\hour=0 \hour=12 \fi
        \number\hour:\ifnum\minute<10 0\fi\number\minute\the\amorpm}}
\edef\militarytime{\number\hour:\ifnum\minute<10 0\fi\number\minute}
\def\draftlabel#1{{\@bsphack\if@filesw {\let\thepage\relax
   \xdef\@gtempa{\write\@auxout{\string
      \newlabel{#1}{{\@currentlabel}{\thepage}}}}}\@gtempa
   \if@nobreak \ifvmode\nobreak\fi\fi\fi\@esphack}
        \gdef\@eqnlabel{#1}}
\def\@eqnlabel{}
\def\@vacuum{}
\def\draftmarginnote#1{\marginpar{\raggedright\scriptsize\tt#1}}
\def\draft{\oddsidemargin 0.0truein
        \def\@oddfoot{\sl preliminary draft \hfil
        \rm\thepage\hfil\sl\today\quad\militarytime}
        \let\@evenfoot\@oddfoot \overfullrule 3pt
        \let\label=\draftlabel
        \let\marginnote=\draftmarginnote
   \def\@eqnnum{(\theequation)\rlap{\kern\marginparsep\tt\@eqnlabel}%
\global\let\@eqnlabel\@vacuum}  }
\catcode`@=12
\newcommand{\TeV}{\,\mathrm{TeV}}
\newcommand{\GeV}{\,\mathrm{GeV}}

\newcommand{\eV}{\,\mathrm{eV}}

\newcommand{\fracwithdelims}[4]{\left#1 \frac{#3}{#4} \right#2}
\newcommand{\ord}[1]{\mathcal{O}\left( #1 \right)}

\newcommand{\Dm}[1]{{\Delta m^2_{#1}}}
\newcommand{\Fig}[1]{Fig.~\ref{fig:#1}}

\newcommand{\Sect}[1]{Sec.~\ref{sec:#1}}

\newcommand{\Eq}[1]{Eq.~(\ref{eq:#1})}
\newcommand{\eq}[1]{eq.~(\ref{eq:#1})}

\newcommand{\eqs}[1]{eqs.~(\ref{eq:#1})}

\newcommand{\nohyphens}%
        {\hyphenpenalty=10000\exhyphenpenalty=10000\relax}
\definecolor{amethyst}{rgb}{0.6, 0.4, 0.8}
\definecolor{antiquefuchsia}{rgb}{0.57, 0.36, 0.51}
\definecolor{airforceblue}{rgb}{0.36, 0.54, 0.66}
\definecolor{ao}{rgb}{0.0, 0.5, 0.0}
\definecolor{asparagus}{rgb}{0.53, 0.66, 0.42}
\definecolor{auburn}{rgb}{0.43, 0.21, 0.1}
\definecolor{brightcerulean}{rgb}{0.11, 0.67, 0.84}

\newcommand{\rep}{U}

\DeclareMathOperator{\diag}{Diag}

\begin{document}

\title{Lepton Flavour Symmetries}

\author{Ferruccio Feruglio}
\affiliation{Dipartimento di Fisica e Astronomia `G.~Galilei', Universit\`a di Padova \\
INFN, Sezione di Padova, Via Marzolo~8, I--35131 Padua, Italy}
\author{Andrea Romanino} 
\affiliation{SISSA and INFN, \mbox{Via Bonomea 265, I--34136}}
\affiliation{ICTP, \mbox{Strada Costiera 11, I--34151}, \mbox{Trieste, Italy}}

\begin{abstract}
We provide a general classification of flavour symmetries according to their interplay with the proper Poincar\'e and gauge groups and to their linear or nonlinear action in field space. We focus on the lepton sector and we review the different types of symmetries describing neutrino masses and the lepton mixing matrix. For each type of symmetry we present several illustrative examples and we discuss specific strengths and limitations. 
\end{abstract}

\maketitle
\tableofcontents

\newpage

\section{Introduction}
The replica of fermion families, their masses and intergenerational properties constitute one of the most
fascinating mysteries of particle physics. While gauge symmetry strongly restricts matter interactions
mediated by spin one particles, it leaves essentially unconstrained scalar-fermion interactions, responsible for
fermion masses and mixing angles. In the flavour sector of the Standard Model (SM) there are as many independent parameters as
the number of charged fermion masses and quark mixing parameters. The toll raises to 22 if we include,
in a general low-energy description, 
neutrino masses and lepton mixing parameters. We are facing a puzzle with many known
pieces, that we are still unable to put together in a coherent picture.
The discovery of neutrino oscillations has brought great hopes for the solution of this puzzle.
Neutrinos are extremely light, calling for a different origin of their masses, potentially related to
new undiscovered properties of particle interactions. Moreover atmospheric and solar neutrino oscillations
require large lepton mixing angles, a completely unexpected feature, clashing against the properties of the quark sector. As we briefly review here, many of these properties have been determined to a good precision and there are excellent prospects for future improvements aimed to pin down the few unknown aspects. Nevertheless, while neutrino data stimulated a great deal of theoretical activity, they also heightened the mystery of fermion masses, in that no compelling underlying principle to describe this aspect of elementary particles has uniquely emerged so far. Neutrinos and charged leptons possess special features, that are the focus of the present review, although
any description applicable to this sector alone should only be viewed as a partial answer to the general problem
of fermion masses.

The observation and study of neutrino oscillations have established that neutrinos are massive. Two independent squared
mass differences and three lepton mixing angles have been determined with an accuracy approaching the percent level, moving the whole field into a precision era. Most of the experimental results can be coherently
interpreted in the context of three light active neutrinos and CPT invariance. Experiments sensitive to solar, 
atmospheric, reactor and accelerator neutrinos provide a consistent picture supported by many redundant tests. 
In table \ref{tableData} we report the results of recent fits to the oscillation parameters. Notation and conventions are those of the Review of Particle Physics by the Particle Data Group (PDG) \cite{PhysRevD.98.030001}, unless otherwise stated. Most remarkably, the mixing pattern in the lepton sector appears to be 
totally different from that in the quark sector, with two large mixing angles and a third one similar in size to the Cabibbo angle.

Very interestingly, global analysis start to be sensitive both to the mass ordering and to the Dirac CP violating phase
$\delta$. A preference for the normal mass ordering (NO) over the inverted one (IO) is emerging from the data, at the level of about 3 $\sigma$.
The best fit value for the Dirac CP violating phase is $\delta \approx (1.2\div 1.3) \pi$ (for NO), but uncertainties are large and CP conservation is still allowed within $2\sigma$. 
\begin{table}[h!] 
\centering
\begin{tabular}{|c|c|c|}
\hline
&{\tt Normal Ordering}&{\tt Inverted Ordering}\rule[-2ex]{0pt}{5ex}\\
\hline
$\sin^2\theta_{12}$&$0.310^{+0.013}_{-0.012}$&$0.310^{+0.013}_{-0.012}$\rule[-2ex]{0pt}{5ex}\\
\hline
$\sin^2\theta_{23}$&$0.563^{+0.018}_{-0.024}$&$0.565^{+0.017}_{-0.022}$\rule[-2ex]{0pt}{5ex}\\
\hline
$\sin^2\theta_{13}$&$0.02237^{+0.00066}_{-0.00065}$&$0.02259^{+0.00065}_{-0.00065}$\rule[-2ex]{0pt}{5ex}\\
\hline 
$\delta/\pi$&$1.23^{+0.22}_{-0.16}$&$1.57^{+0.13}_{-0.14}$\rule[-2ex]{0pt}{5ex}\\
\hline
$\Delta m^2_{21}/10^{-5}$eV$^2$&$7.39^{+0.21}_{-0.20}$&$7.39^{+0.21}_{-0.20}$\rule[-2ex]{0pt}{5ex}\\
\hline
$\Delta m^2_{3\ell}/10^{-3}$eV$^2$&$2.528^{+0.029}_{-0.031}$&$-2.510^{+0.030}_{-0.031}$\rule[-2ex]{0pt}{5ex}\\
\hline
\end{tabular}
\caption{Best fit values and 1$\sigma$ errors of the three-flavour oscillation parameters in the global analysis of ref. \cite{Esteban:2018azc}. The results include data on atmospheric neutrinos provided by the Super-Kamiokande collaboration.  There is a difference of $\Delta\chi^2(\text{IO}-\text{NO})=10.4$ between 
inverted ordering (IO) and normal ordering (NO). Note that $\Delta m^2_{3\ell}=\Delta m^2_{31}>0$ for NO and $\Delta m^2_{3\ell}=\Delta m^2_{32}<0$ for IO. For other recent global analysis, see refs. \cite{Gariazzo:2018pei,deSalas:2018bym,Capozzi:2019vbz}.}
\label{tableData}
\end{table}

Dedicated experiments have been planned to determine the mass ordering and $\delta$.
Mass ordering measurements with an individual significance of more than 3$\sigma$ could be realized with several different technologies and methods, exploiting atmospheric (KM3NeT/ORCA, PINGU, INO), reactor (JUNO) and accelerator (DUNE, Hyper-K) neutrinos.
DUNE and Hyper-K have planned sensitivities to CP-violation higher than 5$\sigma$ for most of the allowed range, even though a precise determination of $\delta$ around the maximal value will be very challenging.

The absolute neutrino mass scale is still unknown, though well constrained by both laboratory  and cosmological  observations. The current laboratory limit $m_\nu=\sqrt{\sum_i |U_{ei}|^2 m_i^2} < 1.1$ eV (90\% CL), recently set by the KATRIN experiment \cite{Aker:2019uuj}, is expected to be further improved in the future. At present cosmology provides the most stringent bound on the sum of neutrino masses,
$\sum_i m_i<0.12\div 0.68$ eV, though subject to uncertainties inherent to the adopted cosmological model, the number of free parameters used to fit observations and the actual set of data included in the analysis \cite{PhysRevD.98.030001}. Upper bounds on neutrino masses become weaker when the data are analyzed in the context of extended cosmological models, or when a conservative set of data is used, but not considerably weaker. 
These bounds are expected to improve significantly over the next years, thanks to the new planned experiments.
If the $\Lambda$CDM model of the universe is confirmed, and if neutrinos have standard properties, non-vanishing neutrino masses should be detected at the level of at least 3$\sigma$ \cite{PhysRevD.98.030001}.

The impressive suppression of neutrino masses is quite peculiar, even compared to that of the lightest charged fermions. Not only the electron mass is suppressed by ``only'' a factor $\ord{10^{5}}$ but, more important, the latter suppression follows from the inter-family hierarchy displayed by charged fermion masses, with subsequent families separated by only about two orders of magnitudes. Conversely, all the three neutrino families are separated from the electroweak scale by at least 11 orders of magnitude. The striking size of neutrino masses might be related to the possibility that the total lepton number $L$ is violated, though this is certainly not the only possible explanation.
The violation of the individual lepton numbers have been established, but we still do not know whether $L$ is violated or not in Nature. The experimental clarification of this central aspect might shed light on the possible origin of flavour. Indeed from the theory viewpoint the simplest explanation of the smallness of neutrino masses is in term of the violation of $L$ at a very large scale,
possibly not far from the grand unified scale. 

Experimentally, the most promising $L$-violating transition is the neutrinoless double beta ($0\nu\beta\beta$) decay.
If interpreted in the context of three light Majorana neutrinos, the present experiments allow to set an upper bound on $|m_{ee}|=|\sum_i U_{ei}^2 m_i|$,
a combination of neutrino masses, mixing angles and Majorana phases. Despite the uncertainties due to the lack of knowledge 
of absolute masses and Majorana phases, $|m_{ee}|$ can be constrained by neutrino oscillation data alone and, at least for the case
of IO, the allowed region is getting closer and closer to the range explored by the present $0\nu\beta\beta$ decay experiments.
In table \ref{nonenubb} we report some of the most recent experimental results. We refer the interested reader to the recent reviews \cite{Pas:2015eia,DellOro:2016tmg,Vergados:2016hso,Dolinski:2019nrj}.
\begin{table}[h!] 
\centering
\begin{tabular}{|c|c|c|c|}
\hline
{\tt Isotope}& {\tt Lower Bound on} $T_{1/2}^{0\nu}$ ($\rm{yr}$)&{\tt Upper Bound on} $|m_{ee}|$ ($\rm{meV}$)&{\tt Collaboration}\rule[-2ex]{0pt}{5ex}\\
\hline
$^{76}$Ge&$8.0\cdot 10^{25}$ &$120\div260$& GERDA\rule[-2ex]{0pt}{5ex}\\
\hline
$^{130}$Te& $1.5\cdot 10^{25}$&$110\div520$& CUORE\rule[-2ex]{0pt}{5ex}\\
\hline
$^{136}$Xe& $1.07\cdot 10^{26}$&$61\div165$& KAMLAND Zen\rule[-2ex]{0pt}{5ex}\\
\hline
$^{136}$Xe& $3.5\cdot 10^{25}$&$93\div286$& EXO 200\rule[-2ex]{0pt}{5ex}\\
\hline
\end{tabular}
\caption{Lower bound on $T_{1/2}^{0\nu}$ (90\% CL) and upper bound on $|m_{ee}|$ from GERDA \cite{Agostini:2018tnm}, CUORE \cite{Alduino:2017ehq}, KAMLAND Zen \cite{KamLAND-Zen:2016pfg}, EXO 200 \cite{Anton:2019wmi}. The quoted range reflects the uncertainty
in the nuclear matrix elements required to translate the half-life $T_{1/2}^{0\nu}$ into $|m_{ee}|$.}
\label{nonenubb}
\end{table}

Few experimental anomalies are still looking for more observational support or a coherent theoretical interpretation.
These include: i) the so-called reactor anomaly \cite{Mention:2011rk}, i.e.\ the evidence for disappearance of electron antineutrinos 
in short baseline experiments; ii) the Gallium anomaly \cite{Abdurashitov:1998ne,Abdurashitov:2005tb,Kaether:2010ag}, i.e.\ the observed deficit in the Gallium radioactive source experiments;
iii) the indications for $\nu_\mu\to \nu_e$ conversion from the LSND \cite{Aguilar:2001ty} and MiniBoone \cite{Aguilar-Arevalo:2018gpe} experiments. Taken at face value, these effects do not fit the standard
framework with three light neutrinos and explanations invoking a fourth sterile neutrino have been adopted. 
Even in such an extended scheme the anomalies do not find a coherent interpretation,
due to the tensions between appearance and disappearance data \cite{Dentler:2018sju}, indicating either
the need for a less minimal framework or the invalidation of some of the experiments.
While the discovery of a sterile neutrino would represent a major result of the current experimental activity
and a non-trivial challenge for its interpretation in the context of the flavour puzzle, here we will assume a low-energy framework
with three light active neutrinos and CPT invariance. New states are not excluded, but are assumed to be heavy,
allowing for an effective description of the current experiments where only the light degrees of freedom take action.

There are few theoretical tools allowing a quantitative and predictive description of neutrino mass and mixing parameters. The focus of this review is on flavour symmetries, one of the most appealing options, given the role that symmetries have played in accounting for the properties of fundamental interactions. 
The idea that relations among mass parameters can be enforced by symmetries is an old one. The most predictive case is represented by exact symmetries, a prototype of which is gauge invariance in quantum electrodynamics,
guaranteed only if the photon is massless. 
Regrettably, exact symmetries do not apply to fermion masses and mixing angles. 
For example, the SM Yukawa couplings break the large non-abelian global symmetry of the quark gauge interactions, down to the baryon number and to the global hypercharge transformations, which 
provide no restrictions to mass parameters.
The lepton sector follows a similar fate and a realistic description of fermion masses should necessarily rely on approximate symmetries. As a consequence, breaking terms are crucial 
to determine the correct pattern of masses and mixing angles. Moreover, 
in interesting cases, flavour symmetries are realized far from the exact phase, with symmetry breaking effects playing a leading role. This feature makes difficult to single out a baseline model or a unique candidate for
the flavour group. 

For these reasons a large part of this review is devoted to a general discussion of
symmetries and symmetry breaking, independently from their specific realization in model building. 
We provide a general classification of flavour symmetries compatible with a
local, gauge invariant and relativistic quantum field theory. 
We distinguish symmetries acting linearly or nonlinearly in field space. In particular, dealing with
the non-linear case, we go beyond the well-established Callan-Coleman-Wess-Zumino formalism \cite{Coleman:1969sm,Callan:1969sn},
which does not cover the relevant case of discrete symmetries. We offer to the reader 
a more general description, suitable to accommodate all cases of interest. We also distinguish
symmetries commuting with the Poincar\'e and gauge groups, from those that do not. The latter choice includes CP-like flavour symmetries, that got lot of attention in the recent years, especially in connection with discrete symmetry groups. This classification,
meant to cover not only the lepton sector but the whole fermion area,
is particularly relevant to clearly identify the uncharted directions from the already explored ones. 
Moreover, in our view, it should not be viewed as a formal mathematical exercise 
since it reflects 
important physical aspects of the symmetries in question. For example, CP-like
flavour symmetries are especially efficient in constraining physical phases. Symmetries whose action is non-linear can potentially enhance the
predictive power of the model, being able to relate operators of different dimensionality.

We also examine how symmetry breaking can be efficiently described through the use
of spurions, allowing to capture both the case of explicit and spontaneous breaking. 
We discuss how predictions about the mixing matrix can be viewed as solution to a problem of vacuum alignment. When the vacuum arises from the minimization of an energy density functional, general results are encoded in the space of invariants of the theory and in the structure of its boundaries.
We provide, for the first time in the context of flavour symmetries, a concise review of this important
topic, where the problem of symmetry breaking finds its most natural mathematical formulation.
The rest of our review is devoted to summarize the state of the art in model building,
organized according to our general classification of flavour symmetries. Aware that
this part can be easily become obsolete in a short time, we have emphasized more the
general features of model building, limiting the discussion of specific models to few examples
per each category. We also comment on the possibility of extending each type of symmetries
from the lepton sector to the quark one.
The number of possibilities offered to model building is huge and many of them have already been surveyed in excellent reviews \cite{Altarelli:2010gt,Ishimori:2010au,Smirnov:2011jv,King:2013eh,King:2014nza,King:2017guk,Petcov:2017ggy,Xing:2019vks}. 

Of course flavour symmetries do not exhaust all possible quantitative approaches to the flavour puzzle. For example, mass and mixing low-energy parameters can satisfy fixed-point relations, originating from the renormalisation group flow of generic input parameters
defined at a very high energy scale. Infrared stable fixed points of the renormalization
group equations for Yukawa couplings and fermion masses have been studied long ago. In the lepton sector, no acceptable relations among the mixing angles
have been found in the CP-conserving regime \cite{Chankowski:2001mx}, 
while in the CP-violating regime the only viable constraint \cite{Casas:1999tg} requires a strong degeneracy between the closest neutrino masses.

Another possibility is offered by the mechanism of radiative mass generation, when a combination of mass parameters that accidentally vanishes at the classical level, gets a non-vanishing calculable contribution at higher orders of perturbation theory.
In particular, it has been suggested that the lightness of neutrinos might arise in this context from 
loop suppression factors.
States running in the internal lines of the loop can be sufficiently light to be probed at existing facilities, at variance with the typically heavy states of the see-saw mechanism. 
The new states can also lead to lepton flavour violation, potentially observable at present or future high-intensity facilities. 
We briefly comment on such possibility when discussing the mechanism for neutrino masses.

This review consists of seven sections. After recalling the possible origin of neutrino masses in section II, in section III we present a general classification of flavour symmetries and discuss general aspects of symmetry breaking.
The following sections, IV, V and VI provide a more specific
description and several illustrative examples of the type of symmetries classified in section III. Finally in section VII we summarize our personal thoughts on the subject. 
There are many related topics that we have only briefly mentioned or deliberately left out of this work. This list is long and includes 
extension to the quark sector within grand unified theories or string theory, realization in the context of extra dimensions, relation to lepton flavour violation searches and leptogenesis, mathematical aspects such as group theory. We refer the reader to the mentioned literature.

\section{Origin of neutrino masses}
\subsection{Neutrino masses and the Standard Model}
\label{sec:nuSM}

Neutrinos are massless in the Standard Model, according to its usual definition as a renormalizable theory involving left-handed neutrinos only. While such a prediction is certainly at odds with everything we have learned about neutrinos in the past decades, and it represents an incontrovertible reason to extend the SM, it can at the same time be considered as a success of the SM, to the extent to which it offers a basis for the understanding of the peculiar smallness of neutrino masses. 

The SM gauge structure is indeed crucial in forbidding neutrinos from getting a mass. In the effective theory below the electroweak scale, with $\text{SU(3)}_c\times \text{U(1)}_\text{em}$ as gauge group, both the charged fermions and the neutrinos are allowed to get a mass. Therefore, the peculiar size of neutrino masses is not addressed by the gauge structure in this case.

The neutrino mass term allowed in the $\text{SU(3)}_c\times \text{U(1)}_\text{em}$ theory is of Majorana type and, as such, it violates the total lepton number. The fact that such a mass term is not generated by the SM completion can therefore be seen as a consequence of the accidental conservation of lepton number in the SM (or from direct inspection: no renormalizable interaction gives rise to neutrino masses after electroweak symmetry breaking, due to the absence, so far, of right-handed neutrinos).

Accidental symmetries are not imposed by hand, they just happen to be global symmetries of the most general renormalizable Lagrangian invariant under the given gauge transformations. The SM turns out to have four independent accidental symmetries, associated to the conservation of baryon number $B$ and of the three individual lepton numbers $L_i$. The total lepton number $L = \sum_i L_i$ is therefore also accidentally conserved. As we will see, the SM accidental symmetries are a residual subgroup of the $\text{U(3)}^5\times \text{U(1)}_H$ global symmetry that the SM acquires when its Yukawa couplings are set to zero, which in turn underlies the very idea of flavour symmetries. 

The emergence of lepton number as an accidental symmetry is one of the notable features of the SM. On the one hand, it predicts the suppression of lepton number violating processes in Nature (thus providing a nice zeroth order approximation for the smallness of Majorana neutrino masses: $m_\nu = 0$). On the other hand, since lepton number is not postulated to be a fundamental symmetry, small lepton number violating effects are not forbidden. This is welcome, as a tiny (but conceptually and practically important) breaking of lepton (and baryon) number takes place even within the SM, because of non-perturbative effects~\cite{tHooft:1976rip,tHooft:1976snw}. Moreover, it is welcome because it leaves room for a small breaking of lepton number, and in particular for small Majorana neutrino masses, originating from possible UV completions of the SM. Grand Unified Theories (GUTs), for example, explicitly break lepton (and baryon) number and are therefore not  compatible with enforcing the conservation of lepton number by hand.

\subsection{Origin of neutrino masses: standard framework}
\label{sec:standard}

The previous subsection lays the ground for the standard understanding of the origin and size of neutrino masses. Such an understanding is based on the sole hypothesis that the new ingredients needed to be added to the SM in order to account for neutrino masses, whatever they are, lie at a scale significantly larger than the electroweak scale. 

If that is the case, effective field theory (EFT) ensures that it is possible to account for the effect (including neutrino masses) of such new ingredients at lower scales by adding to the SM  Lagrangian additional non-renormalizable, or ``effective'', operators. The non-renormalizable Lagrangian one obtains is called the ``SM effective field theory'' (SMEFT). 

The effective operators are suppressed by powers of the scale of the new physics generating them, the ``cutoff'' $\Lambda$. The perturbative validity of the theory is limited to energies well below the cutoff. There, the impact of an effective operators is suppressed by a factor $(E/\Lambda)^{D-4}$, where $D$ is the dimension of the operator in energy. Therefore the most relevant operators are in principle the lowest dimensional ones. In the $E\ll \Lambda$ regime, the theory can be renormalized with a finite number of counterterms order by order in an expansion in the operators dimension. 

The effective operators contain SM fields only and only need to obey the SM gauge invariance, so that no actual knowledge of the physics originating them is required in order to account for its low energy effect .

Interestingly, the single lowest dimensional operator allowed in the SMEFT, the $D=5$ Weinberg operator~\cite{Weinberg:1979sa}
\begin{equation}
\frac{c_{ij}}{2\Lambda} (l_i H) (l_j H) \;,
\label{eq:Weinberg}
\end{equation}
is precisely what is needed to account for neutrino masses. In the above expression, $l_i$, $i=1,2,3$, are the lepton doublets and $H$ is the Higgs doublet. There, and below, $\text{SU(2)}_L$-invariant contractions of the doublet indices are understood (by the $2\times 2$ antisymmetric tensor in \eq{Weinberg}). The splitting of the coefficient into a dimensionless numerator $c_{ij}$ and a dimensionful denominator $\Lambda$ is of course arbitrary. $\Lambda$ is supposed to represent the scale of the new degrees of freedom whose virtual exchange gives rise to the operator and $c_{ij}$ is supposed to group the coupling, mixings, loop factors involved, which are supposed not be larger than $\ord{1}$ in a perturbative regime and $\ord{4\pi}$ in a non-perturbative one. 

The origin of the operator in \eq{Weinberg} must be associated to lepton number violating physics, as the operator itself breaks lepton number by two units. It also breaks $B-L$, an important ingredient for high scale baryogenesis~\cite{Kuzmin:1985mm}. After electroweak symmetry breaking, the operator gives rise to a neutrino Majorana mass term in the form
\begin{equation}
\frac{m_{ij}}{2} \nu_i \nu_j,
\label{eq:Majorana}
\end{equation}
with
\begin{equation}
m_{ij} = c_{ij} \frac{v^2}{\Lambda} \;,
\label{eq:mWeinberg}
\end{equation}
where $v$ is the electroweak scale, $v = |\vev{H}| \approx 174\GeV$. 

The peculiarity of neutrino masses is now elegantly accounted for by their different dependence on the electroweak scale. While charged fermion masses are linear in $v$, neutrino masses turn out to be quadratic in $v$ and thus suppressed by a factor $v/\Lambda$ with respect to the former. Their suppression is attributed to the heaviness of the scale $\Lambda$ at which lepton number is violated. If $m_h$ is the heaviest neutrino mass and $c_h$ the heaviest eigenvalue of the matrix $c_{ij}$, we have
\begin{equation}
\Lambda \approx 0.5 \times 10^{15} \GeV \, c_h \fracwithdelims{(}{)}{0.05\eV}{m_h} \;.
\label{eq:Lambda}
\end{equation}
The scale $\Lambda$ of the new physics associated to neutrino masses can be as large as $10^{15}\GeV$, hence hinting a possible connection with GUT physics, or much smaller, if the couplings $\lambda_{UV}$ on which $c_h$ depends are small. As usually $c_h$ depends quadratically on $\lambda_{UV}$, UV couplings of order $10^{-2}$ are sufficient to bring $\Lambda$ down to $10^{11}\GeV$.

While the Weinberg operator is the lowest dimensional, and therefore in principle most relevant, effective operator giving rise to neutrino masses, higher order operators may become relevant if the former turns out to be suppressed. On the other hand, higher order operators contributing to neutrino masses just contain additional pairs of conjugated Higgs fields. Therefore, any symmetry suppressing the Weinberg operator would also suppress those higher order operators. Barring an accidental suppression of the former, the latter have hardly a chance to dominate. The situation changes in extensions of the SM Higgs sector by a singlet and/or a second doublet. Then it is possible to define symmetries forbidding the $D=5$ operator, but not higher order ones~\cite{Gogoladze:2008wz,Babu:2009aq,Bonnet:2009ej}. In such cases, neutrino masses turn out to be suppressed by higher powers of $v/\Lambda$, which lowers the needed scale of $\Lambda$. Higher order operators can also involve new fields that do not get a VEV and still contribute to neutrino masses, if the new field lines close into a loop. If the new fields are heavy, and integrated out, this possibility can still be accounted for in terms of the $D=5$ Weinberg operator (see the paragraph below on its radiative origin).

\subsubsection*{The case for right-handed neutrinos}

While the above framework offers a simple and compelling understanding of the size of neutrino masses, it relies on the absence of a ``right-handed'' counterpart of the SM neutrinos. All the left-handed charged fermions contained in the quark and lepton doublets $q_i = (u_i,d_i)^T$, $l_i = (\nu_i,l_i)^T$ have $\text{SU(2)}_L$ singlet partners $u^c_i$, $d^c_i$, $e^c_i$\footnote{The index $c$ in $f^c$ denotes the charge conjugated of the right-handed component of $f$ in the Dirac spinor formalism, or a left-handed field independent of $f$ in the Weyl spinor formalism.} leading to Dirac masses through the Yukawa interactions $\lambda^U_{ij} u^c_i q_j H + \lambda^D_{ij} d^c_i q_j H^* + \lambda^E_{ij} e^c_i l_j H^* + \text{h.c.}$, so why should not the neutrinos $\nu_i$ also be accompanied by a $\text{SU(2)}_L$ singlet partner $\nu^c_i$, leading to a neutrino Dirac mass term through the Yukawa interaction
\begin{equation}
\lambda^N_{ij} \nu^c_i l_j H + \text{h.c.}\;.
\label{eq:nuYukawa}
\end{equation}
And, if so, what would make neutrino masses peculiar? 

Note that the existence of the singlet neutrinos $\nu^c_i$ is predicted in a number of extensions of the SM providing an understanding for the SM gauge quantum numbers, and thus further motivated. This is the case of extensions based on the left-right symmetric gauge group $G_\text{LR} = \text{SU(3)}_c \times \text{SU(2)}_L \times \text{SU(2)}_R \times \text{U(1)}_{B-L}$, on the Pati-Salam  group $G_\text{PS} = \text{SU(4)}_c \times \text{SU(2)}_L \times \text{SU(2)}_R$, or on the grand unification group SO(10). 

The special size of neutrino masses can be accounted for even in the presence of singlet partners for the neutrinos as well, as such singlet neutrinos carry their own peculiarity. In order for them to give rise to a neutrino mass term through gauge invariant Yukawa interactions, the fields $\nu^c_i$ should be singlets under the whole SM group.\footnote{If the field $\nu^c$ is allowed to have more than one component, it could alternatively be a $\text{SU(2)}_L$ triplet. The argument that follows would still go through, as it is only based on $\nu^c$ being the only fermion in a real representation of the SM group, with all the others belonging to a fully chiral representation.} The SM extensions mentioned above also predict them to be SM singlets. Therefore, the neutrino singlets would be the only fermions allowed to have an explicit, gauge invariant (and lepton number violating) mass term
\begin{equation}
\frac{M_{ij}}{2} \nu^c_i \nu^c_j + \text{h.c.} \;.
\label{eq:M}
\end{equation}
Such a mass term has no ties with the electroweak scale, as it survives in the limit in which the electroweak scale vanishes. Hence, there is no reason why it could not be much heavier than the electroweak scale. If that is the case, the singlet neutrinos represent nothing but a specific (and prototypical) realisation of the very framework discussed above: new degrees of freedom lying at a scale significantly larger than the electroweak scale. It must therefore be possible to account for their effect at the electroweak scale (and below) in terms of effective operators. Indeed, integrating them out (as reviewed e.g.\ in \cite{Altarelli:2004za}) precisely generates the Weinberg operator, with, in a matrix notation,
\begin{equation}
\frac{c}{\Lambda} = - \lambda_N^T \, M^{-1} \lambda^{\phantom{T}}_N \;,
\label{eq:matchingnuc}
\end{equation}
where $\lambda_N$ and $M$ are the parameters in \eqs{nuYukawa} and~(\ref{eq:M}) respectively. The light neutrino masses end up being given by the celebrated seesaw formula~\cite{Minkowski:1977sc,GellMann:1980vs,Yanagida:1979as,Glashow:1979nm,Mohapatra:1979ia} 
\begin{equation}
m_\nu = - m^T_D \, M^{-1} m^{\phantom{T}}_D \;,
\label{eq:seesaw}
\end{equation}
where $m_D = \lambda_N v$ is a Dirac-like neutrino mass term. The advantage of the EFT derivation, compared to the diagonalisation of the $6 \times 6$ matrix of the $\nu^{\phantom{c}}_i + \nu^c_i$ system, is that it allows to organise the computation of potentially large, log-enhanced radiative corrections to the seesaw formula by means of the renormalization group equations. The coefficient of the Weinberg operator is calculated from \eq{matchingnuc} at the singlet neutrino scale and subsequently the Weinberg operator is run down to the electroweak scale. Within the SM, gauge interactions and quark Yukawas only affect (at one loop) the overall neutrino mass scale, while flavour dependent effects from lepton Yukawas are negligible. Sizeable flavour corrections can arise in two Higgs doublet schemes in the large $\tan\beta$ regime, in the presence of an ``unstable''~\cite{Domcke:2016mzc} neutrino mass approximate degeneracy, see for example~\cite{Chankowski:2001mx}. If the heavy neutrinos are hierarchical, threshold effects associated to their sequential decoupling may also be important. 

\subsubsection*{Tree-level origin of the Weinberg operator}
\label{sec:treelevel}

We have seen that neutrino singlets, unless unexpectedly light, represent a specific realisation of the general situation in which the new physics needed to account for neutrino masses lies at a scale significantly higher than the electroweak scale. We can then wonder what is the most general form of the heavy new physics giving rise to the Weinberg operator. A simple and complete answer is found in the assumption that the Weinberg operator is generated at the tree level. In such a case, the virtual heavy states can only have three types of SM quantum numbers, corresponding to type I, type II~\cite{Magg:1980ut,Lazarides:1980nt,Mohapatra:1980yp}\footnote{In~\cite{Schechter:1980gr,Schechter:1981cv}, a scalar triplet VEV directly contributes to neutrino masses, with no see-saw suppression by the triplet mass.}, and type III~\cite{Foot:1988aq} seesaw. We list them below, using the notation $(r_3,r_2,y)$ for the SM gauge quantum numbers, where $r_3$ is the $\text{SU(3)}_c$ representation, $r_2$ is the $\text{SU(2)}_L$ representation, and $y$ is the values of the hypercharge (in units in whih the SM Higgs has $y = 1/2$). 

\begin{description}
\item[Type I]
The virtual messengers are fermions $\nu^c$ with SM quantum numbers $(1,1,0)$, i.e.\ they are SM singlets. This is essentially the case discussed above, with the only variation that the number $n$ of singlet neutrinos is not bound to be three. In order to reproduce both the atmospheric and solar squared mass differences, $n \geq 2$ is needed. The relevant high scale Lagrangian is given by \eqs{nuYukawa} and~(\ref{eq:M}), 
\begin{equation}
-\mathcal{L}_\text{I} = \lambda^N_{kj} \nu^c_k l_j H + \frac{M_{kh}}{2} \nu^c_k \nu^c_h + \text{h.c.} \;,
\label{eq:typeIL}
\end{equation}
where the number of singlet neutrinos is now $n$, the Yukawa $\lambda_N$ is a $n\times 3$ matrix, and the mass term $M$ is a $n\times n$ symmetric matrix. The effective Weinberg operator and the neutrino masses are again given by
\begin{equation}
\frac{c}{\Lambda} = - \lambda_N^T \, M^{-1} \lambda^{\phantom{T}}_N 
\quad \text{and} \quad
m_\nu = - m^T_D \, M^{-1} m^{\phantom{T}}_D \; .
\label{eq:matching1}
\end{equation}
\item[Type II]
The virtual messengers are complex scalars $\Delta_k$, $k=1\ldots n$, with SM quantum numbers $(1,3,1)$, i.e.\ they are $\text{SU(2)}_L$ triplets with hypercharge $Y = 1$. The relevant high scale Lagrangian is
\begin{equation}
-\mathcal{L}_\text{II} = \frac{1}{2} \left(y_{ijk} l_i \sigma_a l_j \Delta^a_k + \mu_{k} H \sigma_a H \Delta^{a*}_k + \text{h.c.} \right) + M^2_{kh} \Delta^{a*}_k \Delta^{\phantom{*}a}_h \;.
\label{eq:typeIIL}
\end{equation}
where  the mass matrix $M^2$ is now hermitian and $\Delta^a$, $a=1,2,3$, are the components of the triplets $\Delta$. Integrating them out gives rise to the Weinberg operator and neutrino masses, with
\begin{equation}
\frac{c_{ij}}{\Lambda} = - y_{ijh} (M^2)^{-1}_{hk} \mu_k
\quad\text{and}\quad
(m_\nu)_{ij} = - v^2 y_{ijh} (M^2)^{-1}_{hk} \mu_k \;.
\label{eq:matching2}
\end{equation}
The role of the cutoff $\Lambda$ is now played by the combination $M^2/\mu$, where $\mu^2$ can be expected to be of the same order as $M^2$. Unlike in the type I (and type III) case, one triplet is in principle sufficient to reproduce both the atmospheric and solar squared mass differences. 
\item[Type III]
This case is similar to type I, but the messengers are now $\text{SU(2)}_L$ triplets. I.e.\ they are fermions $T_k$, $k=1\ldots n$, with SM quantum numbers $(1,3,0)$, and again $n\geq 2$. The relevant high scale Lagrangian is
\begin{equation}
-\mathcal{L}_\text{III} = \lambda^T_{ij} T^a_i l^{\phantom{a}}_j \sigma_a H + \frac{M_{ij}}{2} T^a_i T^a_j + \text{h.c.} \;,
\label{eq:typeIIIL}
\end{equation}
where $T^a$, $a=1,2,3$, are the components of the triplets $T$. Integrating them out generates the Weinberg operator and neutrino masses, with
\begin{equation}
\frac{c}{\Lambda} = - \lambda_T^T \, M^{-1} \lambda^{\phantom{T}}_T 
\quad\text{and}\quad
m_\nu = - m^T_T \, M^{-1} m^{\phantom{T}}_T \; ,
\label{eq:matching3}
\end{equation}
where $m_T = v \lambda_T$. 
\end{description}

\begin{figure}
 \centering
 \includegraphics[width=\textwidth]{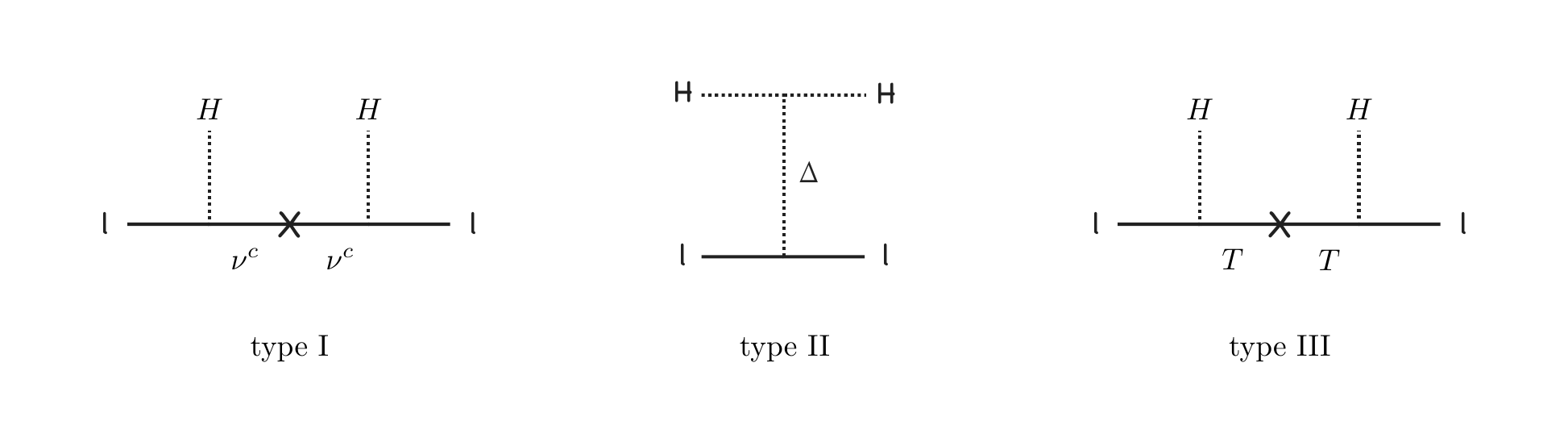}
 \caption{Diagrammatic representation of the three types of seesaw mechanisms. They all give rise to the Weinberg effective operator in \eq{Weinberg} once the intermediate states are integrated out. The crosses denote lepton number violating mass insertions.}
\label{fig:seesaws}
\end{figure}

A simple analysis based on gauge invariance shows that the tree level diagrams in \Fig{seesaws}, corresponding to the three seesaw Lagrangians above, are the only possible ones~\cite{Ma:1998dn}. A complex scalar with quantum numbers $(1,1,1)$, for example, cannot play a role at the tree level, as it does couple to the antisymmetric combination of $l_i l_j$, but not to $h h$. 

\subsubsection*{Radiative origin of the Weinberg operator}

While a tree-level origin of the Weinberg operator is undoubtedly the most appealing option (and the only one with unbroken supersymmetry~\cite{Megrelidze:2016fcs}), the possibility of a radiative origin is not excluded (see~\cite{Cai:2017jrq} for a recent review). Depending on the specific field content of the UV theory, a tree-level origin may not be available, while the Weinberg operator can arise through quantum corrections at the loop level. The topologies of the corresponding Feynman diagrams have been classified up to 2-loop order~\cite{Babu:2001ex,deGouvea:2007qla,Bonnet:2012kz,Angel:2012ug,Sierra:2014rxa} and require at least two new multiplets to play the role of intermediate states~\cite{Law:2013dya}. Once those states are ingrated out, within an effective theory approach, the Weiberg operator is not generated at the tree level. Other lepton number violating operators are, though, and they give rise the the Weinberg one through loops involving SM interactions and fields. The new states can not be far from the EW scale, and the suppression of the neutrino masses compared to the latter is at least partially accounted for by the loop factor $(1/(16\pi^2))^\ell$, where $\ell$ is the loop order at which the diagram arises, if $\ell$ is sufficiently large. 

Such models may be characterized by a possibly interesting phenomenology at colliders and in charged-lepton flavour violation (CLFV) experiments, although their aesthetic appeal does not match the tree-level see-saw one. On the one hand, the suppression of neutrino masses is better accounted for when $\ell$ is relatively large. On the other hand, the increase of $\ell$ leads to a rapid increase of the number of diagrams. The structure and field content of the model is not as constrained as in the tree-level case. On the contrary, a plethora of possibilities are available. Finally, the model parameters often need to be fine-tuned, in order to cope with the present bounds on CLFV and reproduce neutrino masses and mixings. For further information on such class of models, we refer to dedicated reviews~\cite{Boucenna:2014zba,Sugiyama:2015cra}.

\subsection{Lower scale origin of neutrino masses}

As we have seen, effective field theory provides a simple and compelling understanding of the origin and peculiar smallness of neutrino masses, under the sole hypothesis that the new degrees of freedom needed to account for non-vanishing neutrino masses lie significantly above the electroweak scale. Neutrino masses, on the other hand, can also originate well below the electroweak scale. Dirac neutrinos are the prototypical example. The SM neutrinos get in such a case a purely Dirac mass from Yukawa couplings to otherwise massless singlet neutrinos ($M = 0$ in \eq{M}). While the standard framework unavoidably leads to lepton number violating Majorana neutrino masses, Dirac neutrinos conserve lepton number, which offers an opportunity to probe experimentally the origin of neutrino masses. 

Before ending up with $M = 0$ and purely Dirac neutrinos, we shortly consider the intermediate possibility that $M$ does not vanish but it is not significantly larger than the EW scale, so that the SMEFT approach used in \Sect{standard} does not apply. 

If the singlet neutrino masses are not far from the electroweak scale, they can play a role in present of future collider phenomenology~\cite{Deppisch:2015qwa,Antusch:2015mia}. As those masses can in principle be as large as the Planck scale, their proximity to the electroweak scale, about 15 orders of magnitudes smaller, would represent a non-trivial accident. 

In the presence of a single family, a singlet neutrino mass $M \sim \TeV$ requires a neutrino Yukawa coupling as small as
\begin{equation}
\lambda_N \sim 1.3 \times 10^{-6} \left(
\frac{m_\nu}{0.05\eV} \, \frac{M}{\text{TeV}}
\right)^{1/2} \;.
\label{eq:lambdaTeV}
\end{equation}
The smallness of neutrino masses is accounted for by the smallness of $\lambda$, and such a small coupling would make collider effects hardly observable. 

With three families, though, larger Yukawa couplings are allowed if cancellations take place in the seesaw formula. Non-accidental cancellations can be forced by appropriate symmetries, such as lepton number itself,  allowing the large Yukawa couplings while forbidding the neutrino masses~\cite{Kersten:2007vk,Xing:2009in} and can involve additional singlets~\cite{Mohapatra:1986aw,Mohapatra:1986bd,Akhmedov:1995vm,Akhmedov:1995ip,Barr:2003nn,Malinsky:2005bi,Barr:2005ss,Ibarra:2010xw}. The larger Yukawa couplings have then a chance to be probed at colliders. Such symmetric couplings are not anymore directly related to the origin of neutrino masses (and their size), which in this case is instead associated to the symmetry breaking parameters (and their smallness). 

The collider prospects are richer when additional interactions, besides those directly related to neutrino masses, provide additional production or detection channels. This is the case when the heavy states feel gauge interactions. For example, the SM singlet neutrinos can be charged under extensions of the SM group containing an $\text{SU(2)}_R$ factor~\cite{Keung:1983uu,Das:2012ii,Nemevsek:2011hz}. Even sticking to the SM group, the components of $\Delta$ and $T$ (in type II and III seesaw respectively) charged under the SM can enrich the collider phenomenology~\cite{Akeroyd:2005gt,Han:2007bk,delAguila:2008cj}. 

The collider bounds on the charged component of $\Delta$ and $T$ prevent the type II and type III seesaw from being extrapolated below the electroweak scale (barring an unnatural splitting among neutral and charged components). On the other hand, the singlet neutrino mass in type I seesaw can be arbitrarily small, or zero, as argued above. 

In the intermediate regime in which the singlet neutrino masses are lighter than the electroweak scale, but significantly larger than the energy of the relevant neutrino processes, it is still possible to integrate out the singlet neutrinos. As the SM group is badly broken in such a regime, it is appropriate in this case to start from the $\text{SU(3)}_c \times \text{U(1)}_\text{em}$ invariant Lagrangian~\cite{Altarelli:1999gu}
\begin{equation}
m^D_{ij} \nu^c_i \nu_j + \frac{M_{ij}}{2} \nu^c_i \nu^c_j + \text{h.c.} \;,
\label{eq:seesawLE}
\end{equation}
which still leads of course to the seesaw formula in \eq{seesaw}. 

Otherwise, if the singlet neutrinos are light enough to be produced, or not too far from that, a full treatment of the neutrino sector, including the sterile states and their mixing with the active ones, is necessary. In such a regime, the size of neutrino masses requires the relevant parameters to be particularly small. For example, singlet neutrinos in the $\text{eV}$ range (a motivated possibility, see e.g.~\cite{Giunti:2019aiy} for a review) require the Yukawa couplings $\lambda_N$ and the singlet masses $M$ in \eqs{nuYukawa} and~(\ref{eq:M}) to be as small as
\begin{equation}
\lambda_N \lesssim 10^{-11} \;,
\qquad
M \lesssim 10^{-18} M_\text{Pl} 
\label{eq:adhoc}
\end{equation}
(and imply a mild fine-tuning, keeping Dirac and Majorana neutrino masses within one or two orders of magnitude). 

Finally, if the Majorana mass term $M$ is even smaller than the Dirac mass term, solar neutrino experiments force $M$ to be well below the heavier active neutrinos mass range~\cite{deGouvea:2009fp}, and we approach the Dirac neutrino limit, in which $M = 0$. In such a limit, lepton number is conserved in the neutrino sector, and the only role of the sterile fields is to pair to the active ones in the Dirac mass term. The corresponding degrees of freedom can hardly be observed, as their production and detection with an energy $E$ is suppressed by a factor $m_\nu/E$. 

In the cases considered in this subsection, the size of neutrino masses is accounted for by the smallness, often striking, of Lagrangian parameters. While such a smallness may seem quite ad hoc, ideas are available to account for it. The suppression of the Majorana mass term can be associated to the approximate or exact conservation of lepton number. This comes at the price of giving up one of the successes of the SM, as the approximate conservation of lepton number observed in Nature would not be accounted for by accidental symmetries anymore. Lepton number needs to be enforced as a symmetry by hand, with the drawbacks discussed in \Sect{nuSM}. The smallness of the Yukawa couplings can instead be given a dynamical origin. Small, non-zero couplings can arise through the spontaneous breaking of a symmetry forbidding them~\cite{Chikashige:1980ui,Gelmini:1980re,Georgi:1981pg,Chacko:2003dt,Chen:2006hn,Gu:2009hu}, or from a more fundamental theory living in more than four dimensions~\cite{Dienes:1998sb,ArkaniHamed:1998vp,Dvali:1999cn,Mohapatra:1999zd,Barbieri:2000mg,Lukas:2000wn,Lukas:2000rg,Grossman:1999ra,GonzalezGarcia:2002dz}.

\section{Symmetries: general considerations}
\subsection{The flavour puzzle}
\label{sec:basics}

\newcommand{\Gmax}{G_\text{max}}

Having reviewed possible origins of the neutrino masses and their overall scale, we now come to the main subject of this review: the origin, if any, of the pattern of lepton masses and mixings, i.e.\ of the flavour structure of the lepton mass matrices, which is part of the so called SM flavour puzzle. 

The flavour puzzle in the SM, here extended to include a source of neutrino masses, has two aspects. The first one is the existence of three fermion families replicating the same set of gauge quantum numbers. Or, equivalently, the invariance of the SM gauge Lagrangian under a global $\text{U(3)}^5$ global symmetry, where each U(3) factor mixes the three families of fermions with identical gauge quantum numbers: $q_i$, $u^c_i$, $d^c_i$, $l_i$, $e^c_i$, $i=1,2,3$. The Higgs Lagrangian is invariant under a further $\text{U(1)}_H$ rephasing of the Higgs doublet field. Thus $\Gmax \equiv \text{U(3)}^5 \times \text{U(1)}_H$ is the maximal group of global SM field transformations commuting with the actions of the Poincar\'e and gauge groups. It includes the hypercharge global transformations. In SM extensions, $\Gmax$ can be larger, if the matter field content is extended (singlet neutrinos for example, or additional Higgs fields); or smaller, if the gauge group is extended. 

If the source of neutrino masses is neglected, $\text{U(3)}^5 \times \text{U(1)}_H$ is explicitly broken by the SM Yukawa interactions to the four SM accidental symmetries --- the U(1) transformations associated to the individual lepton numbers $L_e$, $L_\mu$, $L_\tau$ and the total Baryon number $B$ --- and to the hypercharge global transformations. The accidental symmetries are anomalous, unless they are combinations of $B-L_e/3$, $B-L_\mu/3$, $B-L_\tau/3$. If neutrino masses are accounted for at the weak scale by the Weinberg operator, the three individual lepton numbers are also broken and only $B$ survives at the perturbative level, though it is anomalous. If neutrino masses are of Dirac type, i.e.\ they are accounted for at the weak scale by Yukawa couplings to otherwise massless right-handed neutrinos, both $B$ and $L$ survive from an initial $\Gmax = \text{U(3)}^6 \times \text{U(1)}_H$, and only the $B-L$ combination is non-anomalous.

The second aspect of the flavour puzzle is the peculiar pattern of fermion masses and mixings originating from the explicit breaking of $\text{U(3)}^5 \times \text{U(1)}_H$. The masses of the three families of charged fermion masses turn out to be hierarchical and the quark mixing is small. Lepton mixing is instead large and at least two neutrino masses are separated by less than an order of magnitude.

The two aspects of the flavour puzzle may be related. The fact that the flavour Lagrangian breaks an underlying $\text{U(3)}^5 \times \text{U(1)}_H$ symmetry, manifest in the gauge Lagrangian, may suggest that it originates from the spontaneous breaking of the above group, or of one of its subgroups $G \subseteq \text{U(3)}^5 \times \text{U(1)}_H$. This is the idea underlying theories based on flavour symmetries~\cite{Froggatt:1978nt}, where $G$ is called the flavour group. The action of $G$ is traditionally assumed, as above, to be linear and to commute with gauge and Poincar\'e transformations. On the other hand, new avenues evading such an assumption have been recently considered. Correspondingly, denoting by $\psi_i$ a generic set of matter fields, in the following we will consider three types of symmetries. 
\begin{itemize}
\item[$\mathbf{1.}$]
The action of $G$ is linear (thus unitary, in order to preserve canonically normalised kinetic terms) and commutes with gauge and proper Poincar\'e transformations:
\begin{equation}
g\in G: \quad
\psi_i(x) \to U_\psi(g)_{ij} \psi_j(x) \;.
\label{eq:generalaction1}
\end{equation}
In such a case, $G$ is a subgroup of $\Gmax$ and $U_\psi(g)$ is a unitary representation of $G$. Such a standard framework will be reviewed below and in section~\ref{sec:flavour}.
\item[$\mathbf{2.}$]
The action of $G$ is linear, but it does not commute with proper Poincar\'e and/or gauge transformations. The case in which flavour and Poincar\'e transformations do not commute leads to symmetries in the form $G=G_f\rtimes \text{CP}$, where $G_f$ is a subgroup of $\Gmax$ as in the previous case:
\begin{equation}
g\in G_f: \quad
\psi_i(x) \to U_\psi(g)_{ij} \psi_j(x) \hspace{2cm} \psi_i(x) \xrightarrow{\text{CP}} X_{ij} \psi^*_j(x) \;.
\label{eq:generalaction2}
\end{equation}
Here $U_\psi(g)_{ij}$ and $X_{ij}$ are unitary representations of $G_f$ and CP, respectively.
This scenario will be reviewed below and in section~\ref{sec:CP}. 
The case in which $G$ commutes with Poincar\'e, but not with gauge transformations has received less attention so far~\cite{Reig:2017nrz}.
\item[$\mathbf{3.}$]
The action of $G$ is nonlinear, it commutes with the gauge group and with proper Poincar\'e transformations. 
$G$ is not necessarily a subgroup of $\Gmax$. In the realization we will consider, the framework includes an additional scalar sector, typically consisting of fields $\tau$ singlet under the gauge group.
\be
g\in G: \quad \tau\to f_g(\tau) \hspace{2cm} \psi_i(x) \to U_\psi(g;\tau)_{ij} \psi_j(x)\;,
\ee
where $f_g(\tau)$ and $U_\psi(g;\tau)_{ij}$ describe the nonlinear realization of $G$ on $\tau$ and $\psi_i(x)$, respectively. This case will be reviewed below and in section~\ref{sec:modular}.
\end{itemize}
A fourth possibility, also discussed in in section~\ref{sec:modular}, arises by combining cases 2. and 3. above.

\subsection{Flavour symmetry group and representation}

We first consider flavour models based on a flavour group $G$ whose action on fields is linear and commutes with Poincar\'e and gauge transformations. $G$ then acts on the flavour indices of each set of fields $\psi_i$ sharing the same Lorentz and gauge quantum numbers:
\begin{equation}
g\in G: \quad
\psi_i(x) \to U_\psi(g)_{ij} \psi_j(x) \;.
\label{eq:generalaction}
\end{equation}
The representation $U_\psi(g)$ is unitary, as the kinetic terms are assumed to be canonically normalised. Moreover, gauge fields must be invariant under $G$, and the action of $G$ on the full set of matter fields can (and will) be assumed to be faithful without loss of generality. Therefore, $G$ can be identified with a subgroup of the unitary internal transformations. More precisely, $\Gmax \subseteq \prod_r U(n_r)$, where $n_r$ is the number of identical copies of each irreducible representations $r$ of the Poincar\'e and gauge groups on matter fields. The Lagrangian is assumed to be invariant under the action of $G$, and this constrains its flavour structure. The symmetry may be spontaneously broken by a set of scalar fields $\phi$ called ``flavons'', or explicitly broken. 

Different types of flavour groups can be considered: $G$ can be a Lie group or a discrete group; abelian or non-abelian; simple or non-simple; it can be assumed to be a symmetry or arise accidentally~\cite{Ferretti:2006df}; it can act rigidly on the fields or it can be gauged. In the case of gauge groups, proper care should be taken of anomalies, possibly cancelling them by adding an appropriate heavy field content. Most often, the scale at which $G$ is spontaneously broken is taken to be significantly higher than the weak scale. As a consequence, the flavons are bound to be SM singlets (they can however transform non-trivially under extensions of the SM group). 

Flavour symmetry breaking at the EW scale or below faces a number of challenges. If $G$ is gauged, constraints from flavour-changing neutral current (FCNC) processes set a lower bound on the mass of the corresponding gauge bosons, and therefore on the breaking scale. If $G$ is a non-anomalous global Lie group, its spontaneous breaking gives rise to massless Goldstone bosons, which must be then sufficiently weakly coupled to SM fields. This is the case for example if the coupling is mediated by sufficiently heavy degrees of freedom. Or, in the effective-theory description, if they couple through non-renormalisable interactions suppressed by a sufficiently heavy scale. The heavy fields mediating flavour breaking can themselves be a source of FCNC. The scale at which $G$ is broken is then again also bound to be correspondingly large. The same argument applies if $G$ is anomalous, unless would-be Goldstone bosons (and therefore the flavour breaking scale) are heavy enough. In the case of the spontaneous breaking of finite groups, a further constraint comes from the need to avoid domain walls~\cite{Riva:2010jm,Antusch:2013toa,Chigusa:2018hhl}. Still, relatively low scales of flavour breaking can be achieved even in the case of gauged models~\cite{Grinstein:2010ve}. The possibility that the flavour symmetry is broken together with the EW symmetry by means of Higgs doublets has also been considered~\cite{Grimus:2003kq,Ma:2007ia,Morisi:2009sc,Morisi:2011pt}. Here we will consider the safest case in which the breaking of the flavour symmetry is due to SM singlets above the EW scale.

\subsection{Exact flavour symmetries}
\label{sec:exact}

We first dismiss the possibility that the flavour symmetry be exact. This is important also because it shows that no (overall) exact unbroken subgroup can survive the breaking of the flavour symmetry. To begin with, we consider the effective description of neutrino masses through the Weinberg operator. The flavour group acts in the lepton sector through two unitary representations of $g\in G$, one on the leptons doublets $l_i$ and one on the lepton singlets $e^c_i$
\begin{equation}
\begin{aligned}
 l_i &\to \rep_l(g)_{ij} l_j \\
 e^c_i &\to \rep_{e}(g)_{ij} e^c_j \;.
\end{aligned} 
\label{eq:repL}
\end{equation}
The Higgs field could in principle also transform under $G$, but its transformation can, without loss of generality, be reabsorbed in those of $l_i$ and $e^c_i$, and we will therefore neglect it.\footnote{This is not necessarily true in extensions of the SM Higgs sector with two or more Higgs fields.} 

If the flavour symmetry was not broken, the invariance of the lepton flavour Lagrangian, 
\begin{equation}
\lambda^E_{0ij} e^c_i l_j H^* + \frac{c_{0ij}}{2\Lambda} (l_i H) (l_j H) \;,
\label{eq:effective}
\end{equation}
would constrain the couplings $\lambda^E_{0ij}$ and $c_{0ij}$, or equivalently the charged fermion and neutrino mass matrices $M^0_E$ and $m^0_\nu$, as follows:
\begin{equation}
\begin{aligned}
M^0_E &= \rep_e(g)^T M^0_E \, \rep_l(g) \\
m^0_\nu &= \rep_l(g)^T m^0_\nu \, \rep_l(g)  
\end{aligned}
\label{eq:mLinvariance}
\end{equation}
for any $g\in G$. The index ``0'' stresses that the lepton couplings and mass matrices are assumed here to be exactly symmetric under $G$. It turns out that the above constraints can lead to fully viable mass matrices (i.e. associated to three non-vanishing charged lepton masses, three non-degenerate neutrinos, and three non-vanishing mixing angles) only if the representation on the lepton doublets is trivial, $U_l(g) = \pm \mathbf{1}$. The representation on the $e^c_i$ fields must also be trivial and identical to the one on the lepton doublets. In other words, the only accidental symmetry of the SM lagrangian augmented by the Weinberg operator is a $\mathbf{Z}_2$. The argument is simple, and is best formulated in the charged lepton mass basis, in which $M^0_E$ is diagonal and positive. The charged lepton masses relegate $G$ to be a subgroup of $\text{U(1)}_e \times \text{U(1)}_\mu \times \text{U(1)}_\tau$, the three lepton number U(1)'s: as $\rep_l$ and $\rep_{e^c}$ must commute with $(M^0_E)^2$, which is non-degenerate, $\rep_l$ and $\rep_{e^c}$ must both be diagonal matrices of phases; as $M^0_E$ is non-singular, \eq{mLinvariance} forces $\rep^{\phantom{\*}}_{e^c} = \rep_l^*$. The PMNS matrix further reduces $G$ to be a subgroup of the total lepton number U(1): inserting $m^0_\nu = U^* (m^0_\nu)_\text{diag} U^\dagger$, where $U$ is the PMNS matrix, in \eq{mLinvariance}, we see that the combination $U^\dagger \rep_l U^{\phantom{\dagger}}$ commutes with $((m^0_\nu)_\text{diag})^2$ and must also be a diagonal matrix of phases. Since all elements of the PMNS matrix are non-vanishing, this means that $\rep_l$ is just an overall phase, i.e.\ $G$ acts as a subgroup of the total lepton number. Finally, the Majorana nature of the neutrino operator only allows the $\mathbf{Z}_2$ subgroup, as it can be shown by substituting $\rep_l = e^{i\phi} \mathbf{1}$ in \eq{mLinvariance}.
Needless to say, a trivial representation such as $\rep_l(g) = \rep_{e^c}(g) = \pm \mathbf{1}$ does not constrain at all lepton masses and mixings, as any $M_E$ and $m_\nu$ would satisfy \eq{mLinvariance}. An accurate non-trivial description of lepton flavour thus requires a (spontaneously) broken flavour symmetry. Moreover, the flavour symmetry should be fully broken. No residual non-trivial subgroup should survive the breaking, except possibly the trivial $\mathbf{Z}_2$ above. The same conclusion holds if the flavour symmetry constrains the renormalizable theory from which the Weinberg operator originates, provided that the heavy fields stay heavy in the exactly symmetric limit (see below). This is because \eq{mLinvariance} still holds, as a consequence of the invariance of the full theory. 

The above assumes a high-scale origin of neutrino masses. In the paradigmatic caveat of Dirac neutrinos masses originating from Yukawa couplings to three right-handed neutrinos, the analysis is different but the conclusion is the same. The only possible exact flavour symmetry in the lepton sector is in this case the total lepton number U(1), or one of its subgroups. As above, such a flavour group would not constrain at all lepton masses and mixings, as any form of the lepton mass matrices would be allowed.

Finally, the above considerations extend to the quark sector. The only allowed exact symmetry is in that case the total Baryon number. The latter however does not provide any constraint on the quark mass matrices.

\subsection{Symmetry Breaking}
\label{sec:SB}
Having to abandon the idea that lepton masses and mixing angles can be inferred 
from an exact flavour symmetry, the usefulness of the whole approach relies very much on 
the knowledge of breaking effects. In general we can distinguish between an explicit breaking,
where the nature of the breaking terms is unrelated to the dynamics of the system,
and a spontaneous breaking originating from the non-invariance of the vacuum state.
Typically the spontaneous breaking offers better chances in terms of predictability, especially
if some dynamical requirement, like the minimization of the energy density of the system,
is invoked to select the vacuum of the theory. There are however exceptions to this
general trend. Also the case of explicit breaking can retain some predictability,
if breaking terms are not completely arbitrary. Actually, to some extent, the two cases can 
be described within the same formalism. Consider, for example, the charged lepton
Yukawa coupling  $\lambda^E_{ij} e^c_i l_j H^* + \text{h.c.}$ and assume that the singlets
$e^c$ and the doublets $l$ transform according to unitary representations $r_{e^c}$ and $r_{l}$
of the flavour group $G$. It is useful to write the Yukawa coupling in the form:
\be
\lambda^E_{ij} e^c_i l_j H^*=\sum_{I\alpha} S^{I}_\alpha (\Gamma^I_{ij\alpha} e^c_i l_j) H^*
\label{SB1}
\ee
where the combinations $(\Gamma^I_{ij\alpha} e^c_i l_j)$ ($\alpha=1,\ldots ,d_I$) transform
in the irreducible representations $r_I$ (of dimension $d_I$) of the group $G$ occurring in the decomposition of the tensor product $r_{e^c}\otimes r_l$. In case of $N_f$ fermion generations we have the obvious constraint $\sum_I d_I=N_f^2$ and $\Gamma^I_{ij\alpha}$ are Clebsch-Gordan coefficients. The Yukawa interaction can be seen as an invariant of the flavour group, provided $S^{I}_\alpha$ are interpreted as spurions transforming in the conjugate representation $\bar r_I$. 
Arbitrary Yukawa couplings $\lambda^E_{ij}$ are traded by arbitrary spurions $S^{I}_\alpha$
and at this stage we have no benefit. However, in model building we can complement the above decomposition by some additional assumptions about the set of allowed spurion representation, their size and relative orientation in flavour space and thus gather information on the pattern of  $\lambda^E_{ij}$, through the
relation $\lambda^E_{ij}=\sum_{I\alpha} S^{I}_\alpha \Gamma^I_{ij\alpha}$. 

In general the model is specified by the gauge group $G_g$ and the flavour group $G$, together with the
field content which includes matter fields, spurions and their representations under $G_g$ and $G$.
To cover the general case
where the fields $S^{I}_\alpha$ in eq.~(\ref{SB1}) are functions of some fundamental $G$-multiplet,
$S^{I}_\alpha=S^{I}_\alpha(\varphi)$, we will denote the set of allowed spurions by $\varphi$.
In the context of flavour symmetries such spurions are nothing but the flavons.
They transform under some (possibly reducible) representation $r_\varphi$ of the group $G$.
A common, but not mandatory, choice is to assume that spurions $\varphi$ are singlets under the gauge group.
The Yukawa couplings $\lambda^E_{ij}(\varphi)$ become functions of the spurions $\varphi$,
constrained by the flavour symmetry. If they can be expanded in powers of $\varphi$, they assume the form:
\be
\lambda^E_{ij}(\varphi)= \lambda^E_{0ij}+ \lambda^{E\alpha}_{1ij} \varphi_\alpha+ \lambda^{E\alpha\beta}_{2ij}\varphi_\alpha \varphi_\beta +\ldots
\label{SB2}
\ee
and the corresponding interactions are given by:
\be
 e^c\lambda^E(\varphi)l H^*= (e^c l)_1 H^*+ (e^c l \varphi)_1 H^*+(e^c l \varphi\varphi)_1 H^* + \ldots 
\label{SB3}
\ee
where flavour indices are understood and $(\cdot)_1$ stands for a $G$-invariant combination:
$(e^c l)_1=e^c_i \lambda^E_{0ij} l_j$, $(e^c l\varphi)_1=e^c_i \lambda^{E\alpha}_{1ij} \varphi_\alpha l_j$
and so on. This type of description is equally good for both non-dynamical spurions
and for new dynamical degrees of freedom described by the fields $\varphi$. In the first case we reproduce an
explicit breaking of $G$, while in the second case the breaking is spontaneous, being related to the
VEV of $\varphi$. In the above description $\varphi$ are dimensionless.
Fields with canonical dimensions are easily recovered by the replacement $\varphi\to\varphi_\text{CD}/\Lambda$, where $\Lambda$ stands for a new physical scale related to flavour dynamics.
Then the expansion of eq.~(\ref{SB3}) contains operators of growing dimensionality providing, in the spirit of an EFT,
a low-energy description of the flavour sector valid at energy scales much lower than $\Lambda$. 
The scale $\Lambda$ controlling the spurion expansion does not necessarily coincide with 
that introduced in eq.~(\ref{eq:Weinberg}), which breaks the lepton number $L$.
Operators of high dimensions can be helpful to describe
light fermions, if the expansion parameter $\vev{\varphi}$ is sufficiently small.

As an example \cite{Linster:2018avp}
we take $G=\text{U(2)} \sim \text{SU(2)}\otimes \text{U(1)}$ and let the lepton fields transform as in table \ref{tab_ex1_1}.
\begin{table}[h!] 
\centering
\begin{tabular}{|c||c|c|c|c||c||c|c|}
\hline
$G$&$e^c_3$&$e^c_a$&$l_3$&$l_a$& $H$&$\varphi_1$&$\varphi_2$\rule[-2ex]{0pt}{5ex}\\
\hline
SU(2)$\times$U(1)
& $(1,0)$&$(2,1)$&$(1,1)$&$(2,1)$&$(1,0)$&$(1,-1)$&$(2,-1)$
\rule[-2ex]{0pt}{5ex}\\
\hline
\end{tabular}
\caption{Representation of leptons, Higgs and spurions under $G$=SU(2)$\times$U(1), ($a=1,2$).}
\label{tab_ex1_1}
\end{table}
The product $r_{e^c}\otimes r_{l}$ decomposes as $(1,1)\oplus (2,1) \oplus (2,2) \oplus(1,2) \oplus(3,2)$. The corresponding
combinations $(\Gamma^I_{ij\alpha} e^c_i l_j)$ are given in table \ref{tab_ex1_2}.~
\begin{table}[h!] 
\centering
\begin{tabular}{|c||c|c|c|c||c|c|}
\hline
$G$&$e^c_3 l_3$&$e^c_3 l_a$&$e^c_a l_3$&$(e^c_1 l_2-e^c_2 l_1)/\sqrt{2}$&$(e^c_1 l_1,(e^c_1 l_2+e^c_2 l_1)/\sqrt{2},e^c_2 l_2)$\rule[-2ex]{0pt}{5ex}\\
\hline
SU(2)$\times$U(1)
& $(1,1)$&$(2,1)$&$(2,2)$&$(1,2)$&$(3,2)$
\rule[-2ex]{0pt}{5ex}\\
\hline
\end{tabular}
\caption{Combinations $(\Gamma^I_{ij\alpha} e^c_i l_j)$ and their transformation properties under $G$=SU(2)$\times$U(1), ($a=1,2$).}
\label{tab_ex1_2}
\end{table}
The elements of a generic Yukawa coupling $\lambda^E_{ij}$ are classified as
$\lambda^E_{33} \sim (1,-1)$, $\lambda^E_{3a} \sim (2,-1)$, $\lambda^E_{a3} \sim (2,-2)$, $(\lambda^E_{12}-\lambda^E_{21})/\sqrt{2} \sim (1,-2)$ and  $(\lambda^E_{11},(\lambda^E_{12}+\lambda^E_{21})/\sqrt{2},\lambda^E_{22}) \sim (3,-2)$. 
In the absence of any indication about the type, size and orientation of the
spurions, this decomposition brings no useful information. We now assume that the only allowed spurions are, for example,  $\varphi_1$ and $\varphi_2$, transforming as $\varphi_1 \sim (1,-1)$ and $\varphi_2 \sim (2,-1)$ under SU(2)$\times$U(1) and with the VEV orientation $\vev{\varphi_2}^T =(\vev{\varphi_{21}},0)$, both invariant under the gauge group.
The choice of this direction in flavour space is not restrictive if the spurions describe vacuum configurations of dynamical fields, since options related by $G$ transformations lead
to equivalent physical systems. In this case, if we only consider terms linear in spurions,
the only non-vanishing entries of $\lambda^E_{ij}$ are $\lambda^E_{33}=\vev{\varphi_1}$ and
$\lambda^E_{32}=\vev{\varphi_{21}}$. To fill the matrix $\lambda^E_{ij}$ we need terms of higher
order. To second order we get:
\be
\lambda^E=
\left(
\begin{array}{ccc}
0&a\, \vev{\varphi_1}^2&0\\
-a\, \vev{\varphi_1}^2&b\, \vev{\varphi_{21}}^2&c\, \vev{\varphi_1} \vev{\varphi_{21}}\\
0&\vev{\varphi_{21}}&\vev{\varphi_1}
\end{array}
\right)
\ee
where the coefficients $a$, $b$ and $c$ are parameters related to independent invariant combinations. 
The vanishing entries of $\lambda^E$ can be filled by invariants of higher order. An assumption about the relative size of $\vev{\varphi_1}$ and $\vev{\varphi_{21}}$ can further shape the pattern of $\lambda^E$.

The set up we illustrated is based on an effective description of the flavon interactions with the SM fields, and is sufficient for most of our purposes. We now briefly discuss the possible UV origin of such a setup. This parallels the discussion of the UV origin of the Weinberg operator in \Sect{treelevel}. 

Consider for simplicity a $D=5$ operator involving a single flavon, in the form 
\begin{equation}
\frac{c_{ij}}{\Lambda} \, \varphi f^c_i f_j H \;.
\label{eq:D5yukawa}
\end{equation}
The latter contributes to the Yukawa interaction $\lambda_{ij} f^c_i f_j H$ for the charged leptons and neutrinos, $f = l$, $f^c = e^c,\nu^c$ (or for the quarks, $f = q$, $f^c = d^c,u^c$). As for the Weinberg one, there are only three possible UV renormalisable origins of the operator in \eq{D5yukawa}. They correspond to the exchange of heavy vectorlike messengers with the same SM quantum numbers as $f$, $f^c$, or $H$. We consider for example the exchange of $n$ vectorlike messengers with the quantum numbers of $f$: $F_\alpha + \bar{F}_\alpha$, $\alpha = 1\ldots n$. The renormalisable lagrangian contains
\begin{equation}
-\mathcal{L}_F = \eta_{\alpha i} \bar{F}_\alpha f_i \varphi + y_{i\alpha} f^c_i F_\alpha H + M_{\alpha\beta} \bar{F}_\alpha F_\beta +\text{h.c.} \;,
\label{eq:lF}
\end{equation}
where the couplings are constrained by the flavour symmetry. Integrating out the $F$, $\bar{F}$ fields generates the operator in \eq{D5yukawa}, with (cfr.\ \eq{matchingnuc})
\begin{equation}
\frac{c}{\Lambda} = -y\, M^{-1} \eta \;.
\label{eq:seesawFN}
\end{equation} 

Note that in the presence of a single family of messengers the Yukawa couplings generated by \eq{lF} have rank one: $\lambda_{ij} = - y_i \eta_j (\vev{\varphi}/M)$. The first two charged fermion families vanish in this limit, and can be generated by sub-leading effects involving heavier messengers. This way, hierarchical charged fermion masses (and a viable mixing pattern for quarks and leptons) can be accounted for without imposing any flavour symmetries~\cite{Ferretti:2006df}. At the same time, a $\text{U(2)}_{f^c} \times \text{U(2)}_f$ symmetry arises accidentally in the limit in which additional contributions to the Yukawas from heavier messengers are neglected. 

\subsubsection{Vacuum Alignment}
Lepton mixing angles and phases can only be determined once both the neutrino and the charged lepton sectors are specified. For instance, when the lepton number $L$ is violated,
at low energy the relevant Lagrangian is
\be
e^c\lambda^E(\varphi)l H^*+\frac{1}{2\Lambda}(l H)c(\varphi)(l H)+ \text{h.c.}
\label{LEL}
\ee
where now the matrices $\lambda^E$  and $c$ are functions of the fields 
$\varphi$ \footnote{If $c(\varphi)$
originate from the exchange of heavy degrees of freedom whose mass depends on $\varphi$, it might 
be singular as $\varphi$ vanish and a series expansion like the one in eq.~(\ref{SB2}) might not be possible.}
and the Lagrangian is invariant under the group $G$.
The mixing matrix is given by 
\begin{equation}
U =U_e^\dagger U^{\phantom{\dagger}}_\nu \;, 
\label{eq:PMNS}
\end{equation}
where $U_e$ and $U_\nu$ are the unitary matrices that diagonalize the combination $\lambda^{E\dagger}\lambda^E$ and $c$, respectively:
\be
U_e^\dagger \lambda^{E\dagger}\lambda^E U_e=(\hat \lambda^E)^2 \;, \qquad U_\nu^T c\, U_\nu=\hat c \;.
\ee
Here $\hat \lambda^E$ and $\hat c$ are non-negative diagonal matrices and their eigenvalues
have been properly ordered, also accounting for the type of neutrino mass spectrum. After suitable
rephasing of the combination $U_e^\dagger U_\nu$, we can put the mixing matrix in a conventional
form, for instance the one used by the PDG, and read the physical parameters. The latter
follow necessarily from the interplay of both neutrinos and charged leptons. 

Such a trivial observation has important implications on model building. Since both $\lambda^E(\varphi)$
and $c(\varphi)$ depend on $\varphi$, a realistic pattern of lepton masses and mixing angles
can only be achieved if the VEVs of the fields $\varphi$
have the right size and orientation in flavour space. If these fields are dynamical, the problem of deriving
the desired VEV from the minimization of the energy density is called vacuum alignment problem. 
Though the group $G$ is completely broken in the low-energy regime,
it might be that separately the charged lepton sector and the neutrino sector possess an exact
or approximate residual symmetry under subgroups $G_e$ and $G_\nu$, respectively.
Actually this scenario has been extensively studied in the context of discrete flavour symmetries
to predict or constrain the lepton mixing angles. This special case of vacuum alignment can be
implemented by separating $\varphi$ into two sets, $\varphi=(\varphi_e,\varphi_\nu)$,
such that $\lambda^E$ and $c$ mainly depend on $\varphi_e$ and  $\varphi_\nu$, respectively.
The desired residual symmetries are obtained if the VEV of $\varphi_e$ is invariant under $G_e$
and that of $\varphi_\nu$ under $G_\nu$. This possibility will be discussed in greater detail in \Sect{discrete_sequestering}. 

The above discussion already shows
advantages and limitations of the considered setup. The perspective
that fermion masses and mixing angles are determined by some dynamical principle
is certainly very fascinating and makes contact with more fundamental theories like string theory,
where in principle Yukawa couplings are calculable functions of a set of fields describing the
vacuum configuration. A drawback of the approach is exhibited by eqs.~(\ref{SB2},\ref{SB3}).
If a realistic description of fermion masses and mixing angles requires the presence
of several terms in the expansion, a large number of free parameters might be required,
to the detriment of predictability. The predictions can also be affected by the uncertainty
related to the whole tower of higher-dimensional operators, unless the expansion parameters
$\vev{\varphi}$ are very small. Moreover, if we insist in deriving the appropriate pattern of VEV
for the fields $\varphi$ from the minimization of the energy density, the solution of the vacuum alignment problem might require very complicated constructions, with many auxiliary fields
that do not play any role in shaping $\lambda^E$ and $c$ and additional symmetries
to forbid unwanted terms in the scalar potential.
To avoid or reduce the complexity of the vacuum alignment problem, we can give up the possibility that symmetry breaking is dynamically determined. This is a frequent option in models realized in the presence of extra dimensions, 
where the symmetry breaking can be achieved through an appropriate set of boundary
conditions. Examples of this type of breaking for models of neutrino masses can be found in refs.~\cite{Csaki:2008qq,Kobayashi:2008ih,Hagedorn:2011un,Hagedorn:2011pw}.

It is worth noticing that the above formalism is covariant under a general change of basis in the field space, provided
both charged lepton and neutrino sectors are consistently addressed. Let the group $G$ act, in the original basis, as
\be
\psi\to U_\psi(g)\psi \hspace{2cm} \psi=(e^c,l,\varphi) \;,
\ee
$U_{\psi}(g)$ being unitary matrices depending on the generic element $g$ of the group.
If we perform an arbitrary change of basis described by a set of unitary matrices $\Omega_{\psi}$:
\be
\psi\to\psi'=\Omega_\psi\psi \;,
\ee
we end up with new matrices
$(\lambda^E)'=\Omega_{e^c}^T \lambda^E \Omega_l$ and $c'=\Omega_l^T c\,\Omega_l$ in our Lagrangian.
The matrices that diagonalize $(\lambda^{E\dagger}\lambda^E)'$ and $c'$ are now
$U_{e^c}'=\Omega_l^\dagger U_{e^c}$ and $U_{\nu}'=\Omega_l^\dagger U_{\nu}$. 
All the physical parameters are unchanged.
In the new basis the group $G$ acts as
\be
\psi'\to U'_\psi(g)\psi' \;, \hspace{2cm} U'_\psi(g)=\Omega_\psi U_\psi(g)\Omega_\psi^\dagger \;.
\ee

A feature which is not captured by the previous formalism is the possibility
that the flavour symmetry is non-linearly realized. In this case the various terms of the expansion in eq.
(\ref{SB3}) are not expected to be individually invariant under $G$-transformation, 
as occurs above as a result of assuming linear unitary representations. This means that the coefficients $\lambda^E_{0ij}$,$\lambda^{E\alpha}_{1ij}$, $\lambda^{E\alpha\beta}_{2ij},\ldots$, might all be related to provide
a Yukawa interaction invariant under the group $G$. This case might present the advantage of
requiring less free parameters and thus being more predictive.
\subsubsection{Kinetic terms}
In general the breaking of the flavour symmetry affects not only the Yukawa interactions as in eqs.~(\ref{SB3},\ref{LEL}), but also the kinetic terms, leading to additional contributions to mass/mixing parameters.
The kinetic terms read:
\be
i \bar e^c \bar \sigma^\mu K^{e^c}(\varphi)\partial_\mu e^c+i \bar l \bar \sigma^\mu K^{l}(\varphi)\partial_\mu l+ \ldots 
\ee
where the dots stand for terms including $\partial_\mu K^{f}(\varphi)$ $(f=e^c,l)$, required by a hermitian Lagrangian and 
$K^{f}(\varphi)$ are positive-definite hermitian matrices in flavour space, depending on the flavon fields, here assumed to be real. In the spirit of effective field theories and in linearly realized flavour symmetries, $K^{f}(\varphi)$ can be expanded in powers of $\varphi$. Assuming
a choice of basis where $K^{f}(0)=\mathbb{1}$, we have:
\be
K^{f}(\varphi)=\mathbb{1}+K^{f\alpha}_1\varphi_\alpha+K^{f\alpha\beta}_2\varphi_\alpha\varphi_\beta+ \ldots
\ee
where $K^{f\alpha_1\ldots \alpha_p}_p$ are numerical matrices constrained by the requirement of $G$ invariance. When flavons
acquire a VEV, canonical kinetic terms are recovered through the transformations:
\be
f\to(\mathbb{1}-\frac{1}{2}K^{f\alpha}_1\varphi_\alpha+ \ldots)\, f \hspace{2cm} (f=e^c,l) \;,
\ee
and Yukawa interactions are modified accordingly. For instance the charged lepton Yukawa couplings
become:
\be
\lambda^E(\varphi)\to \lambda^E(\varphi)-\frac{1}{2}K^{{e^c}\alpha}_1\varphi_\alpha\lambda^E(\varphi)
-\frac{1}{2}\lambda^E(\varphi)K^{l\alpha}_1\varphi_\alpha+ \ldots
\label{clchange}
\ee
The consequences of such a change are different whether we are dealing with a supersymmetric or a non-supersymmetric theory. In a non-supersymmetric theory, the transformation (\ref{clchange}) merely results in a redefinition of the parameters of the Yukawa matrix $\lambda^E(\varphi)$, since $\lambda^E(\varphi)$ exhausts
all the polynomial invariants depending on the flavons $\varphi$ and describing charged lepton Yukawa couplings.
In the supersymmetric case, $\lambda^E(\Phi)$ are holomorphic functions of chiral multiplets $\Phi$, while in the kinetic terms we should distinguish holomorphic and anti-holomorphic variables. The function $K^{f}(\Phi,\Phi^\dagger)$ depends on both of them:
\be
K^{f}(\Phi,\Phi^\dagger)=\mathbb{1}+K^{f\alpha}_1\Phi_\alpha+K^{f\alpha\dagger}_1\Phi^\dagger_\alpha+ \ldots
\ee
The transformation (\ref{clchange}) becomes:
\be
\lambda^E(\Phi)\to \lambda^E(\Phi)-\frac{1}{2}\left[K^{{e^c}\alpha}_1\Phi_\alpha+K^{{e^c}\alpha\dagger}_1\Phi^\dagger_\alpha \right]\lambda^E(\Phi)
-\frac{1}{2}\lambda^E(\Phi)\left[K^{l\alpha}_1\Phi_\alpha+K^{l\alpha\dagger}_1\Phi^\dagger_\alpha\right] \ldots 
\label{clchangesusy}
\ee
which induces a non-holomorphic dependence of the physical Yukawa couplings on the flavons.
In general this entails additional parameters to the description of masses, mixing angles and phases.
Such effects have been analyzed in ref.~\cite{Dudas:1995yu,Dudas:1995eq,Binetruy:1996xk,Dreiner:2003hw,Jack:2003pb,Dreiner:2003yr} for abelian flavour symmetries, in ref.~\cite{King:2003xq,Ross:2004qn} for nonabelian continuous flavour symmetries, in ref.~\cite{Hamaguchi:2002vi,Chen:2012ha,Chen:2013aya} for nonabelian discrete flavour symmetries, in ref.~\cite{Chen:2019ewa} for modular flavour symmetries. Ref.~\cite{Kakizaki:2003fc} exploits such contribution
to explain the hierarchy between the top and the other quark masses. Ref.~\cite{Kawamura:2018cpu} explores a scenario where the flavour group $G$ remains unbroken in Yukawa interactions and the breaking is entirely due 
to kinetic terms.
A model-independent discussion for linearly realized flavour symmetries and in the supersymmetric case can be found in ref.~\cite{Espinosa:2004ya}. For degenerate neutrinos, the impact of the kinetic term is especially relevant, due to strong dependence of the mixing angles on new contributions. For hierarchical neutrinos the K\"ahler potential is expected to provide a contribution to the mixing of the same order of the contribution from the superpotential. Such effect could be important, for instance to explain the deviations from maximality, possibly enforced by the superpotential, of the solar and atmospheric mixings. In either case the kinetic terms bring additional free parameters, to the detriment of predictability.

\subsubsection{The Space of Invariants}
\label{sec:invariants}
There are general features of the vacuum alignment problem that can be discussed in terms of
the symmetry $G$ and the representation assigned to the fields $\varphi$,
without reference to the explicit form of the energy density functional.
Consider a Lagrangian ${\cal L}(\varphi)$ invariant under the action of a group $G$, depending on a set of scalar fields $\varphi$, transforming in a representation $r_\varphi$ of the group.
The fields $\varphi$ live in a vector space ${\cal M}$, the field space, whose dimension is
$d_\varphi$, the dimension of $r_\varphi$. In non-linear theories, ${\cal M}$ can be a manifold. If the theory is $G$-invariant, two distinct points in 
${\cal M}$ related by a $G$-transformation lead to the same predictions for any physical observable.
In particular, in any of two such points the system has the same residual symmetry, or little group, 
up to a conjugation.
Thus the field space offers a redundant description of the physical system, that can be simplified
by studying the orbits of the group, {\em i.e.} the set of points in the field space ${\cal M}$ that are related by group transformations. The union of orbits having isomorphic little groups forms a stratum. The full field space ${\cal M}$ is partitioned into several strata. For instance the origin of ${\cal M}$ belongs the stratum of type $G$, since for $\varphi=0$ the symmetry is unbroken.
Most of the field space ${\cal M}$ is made of orbits having minimal little group, {\em i.e.} the symmetry $G$ is broken
down to its minimum possible subgroup, which is unique, up to conjugation. This subset of ${\cal M}$ is called principal stratum.

A useful tool is the orbit space, ${\cal M}_I$. ${\cal M}_I$ can be parametrized by the values of invariants, which are constant on the orbits. It is sufficient to consider invariants $I(\varphi)$ that are polynomials in the components of the multiplet $\varphi$. The ring of invariant polynomials is infinite, but it is generated by a finite number of invariants $\gamma_\alpha(\varphi)$, which means that any invariant polynomial can be written as a polynomial in $\gamma_\alpha$. The invariants $\gamma_\alpha$ might be related by a number of algebraic relations, or syzygies, ${\cal Z}_S(\gamma)=0$. The space ${\cal M}_I$ is spanned by the values of the invariants $\gamma_\alpha$ of the theory. A whole orbit of ${\cal M}$ is mapped into a single point of ${\cal M}_I$, which completely characterizes the physical properties of the system, including its symmetry breaking pattern.
The crucial property of ${\cal M}_I$ is that while ${\cal M}$ has no boundaries, ${\cal M}_I$ has boundaries that describe the possible breaking chains of the group.
The tools that allow to characterise the orbit space ${\cal M}_I$ are the Jacobian matrix \cite{Cabibbo:1970rza}
\be
J\equiv\frac{\partial \gamma}{\partial\varphi} \;,
\ee
and the so-called ${\cal P}$-matrix
\be
{\cal P}=JJ^T \;.
\ee
The space ${\cal M}_I$ is identified by the requirements that i) $\gamma$ belongs to the surface ${\cal Z}_S(\gamma)=0$ and ii) the matrix ${\cal P}$ is positive semidefinite, resulting in a set of inequalities involving the invariants $\gamma_\alpha$ \cite{Abud:1981tf,Abud:1983id,Procesi:1985hr,Talamini:2006wd}.

As an example, consider the group $G=$SU(3) and the real scalar fields $\varphi=\varphi_a\lambda^a$,
transforming in the adjoint representation of the group, where $\lambda^a$ $(a=1,\ldots ,8)$ are the
Gell-Mann matrices. As independent invariants in ${\cal M}_I({\rm SU(3)})$ we can take $\gamma_1={\tt tr}(\varphi^2)$ and $\gamma_2={\tt det}(\varphi)$. The ${\cal P}$ matrix is 
\be
{\cal P}=
\left(
\begin{array}{cc}
8\gamma_1 & 12 \gamma_2\\
12\gamma_2& \frac{\gamma_1^2}{3}
\end{array}
\right) \;,
\ee
and it is positive semidefinite under the conditions $\gamma_1\ge0$ and ${\tt det}({\cal P})=\frac{8 \gamma_1^3}{3}-144\gamma_2^2\ge 0$. These inequalities define the space of invariants ${\cal M}_I({\rm SU(3)})$, spanned by $\gamma_{1,2}$. The space ${\cal M}_I({\rm SU(3)})$ is bi-dimensional and its interior corresponds to the point satisfying $\gamma_1^3-54\gamma_2^2 > 0$ and $\gamma_1>0$. In any point of the interior the matrices $J$ and ${\cal P}$ have rank 2 and the group SU(3) is broken down to a subgroup isomorphic to U(1)$\times$U(1). The one-dimensional boundary is defined by $\gamma_1^3-54\gamma_2^2 =0$ and $\gamma_1>0$, and consists of the two branches $\gamma_2=\pm \sqrt{\gamma_1^3/54}$.
Here the matrices $J$ and ${\cal P}$ have rank 1 and the group SU(3) is broken down to its subgroup
SU(2)$\times$U(1). Finally the two branches meet in $\gamma_1=0$, a zero-dimensional boundary where
$J$ and ${\cal P}$ have rank 0 and the group SU(3) is unbroken.
\begin{figure}[h!]
 \centering
 \includegraphics[width=0.4\textwidth]{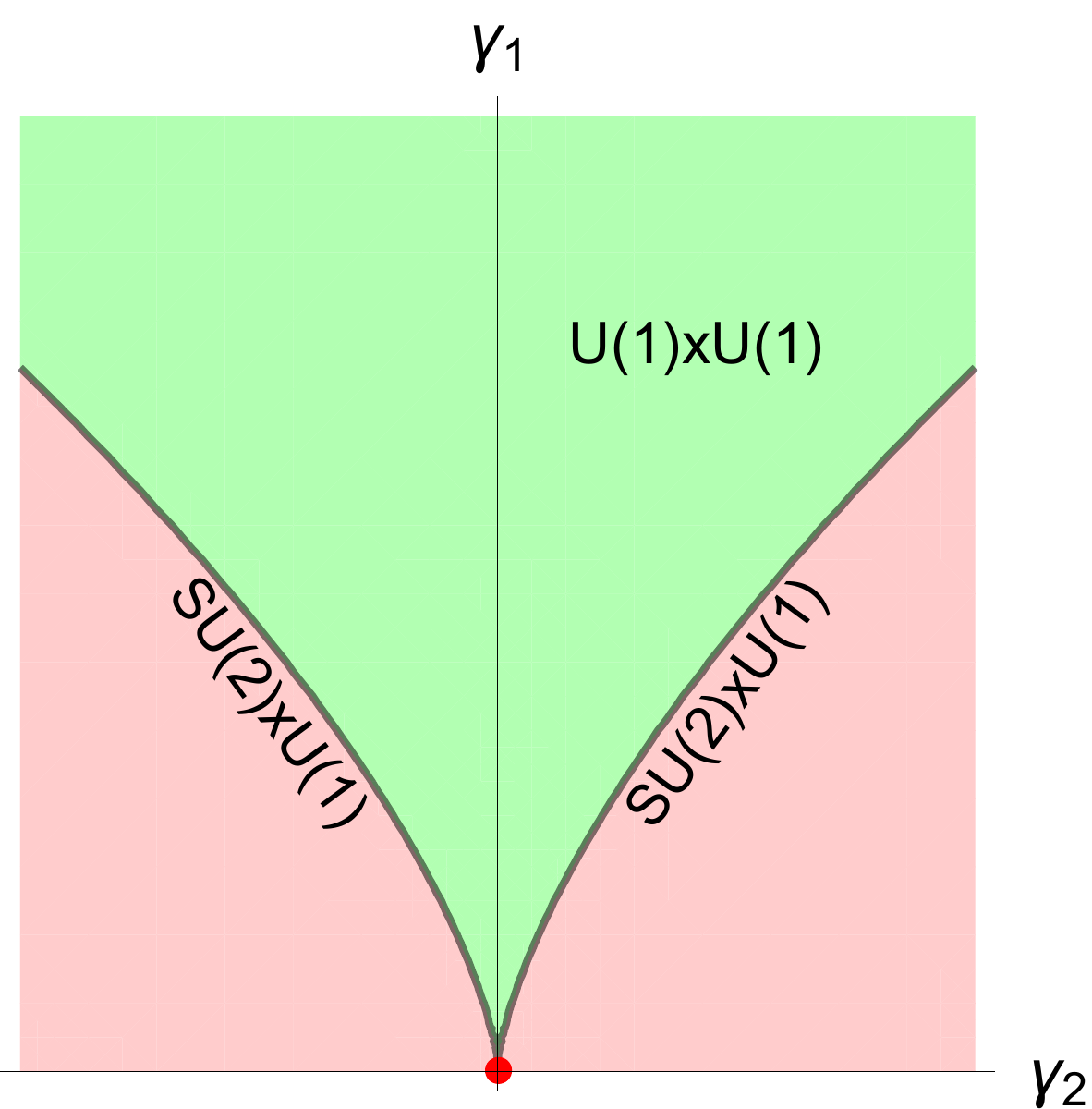}
 \caption{Space of invariants for $G$=SU(3) and $\varphi$ in the real adjoint representation.
 The green region is the interior, defined by $\gamma_1^3-54\gamma_2^2 > 0$ and $\gamma_1>0$. The red point, where the full SU(3) symmetry is unbroken, is the intersection of the one-dimensional boundaries.}
\label{fig:MSU3}
\end{figure}

It can be shown that such decomposition of ${\cal M}_I$ is completely general. The boundaries of ${\cal M}_I$ can be found by studying the rank of $J$.
In the interior of ${\cal M}_I$ the matrix $J$ has maximum rank $r_{max}$. In this region $G$ is broken
down to the smallest residual symmetry group $G_{min}$.
On the boundaries ${\cal P}$ has some vanishing eigenvalue
and the rank of $J$ is reduced.  If the dimension of ${\cal M}_I$ is $d$, in general we have $(d-1)$-dimensional boundaries where ${\tt rank}(J)=r_{max}-1$. Along these boundaries $G$ is broken down to groups containing $G_{min}$.
These boundaries meet along $(d-2)$-dimensional spaces, where ${\tt rank}(J)=r_{max}-2$. 
Here the residual symmetry further increases.
And so on, until the 1-dimensional boundaries meet in a point where the entire group $G$ is preserved. 

The above consideration can be useful when looking for the extrema of a generic smooth function $V(\varphi)$, invariant under $G$. Such a function depends on $\varphi$ through the invariants
$\gamma_\alpha(\varphi)$ and the extrema lay on orbits of the group. The extrema of $V(\varphi)$
are defined by the equations:
\be
\frac{\partial V}{\partial\varphi_i}=\frac{\partial V}{\partial\gamma_\alpha}\frac{\partial\gamma_\alpha}{\partial\varphi_i}=\frac{\partial V}{\partial\gamma_\alpha}J_{\alpha i}=0
\label{eqextrema}
\ee
Consider the previous example with $G=$SU(3) \cite{Michel:1973ug}. Along the orbits of the principal stratum, mapped in the interior of ${\cal M}_I$, $J$ has rank 2 and the derivatives $(\partial V/\partial\gamma_1,\partial V/\partial\gamma_2)$ should satisfy:
\be
\frac{\partial V}{\partial\gamma_1}=\frac{\partial V}{\partial\gamma_2}=0 \hspace{2cm}
(\gamma_1^3-54\gamma_2^2 >0,\gamma_1>0) \;.
\ee
Here SU(3) is broken down to the smallest residual symmetry, U(1)$\times$U(1). Along the orbits satisfying
$\gamma_1^3-54\gamma_2^2 =0$ and $\gamma_1>0$, providing the one-dimensional boundary of
${\cal M}_I$, $J$ has rank 1 and eq.~(\ref{eqextrema}) is solved by requiring $(\partial V/\partial\gamma_1,\partial V/\partial\gamma_2)$ to be one eigenvector of $J^T$ corresponding to the vanishing eigenvalue. This condition reads:
\be
J_{11}\frac{\partial V}{\partial\gamma_1}+J_{21}\frac{\partial V}{\partial\gamma_2}=0 \hspace{2cm}
(\gamma_1^3-54\gamma_2^2 =0,\gamma_1>0) \;.
\ee
Here the group SU(3) is broken down to its maximal subgroup SU(2)$\times$U(1). Finally, the orbit $\gamma_1=0$ corresponds to a vanishing $J$. There are no further conditions on the derivatives $(\partial V/\partial\gamma_1,\partial V/\partial\gamma_2)$ and the symmetry is unbroken. 

From this example we see that the extrema along the boundaries of ${\cal M}_I$ are more natural
than the extrema in the interior, since they require less conditions on the scalar potential $V$.
The extremum where $G$ is unbroken is always present, independently on the specific form of the
$G$-invariant function $V$. The corresponding orbit is isolated, that is in a sufficiently small neighborhood 
we find no other orbits with the same little group. Any such orbit is always an extremum, irrespectively
of the form of $V$ \cite{Michel:1971new,Michel:1971th}. 

Moreover, if the extremum is subject to the condition that $\varphi$ is non-vanishing and bound to a compact manifold, $V$ has always extrema having a maximal little group~\cite{Michel:1971new,Michel:1971th}.
In order to reduce the vector space $V$ where the flavons $\varphi$ live to a compact space, we need to minimize first with respect to the overall normalisation of the flavon fields. An assumption is then needed on the scalar potential: given any direction in the flavon space, the overall normalisation has a non-zero, symmetry breaking, local minumum; such minima form at least one smooth submanifold $M$ (hence compact, and invariant) in $V$. Michel's theorem can now be applied. The little groups found on $M$ are the same as the ones in $V$, except for $G$ itself, which is found in $V$ ($\varphi = 0$) but not in $M$ (flavour singlets can be neglected without loss of generality). This is welcome, as the trivial minimum $\varphi = 0$ is not relevant here. The extrema of $V$ guaranteed by the theorem are then those corresponding to the maximal little groups of $M$, i.e.\ to the little groups in $V$ not contained in any larger little group but $G$ itself. As an example, consider the SU(3) example above. The renormalisable scalar potential is given by 
\be
V=\mu_1^2 \gamma_1+\mu_2 \gamma_2+\lambda \gamma_1^2 \;.
\label{eq:VI}
\ee
The condition for the flavour group to be broken in any direction in flavour space is, not surprisingly, $\mu^2_1 < 0$. Under such condition, a critical point corresponding to the breaking of SU(3) to the maximal little group $\text{SU(2)} \times \text{U(1)}$ is guaranteed to exist. Clearly, this is not the case if $\mu^2_1 > 0$.

Extrema on orbits of the principal stratum might be compatible only with specific forms of $V$. For instance, in the example of \eq{VI}, extrema with little group U(1)$\times$U(1) are allowed only if $\mu_2=0$. For a non-vanishing $\mu_2$, the only allowed little groups of the extrema are SU(3) or SU(2)$\times$U(1). A clear limitation of this approach is that, without further inputs, we do not know whether the extrema are maxima or minima or saddle points of $V$.

\subsection{The role of $\mathbf{CP}$}
\label{roleCP}
In the previous Section, we have considered flavour groups commuting with the proper Poincar\'e group and with gauge transformations. We now relax this hypothesis. We want to argue that, under mild hypotheses, parity-like transformations are the only possible alternative. Indeed, by the Coleman-Mandula theorem \cite{Coleman:1967ad}, any symmetry of the scattering matrix should provide an automorphism of the Poincar\'e algebra. Up to Poincar\'e transformations, i.e.\ changes of reference frame, and dilatations, which require the theory to be conformally invariant in the symmetric limit, there are only two independent non-trivial automorphisms: parity and time-reversal. The action of both on the Poincar\'e algebra is involutive: it squares to the identity. Dilatations are only allowed if the theory is scale-invariant to begin with, which is not a case we are interested in. Because of the CPT theorem, it suffices to consider parity-like automorphisms. 

Consider now the action of such symmetry on the whole collection of matter fields, bosonic and fermionic, including conjugates, denoted by $\Phi_i$:
\begin{equation}
\Phi_i (x)\to  X_{ij} \Phi^\dagger_j (x_P) \;,
\label{eq:Pfields}
\end{equation}
where $(x_P)_\mu=x^\mu$.
It follows that left-handed Weyl spinors $f_a$ transform into right-handed ones: $f_a \to X_{ab} \bar f_b$.\footnote{More precisely, the full CP transformation on Weyl spinors reads: $f_a \to X_{ab} (\epsilon f^\dagger_b)$.} While the action of this symmetry on the Poincar\'e algebra is involutive, it does not have to be involutive on the fields $\Phi_i (x)$ and, in general $X X^*$ corresponds to a standard flavour transformation, not necessarily
equal to the identity.
Additional conditions hold in a gauge theory, where a gauge group $G_g$ acts on the fields $\Phi_i$ through its unitary representation $\rho_{ij}(g)$. In order for the parity-like transformation to be consistent, equivalent field configurations (related by gauge transformations) should be transformed by the parity-like action into equivalent field transformations. Moreover, the gauge interactions should be invariant. The two previous requirements leads to the following two consistency conditions~\cite{Grimus:1995zi}.
\begin{itemize}
\item There must exist an automorphism $g\in G_g \to g' \in G_g$ such that 
\begin{equation}
X \rho(g)^* X^{-1} = \rho(g') \;.
\label{eq:cc1}
\end{equation}
\item The parity-like transformation must transform the gauge fields $A^\mu(x) = A_a^\mu(x) t_a$, where $t_a$ are the gauge group generators as 
\begin{equation}
A^\mu(x) \to A'_\mu(x_P) \;,
\label{eq:cc2}
\end{equation}
where $t_a \to t'_a$ is the generator automorphism induced by $g \to g'$. 
\end{itemize}
The existence of a parity-like transformation inverting the sign of commuting gauge charges is guaranteed~\cite{Grimus:1995zi} in any gauge theory. This is, by definition, a CP transformation. On the other hand, a parity transformation commuting with gauge transformations can only exist if the fermions are not chiral, as well known. Other types of interplay with gauge invariance, other that the ones defining $P$ and $CP$, are in principle also possible. 

Under a CP transformation gauge interactions are automatically invariant, which is not necessarily the case for Yukawa interactions. Indeed when we turn off the Yukawa couplings of the SM, the theory becomes also invariant under CP transformations, whose action in flavour space is usually assumed to be trivial and thus irrelevant as flavour symmetry. However, generalizations of this action
are possible~\cite{Ecker:1987qp,Neufeld:1987wa}. 
We consider a theory with a ``conventional'' (commuting with Poincar\'e and gauge) \emph{global}  flavour symmetry group $G_f$ \footnote{Recent reviews on the combination of global and CP symmetries are \cite{Trautner:2016ezn,Trautner:2017vlz,Chen:2019iup}.}. If $G_f$ includes all flavour transformation leaving the theory invariant, a meaningful action of CP is guaranteed only for special choices of the flavour group and/or its representations.
Indeed, in the presence of a global symmetry $G_f$, CP transformations should satisfy a set of consistency conditions~\cite{Feruglio:2012cw,Holthausen:2012dk} similar to the one in eq.~(\ref{eq:cc1}). In such a theory the transformations of the fermion fields $f$ read:
\be
f\to U(g) \, f \hspace{2cm} f\to X_\text{CP}\,\bar f \;,
\ee
where $U(g)$ is a unitary representation of $G_f$, $g$ is a generic element of $G_f$ and $X_\text{CP}$ a unitary
matrix representing the action of CP in flavour space.
Under the combination of a CP transformation, followed by a $G_f$ transformation and an inverse CP transformation, the theory remains invariant. This implies that for each $g\in G_f$ an element $g'\in G_f$ should
exist such that:
\be
X_\text{CP} U^*(g) X_\text{CP}^{-1}=U(g') \;.
\label{conCP}
\ee
The map $g'=u(g)$, implicitly defined by the previous relation, is an automorphism of the group $G_f$, since it reshuffles the elements of $G_f$ while preserving the composition law.
Moreover, since CP relates particles and antiparticles, 
the function $g'=u(g)$ should map each representation $r$ of the group $G_f$ into its conjugate
$\bar r$. 
We will call such an automorphism a {\em complex conjugation}.
In general, a given group $G_f$ can possess automorphisms
other than complex conjugations. When $G_f$ is a continuous semisimple group, with an appropriate choice of basis in field space, the constraint (\ref{conCP}) can always be solved by $X_\text{CP}= \mathbb{1}$~\cite{Grimus:1995zi}. Moreover,
up to compositions with a transformation of the group $G_f$, $X_\text{CP}= \mathbb{1}$
is essentially the most general solution of (\ref{conCP}). A single exception is provided by the groups SO(2N) $(N\ne 4)$, admitting independent solutions.

The major difference with respect to the case of continuous gauge symmetries is that, if $G_f$ is a discrete group, complex conjugations are not guaranteed to exist. 
It is useful to distinguish between
inner automorphisms of $G_f$ that can be cast in the form $u(g)=h g h^{-1}$ $(h\in G_f)$ and outer automorphisms, that do not allow such a description. The inner automorphisms
map each representation of $G_f$ into an equivalent one, while outer automorphisms can permute
the representations. Thus inner automorphism can describe solutions of (\ref{conCP})
only if the flavour group representation is vectorlike. If it is chiral, the automorphism solving eq.~(\ref{conCP}) should necessarily be a complex conjugation of outer type~\cite{Holthausen:2012dk}.
 It follows that discrete groups $G_f$ can be divided into two classes~\cite{Chen:2014tpa}.
Those not possessing outer complex conjugations are called type I groups.
Theories having this type of flavour symmetry in general do not allow a consistent definition of CP, at least for a generic field content. An example of type I group is $\Delta(27)$.  
To define CP in such theories, we should restrict the field content to a suitable subset of
the available representations, on which an automorphism of the group acts as a complex
conjugation. 
Type II groups possess outer complex conjugation. Theories invariant under such groups
admit a consistent definition of CP. Examples of type II groups are $S_{3,4}$, $A_{4,5}$
and $T'$. Depending on 
the choice of the input parameters, these theories can be CP invariant or not,
exactly as happens for the SM, that admits a consistent action of CP but is CP invariant
only for special values of the parameters. 

A CP transformation is involutive, up to inner automorphisms \cite{Nishi:2013jqa}. This can be seen by applying eq.~(\ref{conCP}) twice, which gives
\be
X_\text{CP}X_\text{CP}^* U(g) X_\text{CP}^{-1*} X_\text{CP}^{-1}=U(u^2(g)) \; , \qquad u^2(g)\equiv u(u(g)) \;.
\ee
Since $X_\text{CP}X_\text{CP}^*$ represents the action of some element $s$ of $G_f$, we have:
\be
U(s) U(g) U(s)^{-1}=U(u^2(g)) \;,
\ee
implying that $u^2(g) = s g s^{-1}$ is an inner automorphism. The relation $u(s) = s$ also follows. If $s^n$ is the identity for some integer $n$, which is always true for finite groups, it follows that $(X_\text{CP}X_\text{CP}^*)^n=\mathbb{1}$.

Finally, if $X_\text{CP}$ is a complex conjugation
solving the constraint (\ref{conCP}), so is also $X_\text{CP}'=U(h)X_\text{CP}$, for any fixed
element $h$ of the group $G_f$. The action of $X_\text{CP}'$ differs from that of $X_\text{CP}$.
For example, we might have a canonical $X_\text{CP}=\mathbb{1}$ and a generalized $X_\text{CP}'$ acting in a nontrivial way. It is important to stress that $X_\text{CP}'$ and $X_\text{CP}$ set the same constraint on the theory, since $X_\text{CP}'$ is the combination of $X_\text{CP}$ with a symmetry transformation. Nevertheless, when considering the breaking of the full flavour symmetry group, it can be useful to exploit generalized CP transformation, to classify the available breaking chains and their features.
Combining a flavour group $G_f$ with CP results in the group $G=G_f\rtimes \text{CP}$, if $\text{CP}^2 = 1$. In general, requiring invariance under $G$ sets additional restrictions among parameters
with respect to only enforcing $G_f$. Physical phases can be constrained or predicted, as discussed in section \ref{sec:CP}.

\subsection{Non-linear flavour symmetries}
The action of the flavour group $G$ on the matter multiplets can also be non-linear. A natural realization of this scenario
involves the introduction of a set of real scalar fields $\varphi^\alpha$, neutral under the SM gauge group, living in a manifold
${\cal M}$ equipped with the metric $g_{\alpha\beta}(\varphi)$. 
Many SM extensions predict the existence of new scalar degrees of freedom. For instance
in string theory components of the metric tensor describing size and shape of the compactified space are scalar in four dimensions. In the present context $\varphi^\alpha$ play the role of flavons. 
Terms with two derivatives read:
\be
{\cal L}_\varphi=\frac{1}{2}g_{\alpha\beta}(\varphi)\partial_\mu \varphi^\alpha \partial_\mu\varphi^\beta \;.
\ee
Under a reparametrization of ${\cal M}$, $\varphi^\alpha\to f^\alpha(\varphi)$, the metric transforms as
\be
g_{\alpha\beta}(\varphi)\to {\tilde g}_{\alpha\beta}(\varphi)=\frac{\partial f^\gamma}{\partial \varphi^\alpha} \, g_{\gamma\delta}(f(\varphi)) \, \frac{\partial f^\delta}{\partial \varphi^\beta} \;,
\ee
and the Lagrangian becomes
\be
{\cal L}_\varphi\to {\cal \tilde L}_\varphi=\frac{1}{2}{\tilde g}_{\alpha\beta}(\varphi)\partial_\mu \varphi^\alpha \partial_\mu\varphi^\beta \;.
\ee
The isometries are reparametrizations leaving invariant the metric and hence the Lagrangian:
\be
 {\tilde g}_{\alpha\beta}(\varphi)=g_{\alpha\beta}(\varphi) \;, \qquad {\cal \tilde L}_\varphi={\cal L}_\varphi \;.
\ee
They form the isometry group $G_I$ of ${\cal M}$. The flavour group $G$ is identified with a subgroup of $G_I$. 
This framework defines a non-linear $\sigma$-model invariant under $G_I$,
to which matter fields of the SM are coupled. For simplicity we consider the SM fermions,
collectively denoted by $\psi^i$, in the limit where gauge interactions are turned off. A minimal coupling
comprises
\be
{\cal L}_\psi=i \, h_{ij}(\varphi)\bar \psi^i\bar\sigma^\mu\partial_\mu\psi^j+k_{ij\alpha}(\varphi)\bar \psi^i\bar\sigma^\mu\psi^j\partial_\mu\varphi^\alpha+ \text{h.c.}
\label{minfer}
\ee
Under a reparametrization of ${\cal M}$, the fermions transform as
$\psi^i\to \chi^i(\varphi,\psi)=\xi^i_j(\varphi)\psi^j+ \ldots$, where dots stand for possible contributions of higher order in $\psi$.
Here we will consider fermion transformations nonlinear in $\varphi$, but linear in $\psi$, the easiest way to guarantee that the transformed fields
have the same gauge quantum numbers as the original ones. Hence a generic reparametrization reads:
\be
\varphi^\alpha\to f^\alpha(\varphi) \; , \hspace{2cm} \psi^i\to \xi^i_j(\varphi)\psi^j \;.
\label{nltr}
\ee
Group properties are guaranteed by the relations
\bea
\varphi&\xrightarrow{g_1}&f_{g_1}(\varphi)\xrightarrow{g_2} f_{g_1}(f_{g_2}(\varphi))= f_{g_1 g_2}(\varphi)\nn\\
\psi&\xrightarrow{g_1}&\xi_{g_1}(\varphi)\psi\xrightarrow{g_2}\xi_{g_1}(f_{g_2}(\varphi))\xi_{g_2}(\varphi)\psi=\xi_{g_1 g_2}(\varphi)\psi \;,
\eea
and
\be
f_e(\varphi)=\varphi \;, \qquad \xi_e(\varphi)=\mathbb{1} \;.
\ee
Under (\ref{nltr}) the metric $h_{ij}(\varphi)$ and the connection $k^i_{j\alpha}(\varphi)\equiv h^{il}(\varphi)k_{lj\alpha}(\varphi)$ transform as \footnote{Indices are lowered and raised by the metric $h_{ij}(\varphi)$ and the inverse metric $h^{ij}(\varphi)$, respectively.}:
\begin{align}
h_{ij}(\varphi)&\to   \xi^{k*}_i \,  h_{kl}(f(\varphi)) \, \xi^l_j\nn\\
k^i_{j\alpha}(\varphi)&\to (\xi^{-1})^i_m \,  k^m_{l\beta}(f(\varphi)) \,
\xi^l_j \,  \frac{\partial f^\beta}{\partial \varphi^\alpha}+i \, (\xi^{-1})^i_l\frac{\partial \xi^l_j}{\partial\varphi^\alpha} \;.
\label{trafer}
\end{align}
If the transformation of eq.~(\ref{nltr}) is an isometry, the metric and connection are required to be invariant. From eq.~(\ref{trafer}) we understand the role of the connection $k^i_{j\alpha}(\varphi)$: even when the isometry of the scalar manifold ${\cal M}$
is realized by global transformations on $\varphi^\alpha$, the fermion transformations
are always local due to the explicit space-time dependence of the functions $\xi^i_j(\varphi)$.
The two terms in eq.~(\ref{minfer}) can be combined into a covariant derivative:
\be
(D_\mu \psi)^i\equiv \left(\delta^i_j\partial_\mu-i\, k^i_{j\alpha}(\varphi)\partial_\mu\varphi^\alpha\right)\psi^j \;,
\ee
which under an isometry transforms as the fermions $\psi^i$:
\be
(D_\mu \psi)^i\to\xi^i_j(\varphi)(D_\mu \psi)^j \;.
\ee
In the case treated in Section \ref{sec:modular} the isometries act on the fermion fields in the following way:
\be
\psi^i\to \left[\det\left(\frac{\partial f}{\partial \varphi}\right)\right]^{-k/2}\rho^i_j\, \psi^j \;.
\label{fst}
\ee
where $k$ is a real number called {\em weight} and $\rho$ is a $\varphi$-independent unitary representation of a compact coset $G/H$, where $G\subseteq G_I$ and $H$ is a normal subgroup of $G$. 
A nice property of the transformation (\ref{fst}) is that it manifestly provides
    a non-linear realization of $G$. Indeed, considering two subsequent isometries we have:
\begin{align}
\psi^i&\xrightarrow{g_1} \left[\det\left(\frac{\partial f_{g_1}}{\partial \varphi}\right)\right]^{-k/2}(\rho_{g_1})^i_j\, \psi^j\nn\\
&\xrightarrow{g_2}
\left[\det\left(\frac{\partial f_{g_1}}{\partial \varphi}\right)\right]^{-k/2}_{\varphi\to f_{g_2}(\varphi)}\cdot
\left[\det\left(\frac{\partial f_{g_2}}{\partial \varphi}\right)\right]^{-k/2}
(\rho_{g_1})^i_k(\rho_{g_2})^k_j\, \psi^j\nn\\
&=\left[\det\left(\frac{\partial f_{g_1 g_2}}{\partial \varphi}\right)\right]^{-k/2}(\rho_{g_1 g_2})^i_j\, \psi^j
\;.
\end{align}
and the group composition property is guaranteed. Invariance of the metric $h_{ij}(\varphi)$
under the isometry (\ref{fst}) requires:
\be
h_{ij}(f(\varphi))=\left[\det\left(\frac{\partial f}{\partial \varphi}\right)\right]^k\,
\rho_i^m\, h_{mn}(\varphi) \, (\rho^\dagger)^n_j \;.
\ee
The law (\ref{fst}) can be generalized by allowing
different pairs $(k,\rho)$ for distinct irreducible representations $\psi_{(I)}$ of the gauge group:
\be
\psi_{(I)}^i\to \left[\det\left(\frac{\partial f}{\partial \varphi}\right)\right]^{-k_I/2}{\rho_{(I)}}^i_j \, \psi_{(I)}^j \;.
\label{fst1}
\ee
Invariance of a fermion bilinear \footnote{For notational convenience we set to 1 the Higgs
multiplet $H$, that can be easily reintroduced in our expressions. Also $H$ can undergo
a transformation of the type (\ref{fst1}).}
\be
{\cal L}_Y=\lambda(\varphi)_{ij}\psi_{(I_1)}^i\psi_{(I_2)}^j+ \text{h.c.}
\ee
requires a Yukawa coupling $\lambda(\varphi)_{ij}$ satisfying:
\be
\lambda(f(\varphi))_{ij}=\left[\det\left(\frac{\partial f}{\partial \varphi}\right)\right]^{(k_{I_1}+k_{I_2})/2} \, [\rho_{(I_1)}]_i^{k*}\, \lambda(\varphi)_{kl}\, [\rho_{(I_2)}^\dagger]^l_j \;.
\ee
The overall Lagrangian
\be
{\cal L}=\frac{1}{2}g_{\alpha\beta}(\varphi)\partial_\mu \varphi^\alpha \partial_\mu\varphi^\beta+
h_{ij}(\varphi)\bar \psi^i\bar\sigma^\mu D_\mu\psi^j+\lambda(\varphi)_{ij}\psi_{(I_1)}^i\psi_{(I_2)}^j+ \text{h.c.} \;,
\ee
is invariant under the non-linearly realized flavour symmetry:
\be
\varphi^\alpha\to f^\alpha(\varphi) \;, \qquad \psi_{(I)}^i\to \left[\det\left(\frac{\partial f}{\partial \varphi}\right)\right]^{-k_I/2}{\rho_{(I)}}^i_j \, \psi_{(I)}^j \;.
\label{fst2}
\ee
Notice that this formalism, at variance with the Callan-Coleman-Wess-Zumino construction
\cite{Coleman:1969sm,Callan:1969sn}, covers both the case of a global flavour symmetry and that of a discrete one.
The purpose of this approach is to select $G$, $G/H$, $\rho_{(I)}$ and $k_{(I)}$ so as to constrain as much as possible the function $\lambda(\varphi)$. In an ideal case, the functional dependence of $\lambda(\varphi)$ on $\varphi$ is completely determined up to an overall constant and all dimensionless parameters such as mass ratios,
mixing angles and physical phases are all fixed functions of $\varphi$, providing a highly constrained system of predictions.
So far this program has been explored in the context of a supersymmetric $\sigma$-model
where the flavour group $G$ is the modular group $\text{SL}(2,Z)$, contained in $G_I=\text{SL}(2,R)$
and $G/H$ is a finite modular group.

\section{Standard flavour symmetries}
\label{sec:flavour}
We will now consider specific flavour symmetry models. We will begin in this section from the ``standard'' case in which the flavour symmetry commutes with the gauge and Poincar\'e transformations, in the context of the standard framework discussed in \Sect{standard}, in which the origin of neutrino masses lies at scales higher than the electroweak scale. We will consider flavour symmetries constraining the effective EW scale Lagrangian containing the Weinberg operator in \eq{Weinberg} and also consider flavour symmetries constraining its possible renormalizable high scale origins (and comment on the equivalence of the two approaches). We will also classify models according to whether the symmetry breaking affects mildly or prominently the flavour observables. 

We have seen in section~\ref{sec:SB} that a viable flavour symmetry must be broken by a set of flavon/spurion fields $\phi$, transforming under a representation $U_\phi$ of $G$. 
The lepton couplings and mass matrices then acquire a dependence on $\phi$, $M_E = M_E(\phi)$, $m_\nu = m_\nu(\phi)$. Because the full Lagrangian is assumed to be invariant under $G$, the mass matrices satisfy 
\begin{equation}
\begin{aligned}
M_E(\phi) &= \rep_e(g)^T M_E(U_\phi(g)\phi) \, \rep_l(g) \\
m_\nu(\phi) &= \rep_l(g)^T m_\nu(U_\phi(g)\phi) \, \rep_l(g)  
\end{aligned}
\label{eq:mLinvariancePhi}
\end{equation}
for any $g\in G$. 

It is often (but not always) the case that the functions $M_E(\phi)$ and $m_\nu = m_\nu(\phi)$ are continuous for $\phi\to 0$  and they admit an expansion in the flavons and their conjugates around their symmetric forms $M^0_E = M(0)$, $m^0_\nu = m_\nu(0)$ (which satisfy \eq{mLinvariance}). Flavour symmetry models can either be in the ``perturbative'' regime in which the symmetry breaking terms provide a moderate correction to the flavour observables; or in the ``leading order breaking'' regime in which symmetry breaking is necessary even for a leading order understanding of the flavour observables. The latter is the case, for example, when the neutrino or the charged lepton mass matrix vanishes in the symmetric limit. In the next section, we will consider the first possibility. The ``leading order breaking'' case will be discussed in section~\ref{sec:NP}.

\subsection{Perturbative breaking: mild corrections to flavour observables}
\label{sec:perturbative}

We have seen in \Sect{exact} that the symmetric forms $M_E^0$, $m_\nu^0$ of the lepton mass matrices cannot provide a (non-trivial) accurate description of lepton masses and mixings. It is however possible that they provide an approximate description. This is how non-exact symmetries of Nature have often emerged. Pions, for example, are close to an isospin symmetric limit in which the charged and neutral pion masses and couplings are equal. Analogously, one can wonder if lepton flavour observables are close to the symmetric predictions of a flavour theory. If this is the case, we can say that the understanding of the (leading order) pattern of lepton flavour lies in the flavour symmetry itself, and symmetry breaking effects only provide the moderate correction to the observables needed for their accurate description. 

\subsubsection{Flavour symmetries at low scales}

We first consider the case in which neutrino masses are fully described by the Weinberg operator and the flavour symmetry operates on the Lagrangian in \eq{effective}. In such a case, the flavour symmetry constrains the lepton mass matrices as in \eqs{mLinvariance} and a complete study of the perturbative option is possible. In fact, given a mass and mixing pattern considered to be a viable leading order approximation, the full set of flavour groups and representations leading to that pattern in the symmetric limit can be characterised in terms of the structure of the decomposition of $\rep_l$, $\rep_{e^c}$ into irreducible components; namely in terms of the type (real, complex, pseudoreal), dimension, and equivalence of the irreducible components. To be conservative, we consider viable symmetric predictions all those in which
\begin{itemize}
\item[i)] the PMNS matrix is not fully undetermined;
\item[ii)] both the $\theta_{23}$ and $\theta_{12}$ angles are allowed to be non-vanishing;
\item[iii)] the non-vanishing charged lepton masses are not forced to be degenerate.
\end{itemize}
The flavour symmetry models compatible with the above requirements are then those whose representations on the SM leptons have one of the six decompositions listed in table~\ref{tab:irrepdecomp}~\cite{Reyimuaji:2018xvs}. 

\begin{table}
\[
\begin{array}{|l|l|c|c|c|c|}
\hline
\quad \rep_l &  \quad \rep_{e^c}
& (m_\tau m_\mu  m_e) & (m_{3} m_{2} m_{1}) & \text{$\nu$ hierarchy} & \text{PMNS zeros}  \\[1mm]
\hline
\hline
\begin{array}{lll} 1 & 1 & 1 \end{array} &
\begin{array}{lll} 1 & \multicolumn{2}{l}{r \nsupseteq 1} \end{array} &
(A00) & (abc) &
\text{NH or IH} &
\text{none}
 \\
\hline
\begin{array}{lll} \mathbf{1} & \mathbf{1} & \overline{\mathbf{1}} \end{array} &
\begin{array}{lll} \overline{\mathbf{1}} & \multicolumn{2}{l}{r \nsupseteq \mathbf{1},\overline{\mathbf{1}}} \end{array} &
(A00) & (0aa) &
\text{IH} &
\text{none} \, (\mathbf{13}) 
 \\
\hline
\hline
\begin{array}{lll} 1 & 1 & 1 \end{array} &
\begin{array}{lll} 1 & 1 & r \neq 1 \end{array} &
(AB0) & (abc) &
\text{NH or IH} &
\text{none} 
 \\
\hline
\begin{array}{lll} \mathbf{1} & \mathbf{1} & \overline{\mathbf{1}} \end{array} &
\begin{array}{lll} \overline{\mathbf{1}} & \overline{\mathbf{1}} & r \neq \mathbf{1} \end{array} &
(AB0) & (0aa) &
\text{IH} &
\mathbf{13} 
 \\
\hline
\begin{array}{lll} 1 & 1 & 1 \end{array} &
\begin{array}{lll} 1 & 1 & 1 \end{array} &
(ABC) & (abc) &
\text{NH or IH} &
\text{none}
\\
\hline
\begin{array}{lll} \mathbf{1} & \mathbf{1} & \overline{\mathbf{1}} \end{array} &
\begin{array}{lll} \overline{\mathbf{1}} & \overline{\mathbf{1}} & \mathbf{1} \end{array} &
(ABC) & (0aa) &
\text{IH} &
\mathbf{13}, 23, 33 
 \\
\hline
\end{array}
\]
\caption{Classification of flavour groups and representations leading to an approximately viable prediction in the symmetric limit. The Weinberg operator is assumed to describe neutrino masses. The decompositions of the representation on the charged lepton doublets and singlets $l_i$ and $e^c_i$ into irreducible components is shown in the first two columns. The notation shows the dimension and type (boldface = complex, roman = real) of the representation. Identical symbols are associated to equivalent representations, while $\overline{\mathbf{1}}$ is the complex conjugate of $\mathbf{1}$. ``$r$'' denotes a generic, possibly reducible, representation. The predicted charged lepton and neutrino mass patterns are shown in the third and fourth column. The fifth column shows the type of neutrino mass hierarchy (normal or inverted hierarchical). The last column specifies whether the PMNS matrix contains a zero and in which position. In the second line the 13 entry can vanish or not, depending on an unknown ``12'' rotation determined by symmetry breaking effect. In the last line, the position of the zero depends on the relative size of $A$, $B$, and $C$. In the cases corresponding to the last four rows, the hierarchy of charged lepton masses is not explained by the flavour model and is accounted for by a hierarchy among the free parameters $A$, $B$, $C$.}
\label{tab:irrepdecomp}
\end{table}

Note that only abelian representations are allowed (with the only possible exception of a non-abelian two dimensional representation on $e^c_1$, $e^c_2$ when $m_e = m_\mu = 0$). More important, either neutrinos are inverted hierarchical, or the neutrino mass matrix is completely unconstrained. This is because there are only two possibilities for the representation on lepton doublets. When $\rep_l \sim 1 + 1+ 1$,  the three lepton doublets can at most transform by an overall sign under $G$. The neutrino mass matrix is then completely unconstrained, and any neutrino masses and mixings are possible. The flavour symmetry is useless in the neutrino sector, where it leads to anarchy~\cite{Hall:1999sn,Haba:2000be,Hirsch:2001he,Altarelli:2002sg,deGouvea:2003xe} (it may still be useful to explain the charged lepton mass hierarchy). When $\rep_l \sim \mathbf{1} + \mathbf{1}+ \overline{\mathbf{1}}$, the neutrino masses are in the form $(0aa)$ in the symmetric limit, which is close to the inverted hierarchical spectrum. Therefore, if the present hint for normal hierarchy transformed into an evidence, we would conclude that no flavour symmetry can provide a non-trivial approximate understanding of lepton flavour in the symmetric limit. Symmetry breaking effects would then play a leading role in determining (at least some of) the flavour observables. 

In all cases, no precise prediction on any of the lepton observables can be obtained (except possibly $\theta_{13} = 0$, which however is not precise on the experimental side), as the representation $\rep_l$ on lepton doublets is alway found to be abelian and because of the unknown $\ord{1}$ factors involved in each matrix elements. Note that the 1-dimensional representations are always abelian and abelian groups only have 1-dimensional irreducible representations. On the other hand, the 1-dimensional representations in table~\ref{tab:irrepdecomp} can also belong to non-abelian groups. In the case $G = \text{U(1)}$, the 1-dimensional representations are specified by their charges under the U(1). 

Examples of flavour models corresponding to the non-trivial examples in table~\ref{tab:irrepdecomp} have long been known. As mentioned, the three cases corresponding to the trivial representation $\rep_l \sim 1 + 1+ 1$ correspond to anarchical neutrinos. No special prediction is obtained, but the mixing angles and neutrino mass ratios are expected to be all $\ord{1}$. Indeed, the neutrino spectrum does not need mass ratios smaller than a factor 1/5--6 and the smallest mixing element is $|U_{e3}| \sim 1/7$. Moreover, they can arise from moderately small Yukawa couplings in the context of the see-saw, as the neutrino Yukawas are squared in the see-saw formula. The size of $|U_{e3}|$ only had an upper bound when anarchy was first considered. The measurement of a value not far from that bound corroborated the proposal~\cite{deGouvea:2012ac,Altarelli:2012ia}. The three cases have different $\rep_{e^c}$. The use of a non-trivial representation on the $e^c$ fields can forbid the electron and the muon masses in the symmetric limit and can therefore be used to account for the hierarchy of charged lepton masses even in the presence of anarchical neutrinos. 

As for the three non-anarchical cases, they require continuous or discrete groups with a complex 1-dimensional representation, ``$\mathbf{1}$'', and a representation on the lepton doublets decomposing as $\mathbf{1} + \mathbf{1} + \overline{\mathbf{1}}$. A simple choice is $G = U(1)$ with charges $(q^l_1,q^l_2,q^l_3) = (-1,1,1)$ on the three lepton doublets. In all cases, the neutrino mass matrix is in the form 
\begin{equation}
m_\nu = 
\begin{pmatrix}
0 & a & b \\
a & 0 & 0 \\
b & 0 & 0
\end{pmatrix}
+\text{corrections,} 
\label{eq:nuinverted}
\end{equation}
where the corrections are provided by symmetry breaking effects. Depending on whether $\rep_{e^c}$ matches or not $\rep_l$, the lighter charged lepton masses may or may not vanish in the symmetric limit, thus providing a rationale for their hierarchy. One obtains in fact 
\begin{equation}
M^{0}_E = 
\begin{pmatrix}
0 & 0 & 0 \\
0 & 0 & 0 \\
0 & B & A
\end{pmatrix}\;, \quad
\begin{pmatrix}
0 & 0 & 0 \\
0 & D & C \\
0 & B & A
\end{pmatrix}\;, \quad
\begin{pmatrix}
E & 0 & 0 \\
0 & D & C \\
0 & B & A
\end{pmatrix}\;.
\label{eq:Einverted}
\end{equation}
in the three cases of table~\ref{tab:irrepdecomp}, before switching on symmetry breaking effects. 

In the U(1) example, the last pattern of \eq{Einverted} can be reproduced by choosing opposite charges $(q^{e^c}_1,q^{e^c}_2,q^{e^c}_3) = (1,-1,-1)$ for the three $e^c$ fields. This corresponds to a U(1) symmetry with charge $L_\tau+L_\mu-L_e$~\cite{Barbieri:1998mq} (similar symmetries were considered in~\cite{Zeldovich:1952aa,Konopinski:1953aa,Petcov:1982ya}). None of the charged lepton hierarchies $m_e \ll m_\mu \ll m_\tau$ is accounted for. Moreover, the PMNS matrix contains a zero in the symmetric limit that should be identified with the $U_{13}$, but it can appear in the 12 or 33 position, depending on which charged lepton family ends up being lighter. In order to get rid of both such drawbacks, one can depart from $L_\tau+L_\mu-L_e$ by using different representations $\rep_{e^c} \neq \rep_l^*$, forcing $m_e = 0$ and possibly $m_\mu = 0$ in the symmetric limit. In all cases, the solar mixing angle is maximal in the symmetric limit and requires significant corrections from symmetry breaking (see Sect.~\ref{sec:corrections}). 

Normal hierarchy can be obtained in an important class of models that does not appear in table~\ref{tab:irrepdecomp}, in which the neutrino mass matrix is in the form~\cite{Irges:1998ax,Barbieri:1998mq,Grossman:1998jj}
\begin{equation}
m_\nu = 
\begin{pmatrix}
0 & 0 & 0 \\
0 & c & b \\
0 & b & a
\end{pmatrix}
+\text{corrections.} 
\label{eq:nunormal}
\end{equation}
Such a texture is obtained if $\rep_l = 1 + 1 + \mathbf{1}$. We use roman and boldface fonts to denote real and complex representations respectively (see table~\ref{tab:irrepdecomp}). This texture is sometimes called ``semi-anarchy'', as the ``23'' block of the neutrino mass matrix (corresponding to the trivial representation $1+1$), but not the whole one, is now unconstrained. In the symmetric limit, the solar angle is predicted to vanish, $\theta_{12} = 0$, a prediction that is far from the observation. In order for the above texture to be phenomenologically viable, the correction to $\theta_{12}$ from symmetry breaking effects cannot be mild. On the contrary, it must be fully responsible for the observed value of $\theta_{12}$. Therefore, such models belong to the class considered in \Sect{NP}. On the other hand, the $\ord{1}$ correction to $\theta_{12}$ does not need the symmetry breaking effects to be large in size. This is because of another drawback of the texture in \eq{nunormal}: it does not account for the observed suppression $\Dm{12} \ll |\Dm{23}|$. The latter needs an accidental cancellation in the determinant $ac-b^2$, which should vanish up to $\ord{m_2/m_3}$ corrections. Once such a (mild) accident is accepted, subleading $\ord{m_2/m_3}$ symmetry breaking effects are sufficient to generate a $\ord{1}$ solar angle. 

Note that predictions based on accidental relations may be unstable with respect to RGE~\cite{Chankowski:2001mx} or generic corrections~\cite{Marzocca:2014tga,Domcke:2016mzc}. An apparently accidental suppression can be accounted for in the see-saw context, see below. 

The results in table~\ref{tab:irrepdecomp} can be extended to the quark sector. The constraints one obtains there are independent of those discussed above. However, in the context of grand unified theories, unified quarks and leptons cannot be treated separately, as they must be subject to the same flavour representation. This leads to additional constraints. For example, in minimal SU(5) unification, only the anarchical cases in table~\ref{tab:irrepdecomp} turn out to be allowed.

\subsubsection{Flavour symmetries at high scales}

The previous conclusions were based on the assumption that the flavour symmetry constrains the effective EW scale Lagrangian containing the Weinberg operator. The latter however presumably represents the low-scale remnant of a more fundamental higher scale renormalizable Lagrangian. One can then wonder whether the conclusions summarized in table~\ref{tab:irrepdecomp} would still hold if the flavour symmetry was assumed to constrain the higher scale Lagrangian. In particular, one can wonder whether the choice between anarchy and inverse hierarchy is still necessary, if the symmetric predictions are required to be viable. This is part of a more general issue concerning the results obtained in the symmetric limit. Are the predictions obtained when the flavour symmetry acts on a high-scale Lagrangian equivalent to those obtained when the same  symmetry constrains the corresponding effective Lagrangian? The answer is no. On the other hand, the converse is true: given a flavour symmetry constraining the effective Lagrangian, it is always possible to extend its action to a high-scale Lagrangian providing the same predictions. Therefore, while the low-scale effective flavour theory does not capture all the features of the high-scale one, an appropriate high-scale realization always captures the features of the low-scale effective one.

There are two reasons why the high-scale predictions might not coincide with the low-scale ones~\cite{Reyimuaji:xxxxxx}. The most obvious is that the mass of some of the high-scale fields vanishes when the symmetry is exact. This happens if the flavour group representation on the high-scale fields is not vectorlike. In such a case, the heavy fields cannot be integrated out (as some of them are massless) before symmetry breaking effects have been switched on. Once the breaking effects are added, all the high-scale fields acquire a mass, including those whose mass vanished in the symmetric limit. The latter get a mass from sub-leading symmetry breaking effects. Therefore, their mass is expected to be lighter, and as a consequence their exchange dominates the effective Lagrangian and neutrino masses. In the standard see-saw language, this corresponds to the so-called single or sequential right-handed neutrino dominance~\cite{King:1998jw,Barbieri:1998mq,Altarelli:1998ns,King:1999cm,King:1999mb,Antusch:2004gf}, arising also in the context of non-abelian models~\cite{King:2005bj}. 

Even in the cases in which all the relevant heavy fields stay heavy when the symmetry is exact, the high- and low-scale predictions can differ. Consider for definiteness a type-I see-saw Lagrangian (with an arbitrary number of singlet neutrinos) and assume that the singlet neutrinos are non-singular in the limit in which the flavour symmetry is exact. It turns out that there is a precise condition under which the high- and low-scale predictions of the flavour symmetry are equivalent: this is the case if and only if the vectorlike part\footnote{By vectorlike part, we mean the maximal subrepresentation that is vectorlike, i.e.\ made of real representations, pairs of complex conjugated representations, or pairs of equivalent pseudoreal representations.} of the representation on the lepton doublets is contained in the representation on the neutrino singlets. 

We consider the two above possibilities in turn. We start from the case in which the mass of some of the high-scale fields vanishes when the symmetry is exact, in the context of type-I see-saw. The possible equivalence of the high- and low-scale approaches in the symmetric limit can still be investigated when the limit $m_\nu(\phi)$ for $\phi\to 0$ exists and is finite.\footnote{In some cases, the analysis can be extended to the cases in which the limit diverges, by normalising the neutrino mass matrix to the largest entry when taking the limit.} In some cases, the two descriptions can still be equivalent. Consider for example U(1) see-saw models in which the flavons have charges with definite sign, negative, for example, and the leptons have non-negative charges. We also invoke supersymmetry to prevent a positively charged flavon to be mimicked by a conjugated flavon. In such a case, the high- and low-scale descriptions are equivalent in the symmetric limit, independent of whether some of the right-handed neutrinos are massless or not in that limit. Consider for example the case of a single flavon with VEV $\theta$ (in terms of the cut-off scale) with charge -1 and let $q^l_i\geq 0$, $q^{\nu^c}_i \geq 0$ be the lepton doublet and singlet neutrino charges, $i=1,2,3$. Then in the broken phase, the low-scale flavour theory predicts
\begin{equation}
(m_\nu^{\text{LS}})_{ij} = c^\text{LS}_{ij} \, \theta^{q^l_i+q^l_j} \;,
\label{eq:mnuLSFN}
\end{equation}
where $c$ is a generic, unknown $3\times 3$ (dimensionful) matrix. In the high scale theory, we have instead $(m_D)_{ij} = (c_D)_{ij} \theta^{q^{\nu^c}_i+q^l_j}$, $M_{ij} = C_{ij} \theta^{q^{\nu^c}_i+q^{\nu^c}_j}$ for the Dirac and singlet Majorana mass matrices respectively. Therefore, the light neutrino mass matrix is
\begin{equation}
(m_\nu^{\text{HS}})_{ij} = c^\text{HS}_{ij} \, \theta^{q^l_i+q^l_j} \;,
\label{eq:mnuHSFN}
\end{equation}
where $c^\text{HS} = - c^T_D C^{-1} c_D$ is also a generic, unknown $3\times 3$ matrix. Therefore, the high- and low-scale definitions of the flavour theories are equivalent. 

On the other hand, the two descriptions can be inequivalent. Suppose for example that the lepton doublets and singlet neutrinos have charges $(q^l_1,q^l_2,q^l_3) = (0,1,1)$ and $(q^{\nu^c}_1,q^{\nu^c}_2,q^{\nu^c}_3) = (0,0,-1)$ under a U(1). Then, in the unbroken limit, the low- and high-scale versions of the same U(1) model provide quite different results:
\begin{equation}
m_\nu^{\text{LS}} = 
\begin{pmatrix}
a & 0 & 0 \\
0 & 0 & 0 \\
0 & 0 & 0
\end{pmatrix} \;, \quad
m_\nu^{\text{HS}} = 
\begin{pmatrix}
0 & 0 & 0 \\
0 & c & b \\
0 & b & a
\end{pmatrix} \; \text{, with $ac-b^2 = 0$.}
\label{eq:mnuFNSS1}
\end{equation}
The high-scale result follows from the following forms of the unbroken Dirac and Majorana matrices 
\begin{equation}
m_D = 
\begin{pmatrix}
0 & 0 & 0 \\
0 & 0 & 0 \\
0 & B & A
\end{pmatrix} \;, \quad
M = 
\begin{pmatrix}
\alpha & \beta & 0 \\
\beta & \gamma & 0 \\
0 & 0 & 0
\end{pmatrix} \;.
\label{eq:mnuFNSS1b}
\end{equation}
In the ``almost unbroken'' limit, the see-saw is dominated by the exchange of $\nu^c_3$, the only one taken into account in \eq{mnuFNSS1}. This is the single right-handed dominance mechanism mentioned above in its most classical realization, which now accounts for the apparently accidental suppression of the determinant $ab-c^2$ needed in \eq{nunormal}. 

As mentioned above, there is a second case in which the high- and low-scale formulations of the same flavour model are certainly inequivalent, even when all the right-handed neutrinos are allowed to be massive in the unbroken limit (i.e.\ even when the representation of $G$ on them is vectorlike). This is the case if the vectorlike part of the representation on the lepton doublets is not contained in the representation on the neutrino singlets~\cite{Reyimuaji:xxxxxx}. We illustrate the latter possibility with an example~\cite{Altarelli:1998ns}. Suppose that the lepton doublets and singlet neutrinos have charges $(q^l_1,q^l_2,q^l_3) = (n,0,0)$ and $(q^{\nu^c}_1,q^{\nu^c}_2,q^{\nu^c}_3) = (1,-1,0)$ under a U(1), with $n\neq\pm1,0$. Then the unbroken Dirac, singlet, and light neutrino matrices are
\begin{equation}
m_D = 
\begin{pmatrix}
0 & 0 & 0 \\
0 & 0 & 0 \\
0 & B & A
\end{pmatrix} \;, \quad
M = 
\begin{pmatrix}
0 & \beta & 0 \\
\beta & 0 & 0 \\
0 & 0 & \alpha
\end{pmatrix} \;, \quad
m_\nu^{\text{HS}} = 
\begin{pmatrix}
0 & 0 & 0 \\
0 & c & b \\
0 & b & a
\end{pmatrix} \; \text{, with $ac-b^2 = 0$,}
\label{eq:mnuFNSS2}
\end{equation}
as $a = A^2/\alpha$, $b = AB/\alpha$, $c = B^2/\alpha$. Note that the vanishing of the determinant is obtained as a consequence of the see-saw mechanism without the need to invoke the presence of a lighter singlet neutrino.  In the symmetric limit, the model predicts large $\theta_{23}$, $m_1 = m_2 = 0$, $\theta_{12}$ undefined. Enforcing the same flavour symmetry in the low-scale effective theory gives on the other hand \eq{nunormal} with no condition on the determinant. Therefore the unbroken predictions are different: $m_2$ now does not vanish and $\theta_{12} = 0$. 

As discussed, most instances of the ``perturbative'' breaking of flavour symmetries discussed in this subsection are associated to models with abelian symmetries. The latter have been widely studied in the first wave of model-building following the measurement of a large atmospheric angle. Additional examples and further details can be found in earlier reviews.

\subsection{Non-abelian models and leading order breaking}
\label{sec:NP}

In the case of leading order (LO) breaking of the flavour symmetry, the symmetry breaking effects cannot be disregarded even for a leading order understanding of lepton flavour. This happens when the unbroken limit is not a good approximation. According to the conservative definition used in the previous section, this is the case when i) the PMNS matrix is fully undetermined or ii) either the $\theta_{23}$ or $\theta_{12}$ angle is forced to vanish or iii) two of the charged lepton masses are forced to be degenerate and not vanishing in the unbroken limit. Correspondingly, there are three possible way outs from the results in table~\ref{tab:irrepdecomp}. 

Violating condition iii) is not very appealing. Charged lepton masses are hierarchical. Therefore, models with degenerate charged leptons in the unbroken limit require quite a fine-tuned symmetry breaking contribution. 
We will disregard such a possibility. 

If $\theta_{23}$ or $\theta_{12}$ vanishes in the unbroken limit (case ii)), the symmetry breaking corrections must be sizeable enough to strongly modify the symmetric prediction. This is a concrete possibility, whose realisation does not even require large symmetry breaking corrections. As discussed above, subleading corrections may be sufficient, in the presence of the mild accident necessary to account for the $m_2/m_3$ hierarchy 
(which can arise naturally in the see-saw context, see \eqs{mnuFNSS1} and~(\ref{eq:mnuFNSS2})). Such a possibility has been widely considered and discussed in \Sect{perturbative}.

We are left with the possibility that the PMNS is fully undetermined in the symmetric limit (case i)). This Section mainly deals with such a possibility, which arises when either $m^{0}_\nu = 0$ or $m^{0}_E = 0$ (the suffix ``0'' denotes the symmetric limit).

Predictivity is an independent motivation to consider models leading to $m^{0}_E = 0$, as we now discuss. This may seem paradoxical, as the PMNS matrix is completely undetermined in such a case, the poorest possible prediction. In fact, the predictions one gets in such cases have little to do with the symmetry itself and all to do with the details of symmetry breaking.

In order to see how predictive model may lead to $m^{0}_E = 0$, we first remind that non-abelian flavour groups are welcome in order to provide precise predictions. The predictive power of abelian models is limited by the fact that they only admit $d=1$ irreducible representations (here and below $d$ denotes the dimension of the representation). As a consequence, each flavour matrix entry corresponds to an independent invariant Lagrangian operator (see \Sect{SB}), with an independent, unknown dimensionless coupling. In the spirit of flavour models, aiming at providing a dynamical explanation of hierarchies, such couplings can be assumed to be $\ord{1}$. This means however that predictions are typically plagued by $\ord{1}$ uncertainties (barring predictions associated to texture zeros~\cite{Bjoorkeroth:2019ndr}). In the charged fermion sector, characterized by significant hierarchies, a prediction up to an $\ord{1}$ factor is significant. But in the neutrino sector, where most flavour parameters turn out to be themselves $\ord{1}$, a prediction up to an $\ord{1}$ factor is less exciting. In order to avoid systematic $\ord{1}$ uncertainties and attempt at significant predictions in the neutrino sector, $d > 1$ irreducible representations are then needed. The latter allow to correlate different matrix entries through symmetry transformations. From this point of view, the highest predictive power is achieved, in principle, when all the 3 neutrinos, i.e.\ the three lepton doublets, belong to a single $d=3$ irreducible representation. 

We can now appreciate the connection with $m^{0}_E = 0$: the matrix $m^{0}_E$ is forced to vanish if the lepton doublets belong to a $d=3$ irreducible representation $\rep_l$ of the flavour group, in order to avoid to have three degenerate, massive charged leptons in the unbroken limit. In order to prove the latter statement, we note that \eq{mLinvariance} implies $\rep_l ({m^0_E}^\dagger m^0_E) = ({m^0_E}^\dagger m^0_E) \rep_l$. Since $\rep_l$ is assumed to be irreducible, $({m^0_E}^\dagger m^0_E) = \alpha \, \mathbf{1}$ by Schur's Lemma. In order for the charged leptons not to be massive and degenerate, we need $\alpha$ = 0, i.e.\ we need the $m_E$ to be forced to vanish in the unbroken limit. As a corollary, non-abelian models with $m^0_E \neq 0$ require the lepton doublets to transform as doublet + singlet under the flavour group. 

Non-abelian symmetries can be continuous or discrete. Before reviewing in the next sections the case of finite non-abelian groups, we discuss some examples of continuous ones. Continuous (Lie) group models share some of the features of the discrete ones, which  will be discussed in greater details in the following sections. In particular, they can lead to precise predictions for some mixing parameters, with a substantial help from the scalar potential, arranging proper VEV alignments. In practice, this is most often the case in models in which the three families of lepton doublets belong to a single irreducible $d = 3$ representation of the flavour group.

Simple Lie groups with irreducible representations of dimension $d \leq 3$ are $\text{SU(2)} \sim \text{SO(3)}$, SU(3). The simple factors can be combined, with U(1) factors as well, in larger groups. First, consider the simplest possibilities, with the only possible addition of a U(1) factor. The group SU(2) is indeed often combined with a U(1) suppressing the light charged fermion families into $\text{U(2)} = \text{SU(2)} \times \text{U(1)}$~\cite{Barbieri:1995uv,Barbieri:1996ww,Barbieri:1997tu}. Neutrino masses and mixings can also be accounted for~\cite{Raby:2003ay,Linster:2018avp}, see also \Sect{SB}. The SO(3) case can lead to tribimaximal mixing (see below) within what was called ``constrained sequential dominance''~\cite{King:2005bj,King:2006me}, can originate from gauge-family unification in a SO(18) grand unified theory~\cite{Reig:2018ocz} and can underlie $A_4$ models~\cite{Bazzocchi:2007au,Berger:2009tt,Grossman:2014oqa}. The SU(3) group is more ``democratic'' than SO(3). The action of SO(3) in terms of real matrices singles out a real vector subspace in the three family (complex) flavour space. Moreover, SU(3) is, up to a U(1) factor, the maximal flavour group  for fermions with given quantum numbers. In fact, in the case of grand unified SO(10) models, $G_\text{max} = \text{SU(3)} \times \text{U(1)}$. As SU(3) (U(3)) typically forces the Yukawas to vanish in the symmetric limit, it must be strongly broken by the top Yukawa coupling to a weakly broken SU(2) (U(2)). Maximal atmospheric  and large solar mixing can be obtained together with hierarchical charged fermions~\cite{King:2003rf,Ross:2004qn,Antusch:2007re,Bazzocchi:2008rz}. Tribimaximal mixing can also be achieved consistently with SO(10)~\cite{deMedeirosVarzielas:2005ax,deAnda:2018yfp}. This is not as easy as with finite group models, where the flavour quantum numbers are often different within a single family. 

An example of a less minimal, and in fact almost maximal, flavour group is provided by $G = \text{SU(3)}^5\times \text{SO(3)}$. The $\text{SU(3)}^5$ term is, neglecting U(1) factors, the maximal SM flavour group (see \Sect{basics}). If the SM field content is supplemented by three singlet neutrinos $\nu^c_i$, $i=1,2,3$, and $G$ is required to allow a flavour-universal Majorana mass term in the form $M \nu^c_i \nu^c_i /2$, the maximal flavour group also contains a SO(3) factor acting on the $\nu^c_i$ fields. The Yukawa couplings are assumed to arise as VEVs of flavons transforming as $Y_U \sim 3_{u^c} \times 3_q$, $Y_D \sim 3_{d^c} \times 3_q$, $Y_N \sim 3_{\nu^c} \times 3_l$, $Y_E \sim 3_{e^c} \times 3_l$ under $G$, in the spirit of Minimal Flavour Violation~\cite{DAmbrosio:2002vsn}, extended to the neutrino sector~\cite{Cirigliano:2005ck,Alonso:2012fy,Alonso:2013mca}. The structure of the Yukawa couplings then depends on the scalar potential they minimise. The techniques introduced in \Sect{invariants} can be used to study which values of $Y$ can arise as critical points~\cite{Alonso:2011yg,Espinosa:2012uu}.

\subsubsection{Discrete non-abelian symmetries and the sequestering assumption}
\label{sec:discrete_sequestering}

Discrete non-abelian groups can provide precise predictions for lepton mixing.\footnote{For a ``physicist-oriented'' review of discrete group theory see~\cite{Ramond:2010zz,Altarelli:2010gt,Ishimori:2010au,Grimus:2011fk}.} Their study gained considerable momentum when the measured value of the solar angle was found in agreement with the prediction of the tribimaximal (TB) mixing pattern \cite{Harrison:2002er,Harrison:2002kp,Harrison:2003aw}, $\sin^2\theta_{12} = 1/3$. Such a pattern also corresponds to a maximal atmospheric angle, $\sin^2\theta_{23} = 1/2$ and to $\theta_{13} = 0$. The TB pattern, in turn, is predicted by flavour models based on relatively simple discrete groups. Unfortunately, the $\theta_{13}$ angle ended up to be larger than predicted by most of the early models. However, the tools and ideas developed in this context are still useful and widely used.

We have seen in \Sect{exact} that $G$ must be completely broken (up to an irrelevant $\mathbf{Z}_2$) by the full (including breaking effects) lepton mass matrices $M_E$, $m_\nu$. On the other hand, $M_E$ and $m_\nu$ might separately be invariant under non-trivial subgroups $G_e, G_\nu \subseteq G$. A popular model building strategy relies on the following non-trivial assumption: the subgroups $G_e$, $G_\nu$ are non-trivial and rigidly fix, up to phases, the charged lepton and neutrino mass bases. 
As the PMNS matrix is nothing but a measure of the misalignment between the two mass bases, the above requirement unambiguously determines the PMNS matrix in terms of $G_e$, $G_\nu$. Since $G$ must eventually be completely broken (up to an overall sign change of the lepton fields), their intersection must be trivial, $G_e \cap G_\nu \subseteq \mathbf{Z}_2$, where $\mathbf{Z}_2$ acts as an overall sign change. 

The assumption is non-trivial because $G_e$ and $G_\nu$ could well be trivial. In other words, both $M_E$ and $m_\nu$ could individually break $G$ completely so that $G_e$, $G_\nu$ would not carry any information on the PMNS matrix.  
Another possibility, illustrated in \Sect{loose}, is that $G_e$ and $G_\nu$ are non-trivial but they do not fully determine the mass eigenstates. Therefore, while most easily handled and interpreted, the results obtained within the ``rigid PMNS'' assumption do not exhaust all model building possibilities associated to discrete groups.  

As a consequence of $G_e$ and $G_\nu$ rigidly fixing the mass bases, it is possible to choose a basis in flavour space for the $l_i$ and $e^c_i$ fields in which the invariance of $M_E$ and $m_\nu$ forces them to be in the form
\begin{equation}
M_E = 
\begin{pmatrix}
A & 0 & 0 \\
0 & B & 0 \\
0 & 0 & C
\end{pmatrix}, 
\quad
m_\nu = 
U_0^* 
\begin{pmatrix}
a & 0 & 0 \\
0 & b & 0 \\
0 & 0 & c
\end{pmatrix}
U_0^\dagger \;,
\label{eq:rigid}
\end{equation}
with unconstrained complex diagonal entries and fixed $U_0$. The PMNS matrix is then determined up to phases and permutations: $U = P_e U_0 P_\nu \Psi$, where $\Psi$ is a diagonal matrix of Majorana phases and $P_e$, $P_\nu$ are permutation matrices arising 
because the definition of the PMNS matrix assumes lepton masses to have a specific ordering. 

\Eq{rigid} illustrates three general features of models relying on the above assumption: 
Majorana phases are not constrained; the PMNS matrix is predicted up to permutations of its rows and columns (and only one of the possible forms is usually suitable); neutrino and charged lepton masses are unconstrained. In particular the charged lepton mass hierarchies are not accounted for.
As a remedy to the latter drawback, the present approach can be complemented by adding an additional, possibly abelian, group factor $G_\text{FN}$, taking care of the charged lepton hierarchy. The breaking of $G_\text{FN}$ is perturbative, and it is arranged in such a way that the first two charged lepton families get suppressed, through a standard Froggatt-Nielsen (FN) mechanism~\cite{Froggatt:1978nt}.

The spontaneous breaking of $G$ is achieved as usual through the VEV of flavon fields $\phi$, breaking $G$ completely. The above set-up can be implemented if i) there exist subsets $\phi_e$, $\phi_\nu$ (not necessarily disjoint) of the full set of flavons breaking $G$ to $G_e$, $G_\nu$ respectively, and ii) only $\phi_e$ ($\phi_\nu$) enters $M_E$ ($m_\nu$). We will therefore refer to such an assumption as the ``sequestering'' approximation. 

The sequestering can hardly be exact: no ordinary flavour symmetry can prevent $\phi_e$ and $\phi_\nu$ from contaminating both parts of the Lagrangian. It can however happen to hold at some order in a perturbative expansion in the number of flavons. In other words, sequestering is ``accidental'', in the same sense in which lepton and baryon number are accidental in the SM. In order to see that the flavour symmetry cannot prevent contamination, we consider for simplicity the case in which neutrino masses are accounted for by the Weinberg operator. Suppose that only $\phi_e$ ($\phi_\nu$) enters $M_E$ ($m_\nu$), so that neutrino and charged fermion masses follow from the invariant Lagrangian
\begin{equation}
\mathcal{L_\text{seq}} = f(\phi_\nu)_{ij} \, (l_i H) (l_j H) + g(\phi_e)_{ij} \, e^c_i l_j H^* \;,
\label{eq:exactsequestering}
\end{equation}
when $\phi_{\nu,e} \to \vev{\phi_{\nu,e}}$. The dependence on the flavons is often simple, but in order to be general, we consider generic (say polynomial) functions $f$ and $g$. The invariance of the Lagrangian requires  
\begin{equation}
f(\phi_\nu) = U^T_l f(U_{\phi_\nu}\phi_\nu) U^{\phantom{T}}_l \;, 
\qquad
g(\phi_e) = U^T_{e^c} g(U_{\phi_e}\phi_e) U^{\phantom{T}}_l \;,
\label{eq:fgcovariance}
\end{equation}
where $U_{\phi_\nu}$ and $U_{\phi_e}$ are the representations of $G$ on the flavons $\phi_\nu$ and $\phi_e$ respectively. It is then easy to see that terms breaking the sequestering assumption are allowed. As an example, terms such as
\begin{equation}
\mathcal{L'} = (f(\phi_\nu) g^\dagger(\phi_e) g(\phi_e))_{ij} \, (l_i H) (l_j H) + (g(\phi_e) f^\dagger(\phi_\nu) f(\phi_\nu)))_{ij} \, e^c_i l_j H^* 
\label{eq:spoiledsequestering}
\end{equation}
are allowed and can spoil the invariance of $m_\nu$, $M_E$ under $G_\nu$, $G_e$. Therefore, no symmetry argument can  prevent the sequestering to be spoiled at higher orders in the flavon expansion.\footnote{Needless to say, any further symmetry added to take care of the sequestering can be included in $G$, so that the argument would still hold. In the case of supersymmetric models, the holomorphicity of the superpotential prevents the corrections in \eq{spoiledsequestering} from arising within the superpotential. On the other hand, they can still arise in the K\"ahler potential and propagate to the flavour lagrangian once the K\"ahler is brought into its canonical form, see \Sect{SB}.} In particular, if the typical size of symmetry breaking corrections in the neutrino sector is $\epsilon$, the sequestering-breaking corrections in the charged lepton sector can be expected to be at least $\ord{\epsilon^2}$ and viceversa.

Even if generically present, mixed $\phi_\nu$-$\phi_e$ corrections to \eq{exactsequestering} can be negligible. In such a case, it must be possible to account for the exact values of lepton flavour observables in the limit of exact sequestering. In the next \Sect{exact_sequestering} below we review this class of models,\footnote{Such models are also called ``direct''~\cite{King:2009ap}.} assuming $G_e$ and $G_\nu$ rigidly determine the mass bases, while models in which non-negligible corrections are needed in order to fit data will be considered in \Sect{approx_sequestering}. In \Sect{loose}, we will consider the case in which $G_e$ and $G_\nu$ loosely determine the lepton mass bases.

\subsubsection{Exact sequestering, rigid PMNS}
\label{sec:exact_sequestering}

We consider the possibility that the corrections to sequestering are negligible, so that lepton flavour is accounted for, within the present experimental accuracy, by the Lagrangian in \eq{exactsequestering}. 
The VEVs of the flavons $\phi_\nu$ and $\phi_e$ break $G$ to the $G_\nu$ and $G_e$ subgroup respectively, under which the full $m_\nu$ and $M_E$ are invariant. The subgroups $G_\nu$ and $G_e$ are assumed to unambiguously (up to phases) identify the neutrino and charged lepton mass eigenstate directions in flavour space. In this context, a non-vanishing $\theta_{13}$ must be obtained directly from the misalignment of $G_\nu$ and $G_e$. Simple groups such as $A_4$ and $S_4$, leading to $\theta_{13} = 0$, will be considered in \Sect{approx_sequestering}. 

The form of $M_E$ and $m_\nu$ is subject to general constraints. By using a flavour basis in which $M_E$ or $m_\nu$ is diagonal, and assuming that all neutrinos are massive, we see that $G_e \subseteq \text{U(1)}_e \times \text{U(1)}_\mu \times \text{U(1)}_\tau$ and $G_\nu \subseteq \mathbf{Z}_2^3$, where one of the $\mathbf{Z}_2$ in an overall sign change and is therefore irrelevant. On the other hand, in order for the mass basis to be rigidly identified by the residual groups, and assuming that the residual groups are finite, we need $G_e$ to contain either $\mathbf{Z}_n$, with $n$ a prime number and $n \geq 3$, or $\mathbf{Z}_2^2$.\footnote{\label{foot:Z} In order to prove this result, we first observe that $G_e$ must contain at least three elements, otherwise the charged lepton mass basis would not be fully determined. Given a $z\in G_e$, $z\neq 1$, there exists a minimum $n\in \mathbb{N}$ such that $z^n =1$. If $n \geq 3$, the result is proven (if $n=p\times q$ is not prime, one uses recursively that $\mathbb{Z}^{p\times q}$ contains both $\mathbb{Z}^{p}$ and $\mathbb{Z}^{q}$). If $n \leq 2$, then $z^2 = 1$. We call $w\neq 1,z$ a third element of $G$. Again we must have $w^n = 1$ for a minimum $n\in \mathbb{N}$. If $n\geq 3$, the statement is proven. Otherwise, $w^2 = 1$, and $G_e$ contains two $\mathbb{Z}_2$. Moroever, since $w$ and $z$ belong to a (abelian) subgroup of $\text{U(1)}_e \times \text{U(1)}_\mu \times \text{U(1)}_\tau$, $w$ and $z$ must commute, and so the two $\mathbb{Z}_2$. Therefore, in the case $G \supseteq \mathbf{Z}_2 \times \mathbf{Z}_2$, and the statement is proven. Analogously one shows that $G_\nu$ must contain $\mathbf{Z}_2 \times \mathbf{Z}_2$.} On the neutrino side, we need $G_\nu \supseteq \mathbf{Z}_2^2$. Therefore we conclude that 
\begin{equation}
\text{$\mathbf{Z}_n$ ($n \geq 3$ prime) or $\mathbf{Z}_2^2$} \subseteq G_e \subseteq \text{U(1)}_e \times \text{U(1)}_\mu \times \text{U(1)}_\tau
\qquad
\mathbf{Z}_2^2 \subseteq G_\nu \subseteq \mathbf{Z}_2^3 
\qquad \text{(non-zero neutrino masses)} \;.
\label{eq:GEGn}
\end{equation}
Neutrino data is compatible with one vanishing neutrino mass. If one neutrino is massless, the constraint on $G_\nu$ becomes 
\begin{equation}
\mathbf{Z}_n \times \mathbf{Z}_2 \text{ ($n \geq 3$ prime) } \subseteq G_\nu \subseteq \text{U(1)} \times \mathbf{Z}_2^2 
\qquad \text{(one vanishing neutrino mass)} \;.
\label{eq:GEGn0}
\end{equation}
If one neutrino is massless, there is then more freedom in the choice of $G_\nu$, which is otherwise constrained to be the Klein group $\mathbb{Z}_2\times \mathbb{Z}_2$ (up to a third, irrelevant $\mathbb{Z}_2$).\footnote{Another opportunity to enlarge $G_\nu$ arises with Dirac neutrinos~\cite{Esmaili:2015pna}.}

A systematic analysis of the phenomenologically viable PMNS matrices that can be obtained in this context has been carried out in the assumption that all neutrinos are massive and that the group $G$ is finite~\cite{Fonseca:2014lfa}. The only possible viable PMNS matrices are in a ``trimaximal'' form (TM$_{2}$, see \Sect{approx_sequestering}), with $|U_{e2}|^2 = |U_{\mu 2}|^2 = |U_{\tau 2}|^2 = 1/3$, which predicts $\sin^2\theta_{12} \geq 1/3$. More precisely ($(|U|^2)_{ij}\equiv |U_{ij}|^2$),
\begin{equation}
|U|^2 = \frac{1}{3}
\begin{pmatrix}
1 + \re{\sigma} & 1 & 1 - \re\sigma \\
1 + \re{\omega\sigma} & 1 & 1 - \re{\omega\sigma} \\
1 + \re{\omega^2\sigma} & 1 & 1 - \re{\omega^2\sigma}
\end{pmatrix} \;,
\label{eq:sequesteringU}
\end{equation}
where $\sigma = \text{exp}(2i\pi p/n)$ is a root of unity and $\omega = \text{exp}(2\pi i/3)$. The integers $p$ and $n$ can be taken to be coprime, in which case the minimal discrete group leading to a PMNS matrix in the above form is \begin{itemize}
\item
$\Delta(6m^2)$, where $3m$ is the least common multiple of 6 and $n$, if 9 does not divide $n$;
\item
$\left( \mathbb{Z}_m \times \mathbb{Z}_{m/3} \right) \rtimes S_3$, where $m$ is the least common multiple of 2 and $n$, if 9 divides $n$.
\end{itemize}

The definition of these groups can be found for example in~\cite{Ishimori:2010au}. \Eq{sequesteringU} determines the absolute values of the PMNS entries. The Majorana phases are not constrained, as discussed above. The Dirac phase is instead fixed and predicted to be trivial ($\sin\delta = 0$) in all viable cases, which also predict a non-negligible  deviation from maximal $\theta_{23}$. For a given choice of $\sigma$ (hence of the group), \eq{sequesteringU} corresponds to one of the 36 possible permutations of rows and columns that can in principle arise. 

One of the first attempts at achieving $\theta_{13} \neq 0$ directly from the interplay of $G_\nu$ and $G_e$ used the $\Delta(96)$ group ($m=4$, $n=12$, $\sigma = \text{exp}(i\pi /6)$) ~\cite{Toorop:2011jn,Ding:2012xx,Varzielas:2012ss,King:2012in} but overshot the experimental value of $\theta_{13}$. Experimentally viable possibilities were considered in~\cite{Holthausen:2012wt,King:2013vna,Hagedorn:2013nra,Talbert:2014bda}. The smallest viable $\Delta(6m^2)$ group corresponds to $m=22$ ($n=11,22,33,66$) and has order 2904, while the smallest viable $\left( \mathbb{Z}_m \times \mathbb{Z}_{m/3} \right) \rtimes S_3$ corresponds to $m=18$ ($n=9,18$) and has order 648. Needless to say, such groups are more cumbersome than the ones originally proposed to account for the neutrino mixing pattern. Note that a dynamical mechanism to spontaneously break $G$, and preserve an accurate sequestering, also needs to be exhibited. 

As mentioned, neutrino data is compatible with a single neutrino being massless. If that is the case, the rules of the game allow $G_\nu$ to be larger than the Klein group $\mathbb{Z}_2\times \mathbb{Z}_2$, and the structure of the flavour group to be different. In all models studied so far, non-negligible corrections to the leading order (exact sequestering) results are needed in order to obtain a phenomenologically viable model~\cite{Joshipura:2013pga,Joshipura:2014pqa,King:2016pgv}. 

\subsubsection{Approximate sequestering, rigid PMNS}
\label{sec:approx_sequestering}

In this Subsection, we still assume that $G_\nu$ and $G_e$ rigidly determine the lepton mass eigenvectors up to phases, and therefore the PMNS matrix. However, we allow the PMNS matrix thus obtained to be only a leading order approximation of the measured one, and we rely on sub-leading corrections for an accurate agreement. 

Before discussing their origin, we illustrate some possible leading order forms of the PMNS matrix and the size of the needed corrections. Before the measurement of $\theta_{13}$, the model building efforts were mainly based on three forms of the PMNS matrix, all associated to simple discrete flavour symmetries. They all correspond to maximal $\theta_{23}$ and vanishing $\theta_{13}$, and only differ by the value of the solar angle $\theta_{12}$: 
\begin{description}
\item[\textmd{Tribimaximal (TB)}] $\sin^2\theta_{12} = 1/3$, $\sin^2\theta_{23} = 1/2$, $\sin^2\theta_{13} = 0$. \\\cite{Harrison:2002er,Harrison:2002kp,Harrison:2003aw}
\item[\textmd{Bimaximal (BM)}] $\sin^2\theta_{12} = 1/2$, $\sin^2\theta_{23} = 1/2$, $\sin^2\theta_{13} = 0$. \\\cite{Fukugita:1998vn,Barger:1998ta}
\item[\textmd{Golden ratio (GR)}] $\tan^2\theta_{12} = 1/\phi$ or $\cos\theta_{12} = \phi/2$, $\phi = (1+\sqrt{5})/2$ (golden ratio), $\sin^2\theta_{23} = 1/2$, $\sin^2\theta_{13} = 0$. \\ \cite{Datta:2003qg,Kajiyama:2007gx,Rodejohann:2008ir}
\end{description}
In all the three cases, the PMNS matrix, up to external phases, is in the form
\begin{equation}
U = 
\begin{pmatrix}
c_{12} & s_{12} & 0 \\
\displaystyle -\frac{s_{12}}{\sqrt{2}} & \displaystyle \frac{c_{12}}{\sqrt{2}} & -\displaystyle \frac{1}{\sqrt{2}} \\
-\displaystyle \frac{s_{12}}{\sqrt{2}} & \displaystyle \frac{c_{12}}{\sqrt{2}} & \displaystyle \frac{1}{\sqrt{2}} \\
\end{pmatrix} \;,
\label{eq:PMNSdiscrete0}
\end{equation}
with different values of $\theta_{12}$, as specified above. 

We compare the predictions with the experimental values. The present $1\sigma$ ranges of the neutrino mixing angles, as obtained from global fits (see table~\ref{tableData}) are $\sin\theta_{12} = 0.56\pm 0.01$, $\sin\theta_{23} = 0.75\pm0.02$, $\sin\theta_{13} = 0.150\pm 0.002$, while the predictions obtained in the above schemes are $(\sin\theta_{12})_\text{TB} = 0.58$, $(\sin\theta_{12})_\text{BM} = 0.71$, $(\sin\theta_{12})_\text{GR} = 0.59\text{ or }0.62$, $(\sin\theta_{23})_\text{all} = 0.71$, $(\sin^2\theta_{13})_\text{all} = 0$.

Most encouraging is the TB prediction for $\theta_{12}$, in close agreement with the precise experimental determination. Parametrising the corrections to \eq{PMNSdiscrete0} in a power series in $\lambda_C = 0.22$ (the Cabibbo angle, an expansion parameter borrowed from the quark sector), we see that the agreement is so precise that only corrections $\ord{\lambda_C^{2\div 3}}$ or less are allowed. This provided a considerable boost to models accounting for TB mixing, at a time when the $\theta_{13}$ angle was still unknown. Unfortunately, the experiment now shows that $\theta_{13}$ departs from zero by $\ord{\lambda_C}$. If that is the expected size of corrections to \eq{PMNSdiscrete0}, the success of the TB prediction for $\theta_{12}$ should be considered accidental. Within the same $\ord{\lambda_C}$ accuracy, the measured value of $\theta_{12}$ is as well compatible with the BM prediction $\theta_{12} = \pi/4$. It has in fact been observed that the empirical relation $\theta_{12} + \lambda_C \approx \pi/4$ (``quark-lepton complementarity''~\cite{Raidal:2004iw,Minakata:2004xt,Datta:2005ci,Everett:2005ku,Schmidt:2006rb}) approximately holds. Needless to say, the size of the corrections hinted by the value of $\theta_{13}$ in this class of models partly jeopardizes the predictivity motivation. 

We now focus on the TB scheme and illustrate the model building logic underlying it. This will also serve as an illustration of the ideas and techniques underlying more involved models. The tribimaximal form of the PMNS matrix is, up to external phases, 
\begin{equation}
U_\text{TB} = 
\begin{pmatrix}
\displaystyle \sqrt{\frac{2}{3}} & \displaystyle \frac{1}{\sqrt{3}} & 0 \\
\displaystyle -\frac{1}{\sqrt{6}} & \displaystyle \frac{1}{\sqrt{3}} & \displaystyle -\frac{1}{\sqrt{2}} \\
\displaystyle -\frac{1}{\sqrt{6}} & \displaystyle \frac{1}{\sqrt{3}} & \displaystyle \frac{1}{\sqrt{2}}
\end{pmatrix} .
\label{eq:TB}
\end{equation}

The form of the PMNS matrix determines the relative orientation of $G_e$ and $G_\nu$ in $G$, the commutation relations of the corresponding elements in $G$, and consequently the minimal structure of $G$. The procedure to find the minimal $G$ (when it exists --- only specific forms of the PMNS originate from finite groups) is simple. First, we need to specify $G_\nu$ and $G_e$. For $G_\nu$ the choice is essentially unique, as we assume here that all three neutrinos are massive: $G_\nu = \mathbb{Z}_2 \times \mathbb{Z}_2$. We call $u$ and $s$ the non trivial elements of the two $\mathbb{Z}_2$. In a neutrino mass basis, their representation on the lepton doublets is 
\begin{equation}
U_l^\nu = 
\begin{pmatrix}
-1 & 0 & 0 \\
0 & -1 & 0 \\
0 & 0 & 1
\end{pmatrix} \;, 
\qquad
S_l^\nu = 
\begin{pmatrix}
-1 & 0 & 0 \\
0 & 1 & 0 \\
0 & 0 & -1
\end{pmatrix} \;.
\label{eq:Z2nu}
\end{equation}
The choice of $G_e$ is not unique, see \eq{GEGn}. The smallest (in terms of number of elements) option is $\mathbb{Z}_3$. We call $t$ one of its non-trivial elements. Without loss of generality, its representation on the $l$ and $e^c$ fields, in a charged lepton mass basis, is
\begin{equation}
T_l^e = 
\begin{pmatrix}
1 & 0 & 0 \\
0 & \omega & 0 \\
0 & 0 & \omega^2
\end{pmatrix} , 
\qquad
T_{e^c}^e = 
\begin{pmatrix}
1 & 0 & 0 \\
0 & \omega^2 & 0 \\
0 & 0 & \omega
\end{pmatrix} , 
\label{eq:Z3e}
\end{equation}
where $\omega = \text{exp}(2\pi i/3)$. Therefore, with the present choice of $G_\nu$, $G_e$, the full group $G$ must contain the identity, the three elements $u$, $s$, $t$, and all of their products. In the assumption that the representation on the leptons is faithful,
the group elements can be identified with their representations on the lepton doublets, $U_l$, $S_l$, $T_l$. We need however to write them in the same basis. Choose for example a charged lepton mass basis. Then $T_l$ is given by \eq{Z3e}, while $U_l$ and $S_l$ must be rotated from the neutrino basis used in \eq{Z2nu}. The rotation is of course given by the PMNS matrix $U$ (beware of the abuse of the notation ``$U$''): $U_l^e = U U_l^\nu U^\dagger$, $S_l^e = U S_l^\nu U^\dagger$. Here is where the chosen form of $U$ enters. In the TB case, $U = \Psi U_\text{TB} \Phi$, where $\Psi$ and $\Phi$ are diagonal matrices of phases. With a proper choice of the phases  of the charged leptons, $\Psi = \mathbf{1}$, while $\Phi$ cancels in the products, so that $U_l^e = U_\text{TB}^{\phantom{\dagger}} U_l^\nu U_\text{TB}^\dagger$, $S_l^e = U^{\phantom{\dagger}}_\text{TB} S_l^\nu U_\text{TB}^\dagger$. All in all, 
\begin{equation}
U^e_l = -
\begin{pmatrix}
1 & 0 & 0 \\
0 & 0 & 1 \\
0 & 1 & 0
\end{pmatrix} ,
\quad
S^e_l = \frac{1}{3}
\begin{pmatrix}
-1 & 2 & 2 \\
2 & -1 & 2 \\
2 & 2 & -1
\end{pmatrix} ,
\quad
T_l^e = 
\begin{pmatrix}
1 & 0 & 0 \\
0 & \omega & 0 \\
0 & 0 & \omega^2
\end{pmatrix} .
\label{eq:S4generators}
\end{equation}
By taking all possible products of the three matrices above, it is easy to show that the group $G$ generated by them is finite, contains 24 distinct elements, and is isomorphic to $S_4$, the permutation group of 4 elements. 

The $S_4$ group has two $d=3$, one $d=2$ and two $d=1$ irreducible representations, denoted by $3_1$, $3_2$, 2, $1_1$, $1_2$. The $3_1$ representation is defined by $S$, $T$, $U$ in \eq{S4generators} and the $3_2$ has opposite $U$. The 2 representation has 
\begin{equation}
S=\begin{pmatrix}
1 & 0 \\
0 & 1
\end{pmatrix},
\quad
T=\begin{pmatrix}
\omega & 0 \\
0 & \omega^2
\end{pmatrix},
\quad
U=\begin{pmatrix}
0 & 1 \\
1 & 0
\end{pmatrix},
\label{eq:S42}
\end{equation}
and the $1_1$, $1_2$ representations have $S=T=1$ and $U=\pm1$ respectively. 

The $S_4$ option for TB mixing is motivated and has been widely studied~\cite{Mohapatra:2003tw,Ma:2005pd,Hagedorn:2006ug,Cai:2006mf,Zhang:2006fv,Bazzocchi:2008ej,Ishimori:2008fi,Bazzocchi:2009pv,Bazzocchi:2009da,Ding:2009iy,Dutta:2009ij,Dutta:2009bj,Meloni:2009cz,Hagedorn:2010th,Ishimori:2011mt,Morisi:2011pm,BhupalDev:2011gi,BhupalDev:2012nm,Smirnov:2018luj} however its simplest implementation requires a non-trivial fine-tuning to reproduce hierarchical charged leptons, as we now show.\footnote{The fine-tuning is associated to the underlying $\mu$-$\tau$ symmetry~\cite{Fukuyama:1997ky,Mohapatra:1998ka,Ma:2001mr,Balaji:2001ex,Lam:2001fb,Ma:2002ce,Grimus:2012hu,Xing:2015fdg}, corresponding to the $U$ generator of $S_4$. While the need of fine-tuning in the context of the $\mu$-$\tau$ symmetry has been pointed out long ago~\cite{Kitabayashi:2002jd}, the general argument provided here holds in $S_4$ independently of the (viable) choice of the lepton and flavon representations.} In order to implement the $S_4$ symmetry, we first need to assign the lepton fields $l_i$ and $e^c_i$ to $S_4$ representations. \Eq{S4generators} assigns the $l_i$ fields to a $3_1$. The representation on the $e^c_i$ fields should be such that $T^e_{e^c}$ is given by \eq{Z3e}, which in turn requires them to form one of the following four representations: $3_1$, $3_2$, $2+1_1$, $2+1_2$. The $l$ and $e^c$ fields must then couple to a combination of flavon fields, with $T$-preserving VEV, in a $S_4$ invariant Yukawa interaction. All possible such combinations lead to a diagonal charged lepton mass matrix with at least two diagonal elements of equal size (and possibly different sign). In order to obtain hierarchical and non-vanishing charged lepton masses, a fine-tuning of independent contributions to those diagonal entries must then be invoked. The argument is based on the assumption that $T$ in not broken in the charged lepton sector (so that its mass basis is rigidly determined) at leading order. The possibility that $T$ is broken is considered in \Sect{loose}. 

The above fine-tuning can be avoided if $S_4$ arises accidentally in models based on $A_4$~\cite{Ma:2001dn,Babu:2002dz,Hirsch:2003dr,Ma:2004zv,Ma:2004zd,Altarelli:2005yp,Chen:2005jm,Ma:2005sha,Hirsch:2005mc,Ma:2005mw,Zee:2005ut,Altarelli:2005yx,He:2006dk,Adhikary:2006wi,Ma:2006sk,Lavoura:2006hb,Altarelli:2006kg,Ma:2006vq,Morisi:2007ft,Hirsch:2007kh,Yin:2007rv,Bazzocchi:2007na,Grimus:2008tm,Honda:2008rs,Altarelli:2008bg,Adhikary:2008au,Hirsch:2008rp,Lin:2008aj,Bazzocchi:2008sp,Morisi:2009qa,Ciafaloni:2009ub,Lin:2009ic,Altarelli:2009kr,Antusch:2010es,delAguila:2010vg,Kadosh:2010rm,Gupta:2011ct,BenTov:2012tg,Holthausen:2012wz,Morisi:2013eca,Morisi:2013qna,Chu:2016lkb}, its subgroup of even permutations. The latter has 12 elements and is generated by $S$ and $T$ only. The flavour symmetry extends to $S_4$ if the Lagrangian (at some order in the flavon expansion) turns out to be accidentally invariant under the $U$ generator. Such an option is appealing for a number of reasons: the $A_4$ group is even more minimal than $S_4$; the invariance of the Lagrangian under the $U$ transformation is accidental, which allows welcome corrections to TB mixing; and, as mentioned, no fine-tuning is required in order to obtain hierarchical, non-vanishing charged leptons. Both $A_4$ and $S_4$ can arise from continuous non-abelian groups~\cite{Bazzocchi:2007au,Bazzocchi:2008rz,Berger:2009tt,Grossman:2014oqa}, can be related to compactification in models with two extra-dimensions~\cite{Altarelli:2006kg,Kobayashi:2008ih} and to the modular group~\cite{Altarelli:2005yx}, see also \Sect{modular}.

We see how to implement the above ideas in a concrete model based on $A_4$~\cite{Altarelli:2005yp,Altarelli:2005yx}. We first need to specify the $A_4$ representation on the lepton fields $l_i$ and $e^c_i$. The $A_4$ group has one $d=3$ and three  $d=1$ irreducible representations, denoted by $3$, $1$, $1'$, $1''$. The $3$ representation is defined by $S$, $T$ in \eq{S4generators} and the $1$, $1'$, $1''$ representations are defined by $S=1$ and $T = 1$, $\omega$, $\omega^2$ respectively. \Eq{S4generators} assigns the $l_i$ fields to a $3$. The representation on the $e^c_i$ fields should be such that $T^e_{e^c}$ is given by \eq{Z3e}. Hence, either $e^c \sim 3$ or $e^c \sim 1+1'+1''$. The first option is not welcome, as it allows the charged lepton to get degenerate, non-vanishing, leading order masses. In order to avoid it, one chooses $e^c \sim 1+1'+1''$. 

We now need to couple the leptons to flavons in such a way that $G_e$ and $G_\nu$ are preserved (at leading order) by $M_E$ and $m_\nu$. In the $A_4$ case, $G_\nu$ is generated by $S$ and $G_e$ by $T$. $G_\nu = \mathbb{Z}_2$ alone is not sufficient to determine the neutrino mass basis up to phases, but it gets help from the $U$ transformation, under which $m_\nu$ will turn out to be accidentally invariant. In order to break $G$ to $G_e$, $S$ must be broken, but $T$ must not. This can only be achieved by using a flavon triplet $\varphi_T$, as $1$, $1'$, $1''$ are all invariant under $S$. The index $T$ refers to the invariance under $T$, which forces $\vev{\varphi_T} = \epsilon_T (1,0,0)^T$. Nicely, $A_4$ invariance allows $\varphi_T$ to couple to $e^c l$, at the linear level. The most general charged lepton Yukawa Lagrangian, at leading order in the flavon expansion, is then 
\begin{equation}
\mathcal{L}^{(1)}_e = \lambda_1 e^c_1 (\varphi_T l)_1 H^* + \lambda_2 e^c_2 (\varphi_T l)_{1'} H^*  + \lambda_3 e^c_3 (\varphi_T l)_{1''} H^* \;,
\label{eq:A4Elinear}
\end{equation}
where $\varphi_T$ is dimensionless, i.e.\ normalised to some cutoff scale $\Lambda$, and $()_{1,1',1''}$ denote the triplet contractions transforming as $1,1',1''$ under $A_4$. More precisely, if $a$ and $b$ transform as $3$, $(ab)_1 = a_1 b_1 + a_2 b_3 + a_3 b_2$,  $(ab)_{1'} = a_3 b_3 + a_1 b_2 + a_2 b_1$, $(ab)_{1''} = a_2 b_2 + a_1 b_3 + a_3 b_1$, in the basis specified by~\eq{S4generators}.

Before switching to the neutrino sector, we comment on the above result. The Lagrangian in \eq{A4Elinear} generates a diagonal charged lepton mass matrix as desired, with $(M_E)_{ii} = \lambda_i \epsilon_T v$, where $v$ is the Higgs VEV. The identification of the three families with the $e$, $\mu$, $\tau$ mass eigenstates depends on the relative size of the diagonal entries, and might require field permutations.
The mass hierarchy can be reproduced, without fine-tuning, by an appropriate choice of the $\lambda_i$'s, but it is not explained. In order to account for it (and get rid of the permutation ambiguity), an additional U(1)$_\text{FN}$ factor can be added to the flavour group. The latter is assumed to broken by a flavon VEV $\vev{\varphi_\text{FN}} = \epsilon \ll 1$. By a proper choice of their charge under U(1)$_\text{FN}$, the individual monomials in \eq{A4Elinear} can be forced to contain different powers of $\varphi_\text{FN}$. The corresponding diagonal masses will then get suppressed by different powers of $\epsilon$. 

Alternatively, the role of $\varphi_\text{FN}$ can be played by the $A_4$ flavons themselves~\cite{Lin:2008aj,Altarelli:2009kr}. Suppose that $\vev{\varphi_T} = \epsilon_T (0,1,0)^T$. Such a VEV breaks $T$, and in fact the whole $A_4$ and $S_4$. 
Moreover, $\vev{(\varphi^2_T)_3} = \epsilon^2_T (0,0,1)^T$, and $\vev{(\varphi^3_T)_3} = \epsilon^3_T (1,0,0)^T$, where the index ``3'' denotes the component transforming as the 3 of $A_4$. Therefore multiple insertions of $\vev{\varphi_T}$ are associated to different families, and are more and more suppressed by higher powers of $\epsilon_T$: the $A_4$ flavon $\varphi$ effectively plays the role of a FN flavon.

We now come to the neutrino mass matrix and consider for simplicity its description in terms of the Weinberg operator. We first note that $A_4$ allows an invariant term $(lH lH)_1 /\Lambda_L$, corresponding to three degenerate neutrinos. Such an invariant term needs to be of similar size as the symmetry breaking terms, which may be expected to be suppressed, if a perturbative flavon expansion is to be meaningful. The invariant term can be correspondingly suppressed by forcing it to break ad hoc symmetries.
With this in mind, we will allow the ``invariant'' and the symmetry breaking contributions to the neutrino mass matrix to be comparable. 

In order to break $G$ to $G_\nu$, $T$ must be broken, but $S$ must not. $T$ can in principle be broken by a 3, $1'$, or $1''$ representation. In order to have accidental invariance under $U$ (for generic values of the Lagrangian parameters), $G$ should be broken by a flavon triplet $\varphi_S$, 
where the index $S$ refers to the invariance under $S$, which forces $\vev{\varphi_S} = \epsilon_S (1,1,1)^T$. Nicely, $A_4$ invariance allows $\varphi_S$ to couple to the Weinberg operator, at the linear level. The most general neutrino Lagrangian, at the linear order in the flavon expansion, is then 
\begin{equation}
\mathcal{L}^{(1)}_\nu = \epsilon \frac{(lH lH)_1}{2\Lambda_L}  + \varphi_S \frac{(lH lH)_{3S}}{2\Lambda_L} \;,
\label{eq:A4nulinear}
\end{equation}
where, in the notations used above, the symmetric contraction of the lepton indices into a triplet is $(ab)_{3S} = (2a_1 b_1 - a_2 b_3 -a_3 b_2, 2a_2 b_2 - a_3 b_1 -a_1 b_3, 2a_3 b_3 - a_1 b_2 -a_2 b_1)$. The corresponding neutrino mass matrix is
\begin{equation}
m^{(1)}_\nu = 
\begin{pmatrix}
a+2b & -b & -b \\
-b & 2b & a - b \\
-b & a - b & 2 b
\end{pmatrix} 
, \quad
a = \epsilon \, \frac{v^2}{\Lambda_L}
\;,\quad
b = \epsilon_S \frac{v^2}{\Lambda_L} \;.
\label{eq:A4mnulinear}
\end{equation}
The matrix $m^{(1)}_\nu$ is accidentally invariant under $U$, as desired. Moreover, together with $M_E$, it leads to TB mixing. As $m^{(1)}_\nu$ is not the most general matrix invariant under $S$ and $U$, the relation $m_3 \, e^{i\alpha_{31}} = m_1 -2  m_2 \,e^{i\alpha_{21}}$ holds among the neutrino masses and the Majorana phases $\alpha_{21}$, $\alpha_{31}$ (defined as in \eq{majo2018}).\footnote{Which of the three eigenvalues $|3b+a|$, $|a|$, $|3b-a|$ are identified with $m_1$, $m_2$, $m_3$, in their standard ordering, depends on their relative size. In order for the TB form of the PMNS matrix not to be spoiled by permutations of its columns, the identification should give $m_1 = |3b+a|$, $m_2 = |a|$, $m_3 =|3b-a|$. The relation among masses and phases is an example of mass sum rules~\cite{King:2013psa,Gehrlein:2015ena}.} In the context of see-saw models, an analogous relation holds for the inverse masses. 

We have seen that TB mixing can be obtained from the Lagrangian $\mathcal{L}^{(1)}_e + \mathcal{L}^{(1)}_\nu$. Crucial to this result is the fact that the Lagrangian is in the form in \eq{exactsequestering}, with $\phi_e = \varphi_T$ only entering the charged lepton mass matrix and $\phi_\nu = \varphi_S$ only entering the neutrino mass matrix. In order to enforce such a sequestering, $\varphi_T$ and $\varphi_S$ must be given different quantum numbers, under an additional group factor. For example, one can add a $\mathbf{Z}_3$ factor, under which $\varphi_T$ and $e^c l$ are invariant, while $\varphi_S$ and $lH lH$ transform non-trivially (in conjugated representations). This way, the Lagrangian is forced to be in the form $\mathcal{L}^{(1)}_e + \mathcal{L}^{(1)}_\nu$ at the leading order in the flavon expansion. 

A complete model must also account for the specific alignment of the VEVs, $\varphi_T \propto (1,0,0)$, $\varphi_S \propto (1,1,1)$, assumed above. Indeed, the TB prediction crucially depends on such an alignment, more than from the flavour group itself or the choice of the flavon fields. In other words, what actually underlies the TB prediction is the flavon potential determining the flavon VEVs. It can be shown~\cite{Altarelli:2005yx} that the needed alignment can be naturally obtained in supersymmetric models. 

We have illustrated above how TB mixing can be obtained from an $A_4$ flavour group (supplemented with additional symmetry factors and a proper flavon potential), at the leading order in a flavon expansion. Besides $A_4$ and $S_4$, other finite groups can lead to TB mixing, for example PSL$_2(7)$~\cite{Luhn:2007yr,King:2009mk,King:2009tj,Ferreira:2012ri,Chen:2014wiw}, $\Delta(27)$~\cite{deMedeirosVarzielas:2006fc,Luhn:2007uq,Ma:2007wu,Grimus:2008tt,Bjorkeroth:2015uou,Bjorkeroth:2016lzs}, $\mathbb{Z}_7\rtimes \mathbb{Z}_3$~\cite{Luhn:2007sy,Hagedorn:2008bc,Cao:2010mp,Vien:2014gza,Bonilla:2014xla,Hernandez:2015cra}, $\mathbb{Z}_{13}\rtimes \mathbb{Z}_3$~\cite{Kajiyama:2010sb,Hartmann:2011pq,Hartmann:2011dn,Perez:2019aqq}. Other mixing schemes can be obtained closely following the model-building lines outlined above for $S_4$ and $A_4$. For example, BM mixing can be obtained from $S_4$~\cite{Altarelli:2009gn,Meloni:2011fx} and GR schemes can be obtained from $A_5$~\cite{Kajiyama:2007gx,Everett:2008et,Feruglio:2011qq,Hernandez:2012ra,Gehrlein:2014wda,Gehrlein:2015dxa}.

\paragraph*{Origin of the corrections to approximate sequestering.} 
\label{sec:corrections}

~\\[2mm]
The simplest non-abelian finite group models lead to TB, BM, or GR forms of the PMNS matrix and therefore need to be corrected in order to account for $\theta_{13} \neq 0$. Such corrections are also needed in models based on higher order finite groups leading to a $\theta_{13} \neq 0$, but still not in agreement with the experimental value. 
The above predictions are obtained at the LO in the flavon expansion, at which the lagrangian has the form in \eq{exactsequestering}, supplemented by a flavon potential providing the necessary alignment of $\phi_e$ and $\phi_\nu$. The corrections are associated to higher order terms, and can affect the LO predictions by either i) spoiling the sequestering or ii) spoiling the alignment mechanism provided by the leading order potential. Such corrections are usually $G$-invariant, but they can also be non-invariant because i) part of $G$ arises at the LO as an accidental symmetry, or ii) the group $G$ is not gauged and the corrections are of gravitational nature. In order for the latter case to be phenomenologically relevant, the cutoff scale $\Lambda$ characterising the operator expansion should be sufficiently close to the gravity cutoff. While the form of the corrections is model-dependent, a few model-independent considerations can be made. 

\paragraph*{Size of the corrections. }

~\\[2mm]
The range of the corrections is important to assess whether they can lead to viable predictions and how much they spoil the predictivity of the model. 

The corrections to the LO predictions are associated to higher orders in the flavon expansion. There are two expansion parameters, associated to the typical size of the VEVs of the $\phi_e$ and $\phi_\nu$ flavons: $\epsilon_e \sim \vev{\phi_e}$, $\epsilon_\nu \sim \vev{\phi_\nu}$ (we remind that the flavons are dimensionless here, i.e.\ normalised to some cutoff scale $\Lambda$). We expect corrections to the neutrino and charged lepton mass matrices to be at least as large as $\ord{\epsilon_e^2}$ and $\ord{\epsilon_\nu^2}$ respectively, as discussed in \Sect{discrete_sequestering}.

The ranges of $\epsilon_e$, $\epsilon_\nu$ are often loosely constrained at LO. We first focus on $\epsilon_e$ and consider for example the form of $\mathcal{L}^{(1)}_e$ in \eq{A4Elinear}, where $\phi_e \equiv \varphi_T$ and $\epsilon_e \equiv \epsilon_T$. At LO, the tau lepton mass is given by $m_{\tau} = \lambda_\tau \epsilon_e v$, where $\lambda_\tau$ is the largest among the three couplings in \eq{A4Elinear}. The product $\lambda_\tau \epsilon_e$ is fixed by the tau mass, but $\epsilon_e$ is allowed to vary in quite a broad range, $ 10^{-2} \approx (m_\tau/v) \lesssim \epsilon_e \lesssim 1$. The upper bound is required for the perturbative expansion to be meaningful, and the lower bound corresponds to a coupling $\lambda_\tau$ in the perturbative regime $\lambda_\tau \lesssim 1$. The result still holds if the charged lepton mass hierarchy is accounted for by an independent suppression factor $\epsilon_\text{FN}$ associated to an abelian U(1) factor.
In the latter case, $m_{e_i} = \lambda_i \,\epsilon_\text{FN}^{n_i} \, \epsilon_e v$, where $n_i$ is an abelian charge.

The size of $\epsilon_\nu$ may be even less constrained. We consider for example the lagrangian in \eq{A4nulinear}, where $\phi_\nu \equiv \varphi_S$ and $\epsilon_\nu \equiv \epsilon_S$. \Eq{A4mnulinear} shows that $b$ is bound to be of the order of the light neutrino masses, but a small $\epsilon_\nu$ is allowed provided that $\Lambda_L$ (and $\epsilon$) is correspondingly small. For $\Lambda_L \gtrsim \TeV$ and normal hierarchy, one gets $10^{-12} \lesssim \epsilon_\nu \lesssim 1$. 

The above ranges for $\epsilon_e$ and $\epsilon_\nu$ are broad enough to allow the NLO corrections to be negligible or substantial, in either $M_e$, or $m_\nu$, or in both, and in general to allow the expansion parameters and LO corrections to have different sizes in the charged lepton and neutrino sectors. Such qualitative considerations can be refined or modified in a number of ways. For example, the mass matrices can be non-homogeneous in $\epsilon_e$, $\epsilon_\nu$. This is the case for example if the $A_4$ flavons play the role of FN flavons, and $\epsilon_\text{FN} = \epsilon_e$~\cite{Lin:2008aj,Altarelli:2009kr}. In such a case, the size of $\epsilon_e$ is determined by the charged lepton mass ratios. Moreover, additional constraints on the expansion parameters can arise in models accounting for leptogenesis~\cite{Mohapatra:2004hta,Mohapatra:2005wk,Jenkins:2008rb,Lin:2009ic,Branco:2009by,Bertuzzo:2009im,Hagedorn:2009jy,Riva:2010jm,AristizabalSierra:2011ab,Gehrlein:2015dxa}. 

\paragraph*{Structure of the corrections.}

~\\[2mm]
The PMNS matrix gets contributions from both the neutrino and charged lepton sectors, $U = U_e^\dagger U_\nu^{\phantom{\dagger}}$, as in \eq{PMNS}.
Corrections to the LO form of the PMNS matrix can be due to corrections to $M_E$ (affecting $U_e$) and to $m_\nu$ (affecting $U_\nu$). A special case arises when only one of the two corrections is significant. 

First consider the case in which the corrections come from the charged lepton sector~\cite{Giunti:2002ye,Giunti:2002pp,Frampton:2004ud,Altarelli:2004jb,Romanino:2004ww,Antusch:2004re,King:2005bj,Masina:2005hf,Antusch:2005kw}. This can happen, for example, if $\epsilon_e$ is on the lower side of its range, so that the $\ord{\epsilon^2_e}$ corrections to $m_\nu$ are negligible. 

The charged lepton mass matrix is diagonal at LO, due to $G_e$ invariance. Therefore, the leading order form of the PMNS matrix (TB, BM, GR, or else) is $U^0 = U_\nu^0$, where $U_\nu^0$ diagonalises the LO form of $m_\nu$. At higher orders, $M_E$ is non-diagonal and $m_\nu$ is unaffected. Thus, the PMNS matrix gets a correction from the charged lepton sector, $U = U^\dagger_e U^0$. 

The above observation, per se, is not very constraining: any PMNS matrix $U$ can now be obtained by choosing an appropriate $U_e = U^0_\nu U^\dagger$. The study of charged lepton corrections is useful when $U_e$ has a non-generic, motivated pattern. This is indeed often the case, as $U_e$ is in turn constrained by the hierarchy of charged lepton masses, if the latter is to be stable under small perturbations~\cite{Marzocca:2014tga}. If $M^E_{31}$ is not unexpectedly large, $|M^E_{31}|/m_\tau \ll \sin\theta_{13}$, $U_e$ is approximately in the form
\begin{equation}
U_e = R_{23}^T(\theta^e_{23}) R_{12}^T(\theta^e_{12}) \;,
\label{eq:Ue0}
\end{equation}
up to external phases, where $R_{ij}(\theta)$ is a $2\times 2$ rotation by an angle $\theta$ in the $ij$ block and the transpose is conventional. In all cases illustrated in \Sect{approx_sequestering}, $\theta_{13} = 0$ in $U_0$, hence $U^0_\nu$ is in the same form 
\begin{equation}
U^0_\nu = R_{23}(\theta^\nu_{23}) R_{12}(\theta^\nu_{12}) \;,
\label{eq:Unu0}
\end{equation}
up to external phases. The $\theta_{13}$ angle then originates purely from the interplay of 23 and 12 rotations, and the PMNS matrix is given by
\begin{equation}
U = R_{12}(\theta^e_{12}) \Phi R_{23}(\theta'_{23}) R_{12}(\theta^\nu_{12}) \;,
\label{eq:PMNSME}
\end{equation}
up to external phases, where $\Phi = \diag(1,\text{exp}(-i\delta'),1)$. In \eq{PMNSME}, $\theta^\nu_{12}$ corresponds to the LO prediction for $\theta_{12}$ and is fixed by the model ($\sin\theta^\nu_{12} = 1/\sqrt{3}$, $1/\sqrt{2}$ in TB, BM schemes respectively). The precise relations between the parameterisation in \eq{PMNSME} and the standard one can be found in~\cite{Marzocca:2013cr}. In first approximation, $\theta_{23} = \theta'_{23}$ and $\delta = \delta'$, up to $\ord{s^2_{13}}$ and $\ord{s_{13}}$ corrections respectively. Moreover, $\sin\theta_{13} = \sin\theta^e_{12} \sin\theta_{23}$ and $\sin\theta_{12} = \sin\theta^\nu_{12}(1+\sin\theta^e_{12} \cot\theta^\nu_{12} \cos\theta_{23} \cos\delta)$ up to $\ord{s^2_{13}}$ and $\ord{s_{13}}$ corrections respectively.

The relation $\sin\theta_{13} = \sin\theta^e_{12} \sin\theta_{23}$ allows to determine the size of the charged lepton angle $\theta_{12}$, which turns out to be close to the Cabibbo angle, for $\theta_{23} = \pi/4$~\cite{Minakata:2004xt,Raidal:2004iw,Datta:2005ci,Everett:2005ku}). Motivating such an empirical relation within GUT models, while at the same time accounting for the $m_\mu/m_s$ and $m_e/m_d$ ratios, is not straightforward~\cite{Antusch:2011qg,Marzocca:2011dh,King:2012vj,Antusch:2012fb}. Both the deviation of $\theta_{13}$ from zero and the deviation of $\theta_{12}$ from the LO prediction $\theta^\nu_{12}$ are determined by $\theta^e_{12}$, and are therefore expected to be of the same size. Indeed, one gets
\begin{equation}
\theta_{12} = \theta^\nu_{12} + \theta_{13} \cot\theta_{23}\cos\delta + \ord{\theta_{13}^2} \;.
\label{eq:theta1312}
\end{equation}

The above relation is sometimes called ``solar sum rule''. It allows to predict the CP-violating phase $\delta$ for a given LO prediction $\theta^\nu_{12}$. A solution for $\delta$ can be found for TB, GR, and other LO $\theta_{12}$ predictions not too far from the measured values. 

Actually, as the measured value of $\theta_{13}$ is not so small, $\theta_{12}$ is expected to deviate quite significantly from its LO prediction, $\delta\sin\theta_{12}/\sin\theta_{12} \sim 0.15 \cos\delta$. In this context, the success of the TB prediction, corresponding to $\delta\sin\theta_{12}/\sin\theta_{12} \lesssim 0.03$ looks somewhat accidental. Indeed, sizeable CP violation, i.e.\ small $\cos\delta$, is predicted to be necessary in order to accommodate TB mixing in this context~\cite{Marzocca:2013cr}. On the other hand, a measurement of a small $\cos\delta$ would restore the success of the $\theta_{12}$ prediction.

One can wonder whether the charged leptons effect on $\theta_{12}$ is large enough to account for the observed significant deviation from the BM prediction $\theta_{12} = \pi/4$. Unfortunately, the correction in \eq{theta1312} falls short from providing the necessary deviation~\cite{Ballett:2014dua,Girardi:2014faa,Girardi:2015vha}. Further corrections, pushing $\theta_{12}$ in the desired range, can be obtained if $U_e$ is not in the form in \eq{PMNSME}. This can be the case if $M^E_{31}$ is relatively large. A sizeable $M^E_{31}$ may however generate sizeable contributions to the electron and muon masses that need fine-tuned cancellations, unless the charged lepton mass matrix has special structures~\cite{Marzocca:2014tga}. Such a sizeable $M^E_{31}$ can also be used within asymmetric textures to correct the TB prediction, while leading to a prediction for the CP-phase $\delta$ in agreement with the present hints~\cite{Rahat:2018sgs,Perez:2019aqq}.

Sizeable corrections to $\theta_{12}$ from the neutrino sector are more constrained if the neutrino masses are inverted hierarchical. In such a case, a maximal $\theta_{12}$ can be easily obtained from pseudo-Dirac structures in the neutrino mass matrix, which in turn naturally arise within both non-abelian groups (as in the BM case, \Sect{approx_sequestering}) and abelian groups (as in \eq{nuinverted}). In this context, the needed correction to $\theta_{12}$, if arising in the neutrino sector, tends to destabilise the $|\dm_{12}/\dm_{23}| \ll1$ hierarchy, thus leading to fine-tuning~\cite{Domcke:2016mzc}. In order to avoid that, the bulk of the corrections to $\theta_{12} = \pi/4$ should come from the charged lepton sector. 

\subsubsection{Non-rigid determination of the PMNS matrix}
\label{sec:loose}

The discussion in this Section has been based so far on the assumption that $G_e$ and $G_\nu$, the subgroups of $G$ preserved by $M_E$ and $m_\nu$ are non-trivial and they rigidly determine the charged lepton and neutrino mass bases up to phases. Such an assumption allows to unambiguously determine the PMNS matrix directly from $G_e$ and $G_\nu$. While such an approach is powerful and predictive, the assumption on which it relies is non-trivial. The subgroups $G_e$ and $G_\nu$ can well be trivial, in which case they would not lead to the identification of any mass eigenstate. An intermediate possibility is that $G_e$ and $G_\nu$ are non-trivial, but they identify the mass basis only partially. In this subsection, we review such a possibility.\footnote{Such models are sometimes called ``semi-direct''.} In order to realise it, sequestering is still needed, as $G_\nu$ and $G_e$ still need to be different, with a trivial intersection.

The case in which $G_\nu$ does not fully determine the neutrino mass basis, while $G_e$ does, has been widely considered~\cite{Ge:2011ih,Ge:2011qn,Hernandez:2012sk}. In such a case, the only potentially interesting possibility is $G_\nu = \mathbb{Z}_2$. 
The residual symmetries now determine the PMNS matrix up to a $2 \times 2$ rotation and a phase (and Majorana phases and permutations, as before): $U = U_0 \, U_{ij}(\theta,\phi)$, where
\begin{equation}
U_{23}(\theta,\phi) = 
\begin{pmatrix}
1 & 0 & 0 \\
0 & \cos\theta & \sin\theta e^{-i\phi} \\
0 & -\sin\theta e^{i\phi} & \cos\theta
\end{pmatrix} , \quad
\begin{array}{l}
\theta \in [0, \pi/2] \\
\phi \in [0,2\pi]
\end{array} .
\label{eq:Uij}
\end{equation}
Analogously, $U_{12}(\theta,\phi)$ and $U_{13}(\theta,\phi)$ have the $2\times 2$ rotation embedded in the 12 and 13 blocks respectively. 

If $G_\nu$ is a subgroup of a $\mathbb{Z}_2 \times \mathbb{Z}_2$ rigidly determining the neutrino mass basis, as in \Sect{exact_sequestering} and~\ref{sec:approx_sequestering}, and $U_0$ is the PMNS matrix obtained when $\mathbb{Z}_2 \times \mathbb{Z}_2$ is unbroken, the block on which the $2\times 2$ rotation $U_{ij}(\theta,\phi)$ acts depends on which of the three $\mathbb{Z}_2$ subgroups of $\mathbb{Z}_2 \times \mathbb{Z}_2$ survives. If $U$, $S$ are defined as in \eq{Z2nu}, the three subgroups are generated by $U$, $S$, $US$. Correspondingly, the PMNS matrix is given by
\begin{equation}
U = U_0 U_{ij}(\theta,\phi) 
\text{, where }
\begin{cases}
ij = 12 & \text{if $\mathbb{Z}_2$ is generated by $U$} \\
ij = 13 & \text{if $\mathbb{Z}_2$ is generated by $S$} \\
ij = 23 & \text{if $\mathbb{Z}_2$ is generated by $US$} \\
\end{cases} .
\label{eq:loose2}
\end{equation}
Taking into account the diagonal Majorana phases $\Psi$ and the possible permutations $P_e$, $P_\nu$ of the lepton mass bases, one obtains
$U \to P_e U P_\nu \Psi$ in the previous expression.

In practice, \eq{loose2} means that it is possible to loosen the rigid predictions illustrated in \Sect{exact_sequestering} and~\ref{sec:approx_sequestering} by breaking $G$ to a $\mathbb{Z}_2$ subgroup of $\mathbb{Z}_2 \times \mathbb{Z}_2$ in the neutrino sector. Such a possibility is welcome in the models discussed in \Sect{approx_sequestering}, where the $\theta_{13}$ prediction obtained in the rigid case needs to be corrected. The correction is provided by the $U_{ij}$ rotation. In order for the rotation to affect $\theta_{13}$, it should act either in the 13 or in the 23 block. In the 13 case ($S$-preserving), the second column of $U_0$ appears identical in $U$. In the 23 case ($US$-preserving), the first column of $U_0$ appears identical in $U$.  

We apply the above ideas to models leading, in the rigid limit, to TB mixing. We focus on the simple option reviewed in \Sect{approx_sequestering}, with $G=S_4$ arising accidentally at LO from an $A_4$-symmetric sequestered lagrangian. We remind that the accidental $S_4$ invariance arises because no flavon in the $1'$, $1''$ representations is used to break $A_4$, in which case $G_e$ is generated by $T$, $G_\nu$ is generated by $U$ and $S$ in \eq{S4generators}, and $U = U_\text{TB}$ at LO (up to external phases and assuming lepton masses are correctly ordered). 

In order to reduce $G_\nu$ to $\mathbb{Z}_2$ and preserve $S$, it is then sufficient to introduce flavons $\varphi'$, $\varphi''$ in $1'$, $1''$ representations of $A_4$~\cite{Shimizu:2011xg,Ma:2011yi,King:2011zj,Cooper:2012wf}. Since we still want $T$ to be preserved by $M_E$, $\varphi'$, $\varphi''$ should be sequestered in the neutrino part of the LO lagrangian.  Another possibility is that the role of $\varphi'$, $\varphi''$ is played by the $1'$, $1''$ components of $\varphi_S^2$. In such a case, the accidental symmetry breaking, i.e.\ the corrections to TBM mixing, is suppressed by (only) one power of $\vev{\varphi_S} \sim \epsilon_S$~\cite{Lin:2009bw} (compare with the case in which $U$ is not accidental and the corrections are expected to be 
at $\ord{\epsilon_T^2}$). In both cases (breaking by $\varphi'$, $\varphi''$ or by $(\varphi_S^2)_{1'}$, $(\varphi_S^2)_{1''}$), the lagrangian is no longer accidentally invariant under $U$, and $G_\nu$ is generated by $S$. The PMNS matrix is then in the form in \eq{loose2}, with $ij=13$,
\begin{equation}
\arraycolsep=2.5mm
U_{\text{TM}_2} = U_\text{TB} U_{13}(\theta,\phi) \, \Psi = 
\begin{pmatrix}
\displaystyle \sqrt{\frac{2}{3}} c_\theta & \displaystyle \frac{1}{\sqrt{3}} & \displaystyle \sqrt{\frac{2}{3}} s_\theta e^{-i\phi} \\
\displaystyle -\frac{c_\theta}{\sqrt{6}} +\frac{s_\theta}{\sqrt{2}} e^{i\phi} & \displaystyle \frac{1}{\sqrt{3}} & \displaystyle -\frac{c_\theta}{\sqrt{2}} -\frac{s_\theta}{\sqrt{6}} e^{-i\phi} \\
\displaystyle  -\frac{c_\theta}{\sqrt{6}} +\frac{s_\theta}{\sqrt{2}} e^{i\phi} & \displaystyle \frac{1}{\sqrt{3}} & \displaystyle \frac{c_\theta}{\sqrt{2}} -\frac{s_\theta}{\sqrt{6}} e^{-i\phi}
\end{pmatrix} \Psi \;,
\label{eq:TM2}
\end{equation}
where we have now explicitly included the diagonal matrix of Majorana phases $\Psi$. The above form of the PMNS matrix deserves a few comments. A non-vanishing $\theta_{13}$ has been induced by the rotation $\theta$. Being $\theta$ a free parameter, any value of $\sin\theta_{13}\leq (2/3)^{1/2}$ can be obtained. The size of $\sin\theta_{13}$ is controlled by $\vev{\varphi'}$, $\vev{\varphi''}$ and its relative smallness can be accounted for in terms of a mild suppression of $\vev{\varphi'}$, $\vev{\varphi''}$ (or by the extra $\epsilon_S$ insertion, if $\varphi' \sim (\varphi^2_S)_{1'}$, $\varphi'' \sim (\varphi^2_S)_{1''}$). A CP-violating phase is also generated $\delta \approx \phi$. The solar angle is larger than its TB prediction, $\sin\theta_{12} \geq 1/\sqrt{3}$, but only by a $\ord{\sin^2\theta_{13}}$ amount. The maximal $\theta_{23}$ prediction is also modified, at the $\ord{\sin\theta_{13}}$ order. The precise expression of the PMNS parameters in terms of $\theta$, $\phi$ is given in table~\ref{tableCP2}. With four parameters expressed in terms of two, \eqs{TM2} lead to two predictions (``sum rules'')~\cite{Grimus:2008tt}:   
\begin{equation}
1 = 3 \cos^2\theta_{13}\sin^2\theta_{12} \;,
\quad
\cos\delta=\frac{\cos 2\theta_{13}\cot 2\theta_{23}}{\sin\theta_{13}\sqrt{2-3\sin^2\theta_{13}}} \; .
\label{eq:relTM2}
\end{equation}
The first relation is in relatively good agreement with present data, with the central value of the RHS $\approx 0.91$ and a tension at the $2\sigma$ level. In the second relation
the absence of CP-violation ($\cos\delta = \pm 1$) would require $\theta_{23}$ to be significantly non-maximal, at the boundary of its $3\sigma$ range. As $\theta_{23}$ approaches $\pi/4$, $\delta$ approaches $\pm \pi/2$. 

The second column of $U_{\text{TM}_2}$ coincides with that of $U_\text{TB}$, and corresponds to a neutrino $\nu_2 = (\nu_e+\nu_\mu+\nu_\tau)/\sqrt{3}$ with equal components in $\nu_e$, $\nu_\mu$, $\nu_\tau$. Such a pattern is called ``trimaximal'' mixing~\cite{Haba:2006dz,He:2006qd,Grimus:2008tt,Albright:2008rp,Albright:2010ap,Ishimori:2010fs,He:2011gb}. We will adhere to a common convention by denoting the form of the PMNS matrix in \eq{TM2} as ``TM$_2$'', in order to distinguish it from the form ``TM$_1$'' obtained by combining $U_\text{TB}$ with a $U_{23}$ rotation. The index 1, 2 refers to the $U_\text{TB}$ column unaffected by the rotation. Strictly speaking, only when the second column is unchanged (TM$_2$), we actually have \emph{trimaximal} mixing. 

The form TM$_1$ of the PMNS matrix is obtained from rigid TB models when the residual $\mathbb{Z}_2$ is generated by $US$~\cite{Varzielas:2012pa,Grimus:2013rw,Luhn:2013vna}. As $US$ is not part of $A_4$, such a possibility requires larger flavour groups. The $S_4$ group is viable from this point of view.
The PMNS matrix is in the form in \eq{loose2}, with $ij=23$,
\begin{equation}
\arraycolsep=2.5mm
U_{\text{TM}_1} = U_\text{TB} U_{23}(\theta,\phi) \, \Psi = 
\begin{pmatrix}
\displaystyle \sqrt{\frac{2}{3}} & \displaystyle \frac{c_\theta}{\sqrt{3}} & \displaystyle \frac{s_\theta}{\sqrt{3}} e^{-i\phi} \\
\displaystyle -\frac{1}{\sqrt{6}} & \displaystyle \frac{c_\theta}{\sqrt{3}} +\frac{s_\theta}{\sqrt{2}} e^{i\phi} & \displaystyle -\frac{c_\theta}{\sqrt{2}} +\frac{s_\theta}{\sqrt{3}} e^{-i\phi} \\
\displaystyle  -\frac{1}{\sqrt{6}} & \displaystyle \frac{c_\theta}{\sqrt{3}} -\frac{s_\theta}{\sqrt{2}} e^{i\phi} & \displaystyle \frac{c_\theta}{\sqrt{2}} +\frac{s_\theta}{\sqrt{3}} e^{-i\phi}
\end{pmatrix} \Psi.
\label{eq:TM1}
\end{equation}
The solar angle is smaller than the successful TB prediction this time, $\sin\theta_{12} \leq 1/\sqrt{3}$, but only by a $\ord{\sin^2\theta_{13}}$ amount. The first column of $U_{\text{TM}_1}$ coincides with that of $U_\text{TB}$. The expression of the PMNS parameters in terms of $\theta$, $\phi$ is given in table~\ref{tableCP2}, and lead to two predictions (``sum rules'')~\cite{Albright:2008rp}: 
\begin{equation}
2 = 3 \cos^2\theta_{13} \cos^2\theta_{12} \;, 
\quad
\cos\delta=-\frac{(1-5\sin^2\theta_{13})\cot 2\theta_{23}}{2\sqrt{2}\sin\theta_{13}\sqrt{1-3\sin^2\theta_{13}} } \;.
\label{eq:relTM1}
\end{equation}
The first relation is in good agreement with present data, well within $1\sigma$, with the central value of the RHS being $\approx 2.0$. The second relation shows that CP-invariance ($\cos\delta = \pm 1$) is not compatible with the present $3\sigma$ range for $\theta_{23}$.
As $\theta_{23}$ approaches $\pi/4$, $\delta$ approaches $\pm \pi/2$.

Up to external phases, $U_{\text{TM}_1}$  ($U_{\text{TM}_2}$) is the most general unitary matrix with the first (second) column as in $U_\text{TB}$. 

As discussed, the Majorana phases in $\Psi$ are unconstrained in this setup. On the other hand, we will see in \Sect{CP} that flavour symmetries not commuting with the Poincar\'e group may constrain them. A general parameterisation of $\Psi$ that will be useful in \Sect{CP} is 
\begin{equation}
\Psi = 
\begin{pmatrix}
1 & 0 & 0 \\
0 & e^{i\alpha/2} & 0 \\
0 & 0 & e^{i(\beta/2+\phi)}
\end{pmatrix} .
\end{equation}
Note that the Majorana phases are sometimes defined through the following parameterisation of the PMNS matrix 
\begin{equation}
U = 
\begin{pmatrix}
c_{12} c_{13} &
s_{12} c_{13} &
s_{13} e^{-i\delta} \\
-s_{12}c_{23}
-c_{12}s_{23}s_{13}e^{i\delta} &
c_{12}c_{23} 
-s_{12}s_{23}s_{13} e^{i\delta} &
s_{23}c_{13} \\
s_{12}s_{23}
-c_{12}c_{23}s_{13}s^{i\delta} &
-c_{12}s_{23}
-s_{12}c_{23}s_{13}s^{i\delta} &
c_{23}c_{13}
\end{pmatrix}
\begin{pmatrix}
1 & 0 & 0 \\
0 & e^{i\alpha_{21}/2} & 0 \\
0 & 0 & e^{i\alpha_{31}/2}
\end{pmatrix} .
\label{eq:majo2018}
\end{equation}
The relation between their parameterisations in terms of $\alpha$, $\beta$ and in terms of  $\alpha_{21}$, $\alpha_{31}$ is shown in table~\ref{tableCP2}.

\begin{table}[h!] 
\centering
\begin{tabular}{|c|c|c|c|c|c|c|}
\hline
{\tt Pattern}&$\sin^2\theta_{23}$&$\sin^2\theta_{13}$&$\sin^2\theta_{12}$&$\delta$&$\alpha_{21}$&$\alpha_{31}$\rule[-3ex]{0pt}{7ex}\\
\hline
$\text{TM}_1$&$\dd\frac{1}{2}\left(1-\cos\phi\frac{2\sqrt{6}\sin 2\theta}{5+\cos 2\theta}\right)$&$\dd\frac{\sin^2\theta}{3}$&$\dd\frac{\cos^2\theta}{2+\cos^2\theta}$&$\dd\arg\left(2 e^{-i \phi}-3e^{i\phi}\frac{\cos^2\theta}{\sin^2\theta} \right)$&$\alpha$&$\beta+2\delta$ \rule[-3ex]{0pt}{7ex}\\
\hline
$\text{TM}_2$&$\dd\frac{1}{2}\left(1+\cos\phi\frac{\sqrt{3}\sin 2\theta}{2+\cos 2\theta}\right)$&$\dd\frac{2}{3}\sin^2\theta$&$\dd\frac{1}{2+\cos 2\theta}$&$\dd\arg\left(e^{-i \phi}-3e^{i\phi}\frac{\cos^2\theta}{\sin^2\theta} \right)$&$\alpha$&$\beta+2\delta$\rule[-3ex]{0pt}{7ex}\\
\hline
\end{tabular}
\caption{Predictions of the $\text{TM}_1$ and $\text{TM}_2$ mixing patterns as a function of the parameters $\theta\in [0,\pi/2]$ and $\phi,\alpha,\beta \in [0,2\pi]$.}
\label{tableCP2}
\end{table}

We have illustrated the possibility of loosening the predictions of rigid models by reducing $G_\nu$ in such a way that the neutrino mass basis is only partially determined by $G_\nu$. Analogously, one can consider the possibility that the charged lepton mass basis is only partially determined by $G_e$. In such a case, $G_e$ does not necessarily need to contain $\mathbb{Z}_3$ or $\mathbb{Z}_2\times \mathbb{Z}_2$, cfr.\ \eq{GEGn}; it is sufficient that it contains $\mathbb{Z}_2$. The possibility $G_e = {1}$ may also be viable. While in the latter case $G_e$ would not constrain the charged lepton mass matrix at all, a (hierarchical) structure may be enforced by an additional group factor, playing the role of a FN symmetry, or organising the couplings of the flavons in a specific way. 

If $G_e$ is loosened, a rigid prediction $U_0$ is modified by a unitary transformation appearing on the left side of $U_0$, mixing its rows. We are thus in the presence of charged lepton corrections to the PMNS matrix, as in \Sect{approx_sequestering}. A contribution to $\theta_{13}$ can again be induced. If the charged leptons end up having a hierarchical structure, as they should, such corrections are typically too small to fully account for $\theta_{13}$. Note that the charged lepton mass hierarchy can be now achieved in $S_4$ without fine-tuning, since the $T$ generator can be broken (see discussion in \Sect{approx_sequestering}). Even if the charged lepton contribution to $\theta_{13}$ is subleading, it can still be useful when $U_0$ corresponds to a non-vanishing $\theta_{13}$ not too far from its experimental range. The small corrections from the charged lepton sector can then be sufficient to bring $\theta_{13}$ in the experimental range. An example is $\Delta(96)$~\cite{Toorop:2011jn,Ding:2012xx,King:2012in}. The PMNS matrix is in the latter case in the so called ``bitrimaximal'' form, a special case of TM2 mixing corresponding to $\sqrt{2/3} \,\sin\theta = (1-1/\sqrt{3})/2$. Another possibility if $G = \text{PSL}(2, 7)$~\cite{Hernandez:2012sk}, in which a good fit of the mixing angles can be obtained for near-maximal CP-violation, $\delta \sim \pi/2$ or $\delta \sim 3\pi/2$. 

\subsubsection{Extension to quarks}

The approach followed so far aims at understanding lepton flavour observables. On the other hand, a complete theory of flavour should account for the quark sector as well. The extension of the ideas discussed in this section to the quark sector is not straightforward. 

One of the main features of the lepton models considered is that all the charged lepton masses vanish in the symmetric limit, because a $d=3$ irreducible representation is used for the lepton doublets. Such a setting is not suitable for the up quark sector, characterised by a top Yukawa coupling $\lambda_t = \ord{1}$. The size of $\lambda_t$ suggests that the latter is invariant, at least under the flavour group $G$ considered in the lepton sector.\footnote{A large $\lambda_t = \ord{1}$ might arise from the breaking of a larger group $\tilde G$. In such a case, the corresponding flavon VEV needs to be close to the cutoff scale, $\vev{\phi_t} \sim \Lambda$, and $G$ should be identified with a subgroup of $\tilde G$ unbroken by $\vev{\phi_t}$. What follows still holds, if referred to $G$.} Hence, the up quark mass matrix does not vanish in the $G$-symmetric limit. An invariant $\lambda_t$ requires both the third family quark doublet $q_3$ and up quark singlet $t^c$ to be in conjugated $d=1$ representations of the whole $G$. This requirement naturally leads to models in which both the lighter Yukawa couplings are forced to be small because they are not invariant, in contrast to models based on sequestering that, per se, do not constrain the values of the Yukawa couplings. 

The different strategies needed in the quark and lepton sector are not necessarily in conflict. Quarks and leptons can be constrained by different, independent factors of the flavour group, broken by two independent sets of flavons, effectively leading to separate models in the two sectors. It is however worthy to combine those strategies. As mentioned in section~\ref{sec:approx_sequestering}, a FN-type continuous symmetry suppressing light Yukawa couplings can operate in the charged lepton sector, in combination with a discrete one. Moreover, the two strategies can be combined even more effectively within the discrete groups setup, for example by using discrete groups such as the double tetrahedral group $T'$~\cite{Frampton:1994rk}. Being a subgroup of SU(2) with doublet representations, $T'$ contains the necessary ingredients to account for the (2+1) quark structure along the lines of U(2) models~\cite{Aranda:1999kc,Aranda:2000tm}. On the other hand, as $T'$ contains the representations of $A_4$, it also contains the ingredients necessary to reproduce the lepton observables along the lines of $A_4$ models ~\cite{Feruglio:2007uu,Frampton:2007et,Chen:2007afa,Aranda:2007dp,Ding:2008rj,Frampton:2009fw,Everett:2010rd,Aranda:2010im,Carone:2016xsi,Carone:2019lfc}. In the previous example, the quark mixing is correlated to the quark mass hierarchy. One can wonder whether the same residual subgroup techniques introduced to predict the lepton mixing matrix could be extended to the quark sector. This is possible, but not straightforward. The residual subgroups should determine the relative orientations of the up and down quark mass bases. The small quark mixing angles then require a flavour group large enough to contain, among the many, closely aligned subgroups~\cite{Lam:2007qc,Blum:2007jz,Holthausen:2013vba,Araki:2013rkf,Yao:2015dwa,Varzielas:2016zuo,Li:2017abz}.

Once the flavour symmetry is extended to the quark sector, one can aim at a  model compatible with gauge unification. In such a case, the flavour structures of the quark and lepton sectors are necessarily coupled, in a way dictated by the unified group. In grand unified theories such as SU(5), for example, one family of SM fermions is unified into a $5_i + \overline{10}_i$ of SU(5): $5_i \sim (l_i, d^c_i)$, $\overline{10}_i \sim (e^c_i, q_i, u^c_i)$. As the flavour group is assumed here to commute with the gauge group, the flavour quantum numbers of SM fields belonging to the same irreducible SU(5) representation should be the same. Since $q_3$ and $t^c$ both belong to $\overline{10}_3$, $q_3$ and $t^c$ should be in a real $d=1$ representation of $G$ (i.e. they should be invariant up to a sign change). As $\tau^c$ is also unified with $q_3$ and $t^c$, it should also be in a real $d=1$ representation. This is not compatible for example with the $A_4$ and $S_4$ settings in the form illustrated in \Sect{approx_sequestering}, which require $\tau^c$ to belong to a complex representation. A non-standard $A_4$ realisation can however be achieved with one extra dimension~\cite{Altarelli:2008bg}. Unified flavour models have been reviewed in~\cite{King:2017guk}.

\subsubsection{Outlook}
\label{sec:outlookIV}

The model building avenues explored in this section are based on the interplay of two distinct subgroups $G_\nu$ and $G_e$ of $G$. The group-theoretical construction and the very structure of $G$ crucially depend on the choice of $G_\nu$ and $G_e$ and of their relative orientation. We  considered both the cases in which the subgroups fully or partially determine the flavour directions corresponding to the lepton mass eigenstates. In all cases, though, $m_\nu$ and $M_E$ are, by definition, invariant under $G_\nu $ and $G_e$. 

The model building options are far from being exhausted, even within finite non-abelian group models. For example, there is no reason why $G_\nu$ and $G_e$ should be non-trivial and fully, or partially, determine the lepton mass bases. Another non-trivial, and non-indispensable, assumption has to do with the forms of $m_\nu$ and $M_E$, the matrices invariant under $G_\nu$ and $G_e$. The constraints on $G_\nu$ and $G_e$ in \eq{GEGn} and~(\ref{eq:GEGn0}) assume that $m_\nu$ and $M_E$ provide non-vanishing, non-degenerate masses for all the leptons, with the only possible exception of the lightest neutrino. This is not really necessary. In early models, $m_\nu$ and $M_E$ could be identified with the exact mass matrices (the PMNS matrix was still compatible with being exactly in TB form). On the other hand, this is not in line either with the generic theoretical expectation of higher order corrections to sequestering, or with the experimental determination of the mixing parameters (except in the cases discussed in \Sect{exact_sequestering}). Thus, the matrices $m_\nu$ and $M_E$ allowed by $G_\nu$ and $G_e$ should not be identified with the exact mass matrices, in this context. They are approximations, expected to be corrected by higher order effects, in some cases as large as $\ord{\lambda_C}$. The mass eigenvalues are then also expected to be corrected, as the mixing angles are, and there is no reason to demand that $m_\nu$ and $M_E$ provide non-vanishing, non-degenerate masses for all leptons. In fact, they could equally well correspond, for example, to $m_e = 0$ or $m_e = m_\mu = 0$, or $m_{\nu_1} = m_{\nu_2}$. The corrections to such patterns necessary to obtain viable lepton masses are smaller than those commonly assumed to affect the mixing angles. Such a possibility has been considered in connection to partially degenerate neutrinos~\cite{Hernandez:2013vya,Joshipura:2014qaa}. In principle, any mass pattern that can be considered sufficiently close to the observed one could be considered as well, in the spirit of the discussion in \Sect{perturbative}. 

The above shows that the programme based on (linear, Lorentz-scalar) discrete non-abelian flavour groups has not been fully explored. Still, it is fair to say that such a programme has partially fulfilled, so far, the initial expectations. The approach focuses on mixing angles. The predictivity potential of the simplest models, one of their stronger motivations, has been frustrated by the experimental determination of the $\theta_{13}$ angle that, once again, challenged theoretical prejudice. Two opposite strategies can be pursued to accommodate the value of $\theta_{13}$, both leading to a certain loss of predictivity. On the one side, one can stick to relatively simple models, at the price of accepting relatively large corrections, which reduce predictivity. On the other side, one can aim at more involved models with predictions close to the experiment, at the price of scanning a dense landscape of models. The significance of the prediction is then reduced by the correspondingly dense number of alternatives available. On the model building side, the predictions are not really associated to the flavour group, but rather to the symmetry breaking effects --- ultimately to the detailed structure of the potential determining the VEV alignments --- and to a set of auxiliary symmetries and quantum numbers needed to arrange the proper set of couplings in the lagrangian. On the other hand, the theoretical landscape is still broad, as argued, and its exploration will hopefully provide new insights.

\section{CP-like flavour symmetries}
\label{sec:CP}
The main purpose of including CP transformations in the flavour symmetry group is to constrain Majorana phases. There are several dedicated reviews on this topic, such as for example \cite{King:2014nza,Neder:2015wsa,King:2015aea,Hagedorn:2017icd,King:2017guk,Petcov:2017ggy,Coloma:2018ioo}. 
In a theory invariant under both a flavour symmetry group $G_f$ and CP, besides eq.~(\ref{eq:mLinvariancePhi}), the following constraint holds for the lepton mass matrices:\footnote{In the presence of a single Higgs, a possible phase in its CP transformation can be reabsorbed in the transformation of the lepton fields.}
\begin{eqnarray}
M^*_E(\phi) &=& X_{e^c}^T\, M_E(X_\phi\phi^*) \, X_l \nn\\
m^*_\nu(\phi) &=&X_l^T\, m_\nu(X_\phi\phi^*) X_l \;,
\label{eq:mLinvarianceCP}
\end{eqnarray}
where we have denoted with $X_f$ $(f=e^c,l,\phi)$ unitary matrices describing the action of CP on the field $f$ and we have assumed Majorana neutrinos.
In such a theory CP can only be broken spontaneously and the conditions that realize the breaking are
\begin{itemize}
 \item[{\em  i})] $X_\phi\phi^*\ne \phi$ on the vacuum.
 \item[{\em ii})] No other consistent CP transformation leaving 
invariant both the theory and the vacuum exists.
\end{itemize}
\subsection{Sequestering and residual symmetries}
As in the case of a flavour symmetry commuting with the proper Poincar\'e group, to some extent it is possible to analyze the predictions of the theory without referring to an
explicit realization, relying on the residual symmetries enjoyed by the charged lepton sector and by the neutrino sector, if any. Provided $M_E$ and $m_\nu$ depend 
on two separate sets of flavons, $\phi_E$ and $\phi_\nu$, we can contemplate independent residual symmetries for the two sectors:
\be
U_\phi(g_E)\phi_E=\phi_E \hspace{2cm} U_\phi(g_\nu)\phi_\nu=\phi_\nu \;,
\ee 
where $g_E$ and $g_\nu$ run in different subgroups of $G_f$ and these relations hold in the vacuum. To constrain Majorana phases we should further assume that CP is conserved in the neutrino sector:
\be
X_\phi\phi^*_\nu=\phi_\nu \;.
\ee 
By combining eqs.~(\ref{eq:mLinvariancePhi}) and~(\ref{eq:mLinvarianceCP}), we end up with the relations:
\be
(M_E^\dagger M_E)=\rep_l(g_E)^\dagger (M_E^\dagger M_E) \rep_l(g_E) \; , \qquad m_\nu = \rep_l(g_\nu)^T m_\nu \, \rep_l(g_\nu) \;, \qquad m^*_\nu =X_l^T\, m_\nu X_l \;,
\ee
which constrain at the same time the lepton mixing angles and both Dirac and Majorana phases. 
\begin{table}[h!] 
\centering
\begin{tabular}{|c|c|c|}
\hline
$G_e$&$G_\nu$&\#{\tt parameters}\rule[-2ex]{0pt}{5ex}\\
\hline
$Z_{m_1}\times \ldots \times Z_{m_p}$&$Z_2\times Z_2\times \text{CP}$&0\rule[-2ex]{0pt}{5ex}\\
\hline
$Z_{m_1}\times \ldots \times Z_{m_p}$&$Z_2\times \text{CP}$&1\rule[-2ex]{0pt}{5ex}\\
\hline
$Z_2\times \text{CP}$&$Z_2\times Z_2\times \text{CP}$&1\rule[-2ex]{0pt}{5ex}\\
\hline
$Z_2\times \text{CP}'$&$Z_2\times \text{CP}$&2\rule[-2ex]{0pt}{5ex}\\
\hline
$Z_{m_1}\times \ldots \times Z_{m_p}$&$\text{CP}$&3\rule[-2ex]{0pt}{5ex}\\
\hline
$Z_2$&$Z_2\times \text{CP}$&3\rule[-2ex]{0pt}{5ex}\\
\hline
\end{tabular}
\caption{Number of continuous free parameters describing the lepton mixing matrix $U(\theta_{ij},\delta,\alpha_{21},\alpha_{31})$ \cite{Chen:2014wxa,Lu:2016jit}. In the first and second column, the residual symmetries $G_e$ and $G_\nu$
of the charged lepton sector and the neutrino sector, respectively. The cyclic symmetry $Z_{m_1}\times \ldots \times Z_{m_p}$ is assumed to fully distinguish the charged leptons
by their different transformation properties.
The residual symmetry $Z_2\times Z_2\times \text{CP}$ is also equivalent to the one generated by the four allowed CP transformations of the neutrino sector, see the text. In the fourth line, CP and $\text{CP}'$ are in general independent CP transformations.}
\label{tableCP}
\end{table}

This program has been carried out in the context of discrete flavour symmetry groups
$G_f$. A variety of cases arises from the different possible assignments of the residual symmetries.
Assuming three generations, in the neutrino sector the most general group leaving neutrino masses unconstrained is the Klein group $Z_2\times Z_2$.
To avoid mass degeneracies \footnote{Degeneracies in the neutrino mass spectrum in this context have been anayzed in ref.~\cite{Joshipura:2018cpb}.}, the matrix $X_l$ is required to be symmetric: $X_l^T = X_l$ \cite{Feruglio:2012cw}. Since $X_l$ is also unitary, this automatically implies $\text{CP}^2=1$ \footnote{In Section \ref{roleCP} we have seen that $(X^*_l X^{\phantom{*}}_l)^n = 1$ holds for a finite group.}. To guarantee that the action of CP on lepton electroweak doublets is always represented by a symmetric matrix, $X_l$ is required to commute with the four elements of the Klein group. Given the antilinear action of CP, commutation is
expressed through relations of the type: 
\be
X_l \, U(g_K)^*=U(g_K) \, X_l \;,
\label{XZ}
\ee
where $g_K$ stands for an element of the Klein group. Indeed $X'_l=U(g_K) \, X_l$ represents just another CP transformation and eq.~(\ref{XZ}) implies that the matrix $X'_l$
is symmetric.
It also follows that four CP transformations can be selected as residual symmetries of the neutrino sector.
Conversely, given these four allowed CP transformations, the Klein group can be fully reconstructed \cite{Chen:2014wxa,Everett:2016jsk}.
Usually the group $G_e$ consists of a direct product of cyclic symmetries, $Z_{m_1}\times \ldots \times Z_{m_p}$, such that all the charged leptons
are distinguished by their different 
transformation properties. Among the residual symmetries of the charged lepton sector there can also be an accidental CP symmetry, independent from the one acting in the neutrino sector.

\subsection{Parameter counting}

The freedom in the definition of the model gives rise to many cases and, depending on the specific set of assumptions, the PMNS matrix is determined up to a number of continuous free parameters, listed in table \ref{tableCP}. 
These parameters arise as follows. The invariance under CP provides, in a suitable basis,
a reality condition on the neutrino mass matrix, which can be parametrized in terms of three masses and three angles. An additional $Z_2\times Z_2$ symmetry fully determines these angles, while a single parity $Z_2$ leaves one angle unconstrained. The three angles remain free parameters if the only residual symmetry of the neutrino sector is CP.
In the charged lepton sector the choice $G_e=Z_{m_1}\times \ldots \times Z_{m_p}$, when all leptons have different transformation properties, leaves no free parameters beyond masses. One free angle originates from $G_e=Z_2\times \text{CP}$ and one angle and one phase from $G_e=Z_2$. Adding the parameters of the two sectors reproduces table \ref{tableCP}. 
This approach leaves lepton masses unconstrained, and
the PMNS matrix is always determined up to permutations of rows and columns.
Moreover the intrinsic parity of neutrinos, that is the relative sign of their masses, cannot
be established. As a result the physical phases are fixed modulo $\pi$.
\subsection{Examples}
\subsubsection{$\mu-\tau$ reflection symmetry}
A simple example is provided by the so called $\mu-\tau$ reflection symmetry
\cite{Harrison:2002kp,Harrison:2002et,Grimus:2003yn,Harrison:2004he}\footnote{See \cite{Zhou:2014sya,Mohapatra:2015gwa,Joshipura:2015dsa,Zhao:2017yvw,Rodejohann:2017lre,Nishi:2018vlz,Sinha:2018xof} for more recent applications related to the topic of this section.}. In the basis where the charged lepton mass matrix is diagonal and ordered
from smaller to bigger masses, the CP transformation
acting on neutrinos is specified by:
\be
X_l=
\left(
\begin{array}{ccc}
1&0&0\\
0&0&1\\
0&1&0
\end{array}
\right) \;,
\ee
and the constraint $m^*_\nu =X_l^T\, m_\nu X_l$ implies the relations $\sin\theta_{23}=1/\sqrt{2}$, $\sin\theta_{12}\cos\theta_{12}\sin\theta_{13}\cos\delta=0$ and $\sin\alpha_{21}=\sin\alpha_{31}=0$. Data requires $\sin\theta_{12}\cos\theta_{12}\sin\theta_{13}\ne 0$ and this scheme predicts a maximal Dirac CP phase, $|\sin\delta|=1$.
\subsubsection{$G=S_4\rtimes \text{CP}$}
If we assume  $G_e=Z_{m_1}\times \ldots \times Z_{m_p}$ and 
 $G_\nu=Z_2\times \text{CP}$
in the neutrino sector, see the second row of table \ref{tableCP}, the PMNS matrix depends on a continuous parameter.  An example is provided by $G=S_4\rtimes \text{CP}$, $G_e=Z_3$  \cite{Feruglio:2012cw}. Due to the different embedding of the $Z_2$ subgroup in $S_4$, there are five inequivalent choices of $Z_2\times \text{CP}$ transformations leaving the neutrino sector invariant. Four of them, labelled I, II, IV and V, reproduce particular cases of the so-called trimaximal mixing pattern.

Models I and II reproduce $U_{\text{TM}_2}$, while  Models IV and V give rise to $U_{\text{TM}_1}$, with $\phi$, $\alpha,\beta$ (see \Sect{loose}) quantized and assuming only the values shown in table \ref{tableCP1}.
Models I  and IV predict maximal atmospheric mixing angle, maximal Dirac CP violation, trivial CP Majorana phases and provide two realizations of the $\mu-\tau$ 
reflection symmetry enjoying an additional prediction.
Model II and V predict no lepton CP violation of Dirac or Majorana type. The relations (\ref{eq:relTM2}) 
apply to model I(II) with $|\cos\delta|=0(1)$.
\begin{table}[h!] 
\centering
\begin{tabular}{|c|c|c|c|c|c|c|c|c|c|c|}
\hline
{\tt Model}& {\tt Pattern}& $|\sin\phi|$ & $\sin\alpha$ & $\sin\beta$&$\sin^2\theta_{23}$&$\sin^2\theta_{13}$&$\sin^2\theta_{12}$&$|\sin\delta|$&$\sin\alpha_{21}$&$\sin\alpha_{31}$\rule[-3ex]{0pt}{7ex}\\
\hline
I& $\text{TM}_2$& $1$ & $0$ & $0$&$\dd\frac{1}{2}$&$\dd\frac{2}{3}\sin^2\theta$&$\dd\frac{1}{2+\cos 2\theta}$&$1$&$0$&$0$\rule[-3ex]{0pt}{7ex}\\
\hline
II& $\text{TM}_2$&$0$ & $0$ & $0$&$\dd\frac{1}{2}\left(1\pm\frac{\sqrt{3}\sin 2\theta}{2+\cos 2\theta}\right)$&$\dd\frac{2}{3}\sin^2\theta$&$\dd\frac{1}{2+\cos 2\theta}$&$0$&$0$&$0$\rule[-3ex]{0pt}{7ex}\\
\hline
IV& $\text{TM}_1$&$1$ & $0$ & $0$&$\dd\frac{1}{2}$&$\dd\frac{\sin^2\theta}{3}$&$\dd\frac{\cos^2\theta}{2+\cos^2\theta}$&$1$&$0$&$0$ \rule[-3ex]{0pt}{7ex}\\
\hline
V& $\text{TM}_1$&$0$ & $0$ & $0$&$\dd\frac{1}{2}(1\mp\frac{2\sqrt{6}\sin 2\theta}{5+\cos 2\theta})$&$\dd\frac{\sin^2\theta}{3}$&$\dd\frac{\cos^2\theta}{2+\cos^2\theta}$&$0$&$0$&$0$ \rule[-3ex]{0pt}{7ex}\\
\hline
\end{tabular}
\caption{Specific mixing pattern arising in four out of the five independent cases arising from $S_4$ and CP invariance, broken down to $Z_3$ in the charged lepton sector and to $Z_2\times \text{CP}$ in the neutrino sector $G=S_4\rtimes \text{CP}$  \cite{Feruglio:2012cw} as a function of the parameters $\theta\in [0,\pi/2]$.}
\label{tableCP1}
\end{table}
A general property of $\text{TM}_2$ is $\sin^2\theta_{12}>1/3$.
By letting $\sin^2\theta_{13}$ vary in its 3$\sigma$ allowed range, the first relation predicts $\sin^2\theta_{12}=0.340\div0.342$, presently allowed within 3$\sigma$,
but out of the 2$\sigma$ range. 
In model II, $\tan^2\delta=0$ and the 3$\sigma$ allowed range of $\sin^2\theta_{13}$ results in $\sin^2\theta_{23}=\{0.388\div0.398\} \cup \{0.602\div0.611\}$. The prediction falling in the first octant is excluded at 3$\sigma$, whereas the one falling in the second octant is allowed at 2$\sigma$. A vanishing $\sin\delta$ is disfavored by the current data, but it is not excluded at 3$\sigma$. The relations (\ref{eq:relTM1}) for model IV(V) require $|\cos\delta|=0(1)$.
In $\text{TM}_1$ we always have $\sin^2\theta_{12}<1/3$.
By letting $\sin^2\theta_{13}$ vary in its 3$\sigma$ allowed range, the first relation predicts $\sin^2\theta_{12}=0.316\div0.319$, in very good agreement with present data. Model V is ruled out since the second relation with $\tan^2\delta=0$ leads to values of $\sin^2\theta_{23}$ excluded by data. The quoted ranges have been derived from the results of the global fit in \cite{Esteban:2018azc}. The group $G=A_4\rtimes \text{CP}$ leads to the $\text{TM}_2$ mixing pattern shown in table \ref{tableCP2}, with $|\sin\phi|=0$ or $|\sin\phi|=1$ and $\sin\alpha=\sin\beta=0$ as for models I and II \cite{Feruglio:2012cw,Ding:2013bpa,Nishi:2016jqg,Li:2016nap}.
Explicit models have been constructed for this case in ref.~\cite{Ding:2013bpa,Li:2016nap}.
Starting from $G=S_4\rtimes \text{CP}$, the models of ref.~\cite{Ding:2013hpa,Feruglio:2013hia} reproduce a nearly $\text{TM}_2$ mixing pattern while those of ref.~\cite{Li:2013jya}
come close to the $\text{TM}_1$ scheme. Other examples of models within $G=S_4\rtimes \text{CP}$ are those of ref.~\cite{Luhn:2013vna,Li:2014eia,Penedo:2017vtf,Ding:2018tuj}.
\subsubsection{$\Delta(3 n^2)$ and $\Delta(6 n^2)$}
The groups $A_4$ and $S_4$ are particular cases of the series $\Delta(3 n^2)$ and $\Delta(6 n^2)$, respectively, realized with the choice $n=2$.
General results for the whole series have been given in ref.~\cite{Hagedorn:2014wha,Ding:2014ora,Ding:2015rwa,deMedeirosVarzielas:2017ote,Joshipura:2018rit}. For $G=\Delta(3 n^2)\rtimes \text{CP}$ broken into
$G_e=Z_3$ and $G_\nu=Z_2\times \text{CP}$ the mixing pattern is still of $\text{TM}_2$ type and depends on a continuous parameter. When $G=\Delta(6 n^2)\rtimes \text{CP}$ breaks into 
$G_e=Z_3$ and $G_\nu=Z_2\times \text{CP}$, also more complex mixing patterns arise, beyond the trimaximal one. 
The particular cases $G=\Delta(48)\rtimes \text{CP}$ and $G=\Delta(96)\rtimes \text{CP}$ have also been comprehensively studied in ref.~\cite{Ding:2013nsa,Ding:2014hva} and \cite{Ding:2014ssa}, respectively. As an example of an interesting mixing pattern,
we show in table \ref{tabH} the predictions of a particular case arising in $G=\Delta(384)\rtimes \text{CP}$ when choosing $G_e=Z_3$ and $G_\nu=Z_2\times \text{CP}$. On top of one real continuous parameter $\theta$, they depend on two discrete parameters $m$ and $s$, specifying the embedding of the $Z_2$ and CP transformations, respectively, within 
$\Delta(384)$.
\begin{table}[h!]
\centering
\begin{tabular}{|c|c|c|c|c|c|}
\hline
$s$&$\sin^2 \theta_{13}$&$\sin^2 \theta_{12}$&$\sin^2 \theta_{23}$&$\sin\delta$&$|\sin\alpha|=|\sin\beta|$\\
\hline
$s=1$&$0.0220$&$0.318$&$0.579$&$0.936$&$1/\sqrt{2}$\\
&$0.0220$&$0.318$&$0.421$&$-0.936$&$1/\sqrt{2}$\\
\hline
$s=2$&$0.0216$&$0.319$&$0.645$&$-0.739$&$1$\\
\hline
$s=4$&$0.0220$&$0.318$&$0.5$&$\mp 1$&$0$\\
\hline
\end{tabular}
\caption{Results for lepton mixing parameters from $G_f=\Delta (384)$, $m=4$ and different CP transformations $X(s)$
\cite{Hagedorn:2014wha}. 
The continuous parameter $\theta$ has been optimized to reproduce $\sin^2\theta_{13}$. }
\label{tabH}
\end{table}
Good agreement with the mixing angles is obtained if $|\sin\delta|$ is large and $m=4$.
In this case the bound $|\sin\delta|>0.71$ holds. This mixing pattern is of $\text{TM}_1$ type.
For $s = 1$ and $s=2$ the parameter $|m_{ee}|$ relevant for neutrinoless double beta decay has a non-trivial lower bound, whereas for $s=4$
both Majorana phases are trivial and a cancellation cannot be avoided for normal ordering of neutrino masses. Apart from the constraints on CP phases also the lepton mixing angles are strongly restricted, which further sharpen the prediction 
of $|m_{ee}|$. 
\subsubsection{Other examples}
A remnant CP symmetry in combination with texture zeros has been examined in ref.~\cite{Barreiros:2018bju}. In the case of $G=A_5\rtimes \text{CP}$, one-parameter families of PMNS matrices have been studied \cite{Li:2015jxa,DiIura:2015kfa,Ballett:2015wia,DiIura:2018fnk,Lopez-Ibanez:2019rgb}, typically having trivial or maximal Dirac CP phase and trivial
Majorana phases. This study has been generalized in ref.~\cite{Turner:2015uta} to include lepton mixing matrices depending on three parameters. 
Other groups that have been combined with CP invariance include $T'$ \cite{Girardi:2013sza}, $\Delta(27)$ \cite{Nishi:2013jqa}, the series $D^{(1)}_{9n, 3n}$ \cite{Li:2016ppt}, $\Sigma(36\times 3)$ \cite{Rong:2016cpk}, $\text{PSL}_2(7)$ \cite{Rong:2019ubh}. Variants of the above setup exploiting a generalized CP symmetry have been
considered in refs.~\cite{Girardi:2016zwz,Chen:2018lsv,Ding:2018fyz}.
In ref.~\cite{Yao:2016zev} a scan of all groups of order less than 2000 has been performed,
assuming either $(G_e,G_\nu)=(Z_{m_1}\times \ldots \times Z_{m_p},Z_2\times \text{CP})$ or 
$(G_e,G_\nu)=(Z_2\times \text{CP}',Z_2\times Z_2\times CP)$, with physical quantities depending on one continuous real parameter. The lepton mixing matrices in good agreement with data fall into eight different 
categories up to possible row and column permutations. These viable mixing patterns can be 
reproduced starting from the discrete flavour groups $\Delta(6n^2)$, $D^{(1)}_{9n, 3n}$, $A_5$ and $\text{PSL}_2(7)$ combined with $CP$ symmetry. Most of them are of $\text{TM}_2$ type or deformation thereof. Exceptions are those
related to the survival symmetries $(G_e,G_\nu)=(Z_2\times CP',Z_2\times Z_2\times \text{CP})$ or those derivable from $A_5\rtimes \text{CP}$. 

If we assume $G_e=Z_{m_1}\times \ldots \times Z_{m_p}$ and  $G_\nu=Z_2\times Z_2\times \text{CP}$
in the neutrino sector, we potentially end up with the most predictive scenario, as shown in the first row of table \ref{tableCP}.
In this case, after specifying the embedding of the residual groups $Z_{m_1}\times \ldots \times Z_{m_p}$ and $Z_2\times Z_2$ in the full flavour group $G$, the PMNS matrix is fully determined, up to permutations of rows and columns. However, as shown in ref.~\cite{King:2014rwa,Chen:2015nha}, in this case the only viable PMNS matrix can only be of trimaximal $\text{TM}_2$ type
with trivial $\delta$, $\alpha_{31}=0$ and $\alpha_{21}$ a rational multiple of $\pi$. 
The relations (\ref{eq:relTM2}) with $\tan\delta=0$ and the relative comments apply.
The inverse problem of determining the most general residual CP symmetry of the neutrino sector compatible
with the present data has been studied in ref.~\cite{Everett:2015oka,Everett:2016jsk} and, assuming tribimaximal mixing, in refs.~\cite{Chen:2018zbq,Chen:2019fgb}. 
For a generic PMNS matrix, it is however not guaranteed that the residual symmetries of the neutrino and charged lepton sectors fit into a finite group.

The possibility of exploiting invariance under CP to predict or constrain physical phases find a natural application in the
context of leptogenesis. This aspect has been analyzed in refs.~\cite{Chen:2016ptr,Hagedorn:2016lva,Li:2017zmk,Hagedorn:2017wjy,Samanta:2018efa}. 
\subsubsection{Extension to quarks}

Flavour symmetries embedding CP have been also applied to the more general problem of simultaneously describing quarks and lepton masses. 
Indeed, taking quarks into  account is unavoidable. Whereas the latter could in principle be invariant under the action of a standard flavour group operating on the lepton sector, a CP-symmetry must transform all fermion fields. Its spontaneous breaking in the quark sector must also be assured, in order to reproduce the observed CP-violation in the CKM matrix.

Several difficulties arise when trying to extend flavour symmetries embedding CP to the quark sector.
Most of them are common to the general framework of discrete symmetries and not due to the specific inclusion of CP. 
As we have seen, the approach based on selective residual symmetries 
does not make predictions about masses, but only about angles. In this context the correlation between quark masses and mixing angles suggested by data and supported by abelian symmetries is lost. 
Quark mass hierarchies are typically reproduced with the help of parameters poorly related to the mixing
and CP properties.
Moreover, to simultaneously describe both lepton and quark mixing angles, flavour groups of large order 
are generally required. Indeed the small misalignment between up and down quarks calls for sufficiently
close residual symmetries in the two sectors, which usually occurs if the group $G_f$ has a large number
of densely distributed subgroups. For example, when quark and lepton electroweak doublets
are assigned to irreducible triplets of $G_f$, groups as large as $\Delta(294)$ \cite{Li:2017abz,Lu:2018oxc} or $\Delta(384)$ \cite{Hagedorn:2018gpw,Hagedorn:2018bzo} are needed.

Apart from aesthetic considerations, implementing the desired symmetry breaking pattern
in a concrete model requires a large number of flavon representations. This in turn generates
a serious alignment problem implying that additional cyclic symmetries or selection rules have to be invoked in 
order to get only the desired interaction terms.
Explicit examples of these constructions have been realized via a stepwise breaking of $G_f=\Delta(384)$ combined with CP, where charged fermion mass hierarchies are reproduced through operators with different numbers of flavons \cite{Hagedorn:2018bzo}. These examples also show that a direct embedding in GUT is problematic, since matter and flavon representations do not fit GUT multiplets.

To reduce the order of the group, while preserving predictability about phases of the mixing matrices, 
the use of dihedral groups in combination with CP has been suggested.
This approach takes up the old observation that dihedral groups are suitable to accomodate quark mixing angles \cite{Lam:2007qc,Blum:2007jz}.
Dihedral groups do not possess three-dimensional irreducible representations and quarks and lepton
electroweak doublets are assigned to singlets and doublets of the flavour group. By choosing $G_f=D_{14}$,
and by including CP, quark and lepton mixing angles and phases can both be accommodated by adjusting two continuous free parameters in each sector \cite{Lu:2019gqp}. It would be desirable to show that the symmetry breaking pattern invoked in this analysis can be effectively realized within a concrete model.

\subsection{Outlook}
The embedding of CP into the flavour symmetry provides a valuable complement to the setup dealing with
ordinary flavour groups, fully commuting with the proper Poincar\'e transformations. In such more restricted 
framework lepton mixing angles, Dirac and Majorana phases can all be predicted simultaneously, in terms of a single
continuous real parameter in the most realistic and predictive cases. Many explicit models support the viability of such
approach, with similar disadvantages affecting models dealing with ordinary flavour groups:
a complicated symmetry breaking sector, additional auxiliary symmetries and fields to trigger the desired
pattern of symmetry breaking and a limited accuracy of the predictions due to higher dimensional operators.

\section{Non-linearly realized flavour symmetries}
\label{sec:modular}
\subsection{The modular group $\overline \Gamma$}
Non-linearly realized flavour symmetries have been considered in the context of ${\cal N}=1$ supersymmetric theories by adopting as flavour group the modular group $\overline \Gamma$. 
The idea that modular invariance can play a central role in describing Yukawa couplings is an old one, and has been naturally realized in the context of string theory
\cite{Hamidi:1986vh,Dixon:1986qv,Lauer:1989ax,Lauer:1990tm,Erler:1991nr}, in D-brane compactification \cite{Cremades:2003qj,Blumenhagen:2005mu,Blumenhagen:2006ci,Abel:2006yk,Marchesano:2007de,Antoniadis:2009bg,Kobayashi:2016ovu}, in magnetized extra dimensions \cite{Cremades:2004wa,Abe:2009vi,Kobayashi:2018rad}, and in orbifold compactification \cite{Ibanez:1986ka,Casas:1991ac,Lebedev:2001qg,Kobayashi:2003vi}. Modular invariance has also been incorporated in early
flavour models \cite{Brax:1994kv,Binetruy:1995nt,Dudas:1995eq,Dudas:1996aa,Leontaris:1997vw}.
A step forward has been taken by observing that it can be implemented in a bottom-up perspective, relying on the group transformation properties of the building blocks of the theory \cite{Feruglio:2017spp}.

\begin{figure}[h!]
 \centering
 \includegraphics[width=0.25\textwidth]{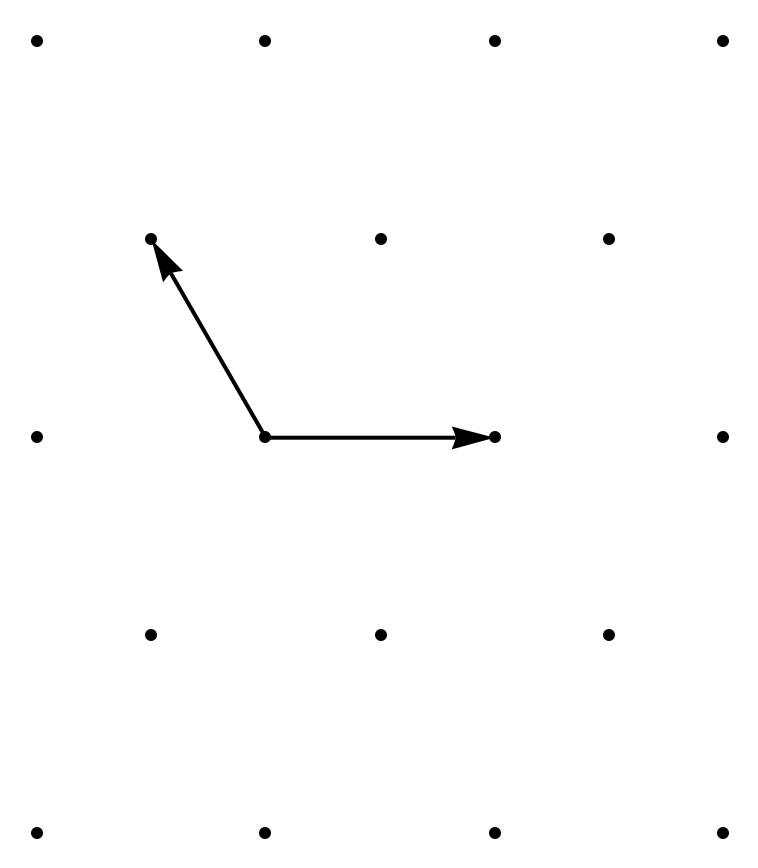}\hspace{3cm}
  \includegraphics[width=0.25\textwidth]{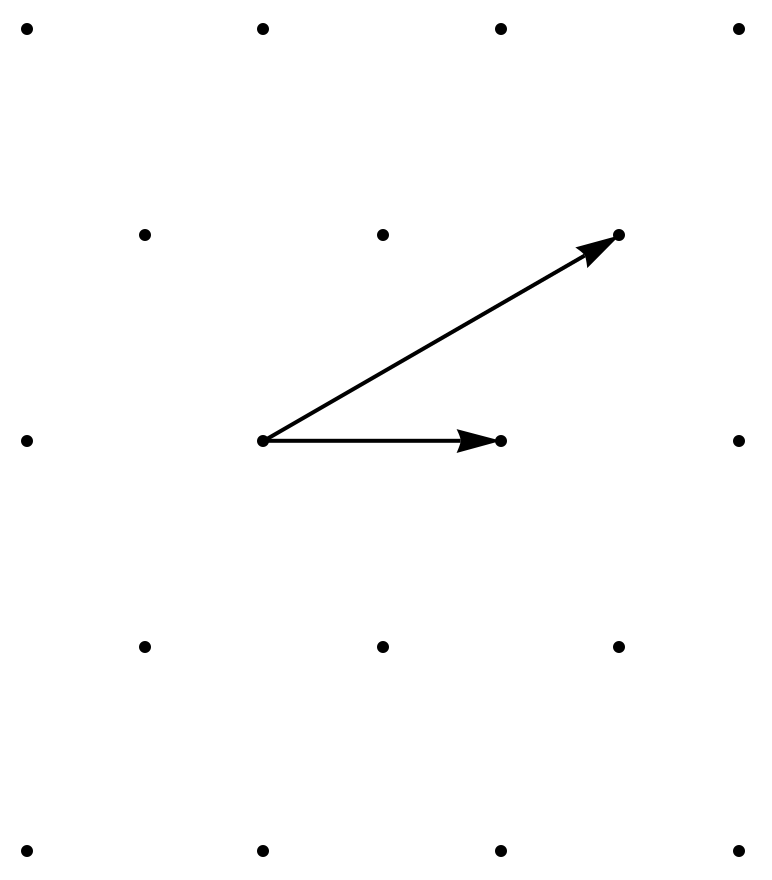}
 \caption{Two equivalent lattices with basis $(\tau,1)$ and $(\tau+2,1)$.}
\label{Lattice}
\end{figure}

\begin{figure}[h!]
 \centering
 \includegraphics[width=0.4\textwidth]{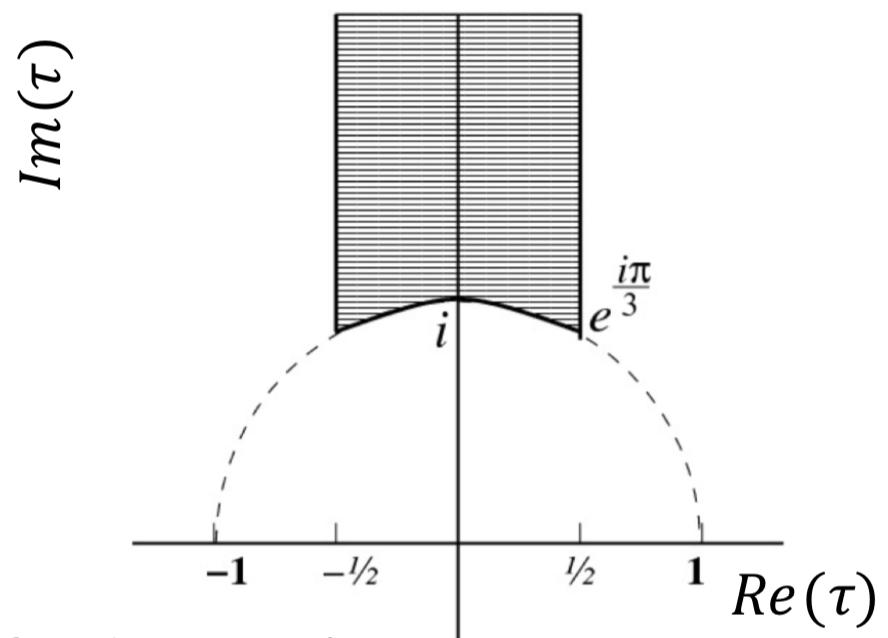}
 \caption{Fundamental region ${\cal F}$: a connected region of ${\cal H}$ such that
each point of ${\cal H}$ can be mapped into ${\cal F}$ by a $\overline\Gamma$ transformation, but no two points in the interior of ${\cal F}$ are related under $\overline\Gamma$.}
\label{FR}
\end{figure}

In ${\cal N}=1$ supersymmetric theories, the field $\tau$, called the modulus, is a chiral supermultiplet, 
whose scalar component is restricted to ${\cal H}$, the upper half of the complex plane. Under $\overline \Gamma$ it
transforms as:
\be
\tau\to \gamma\tau=\frac{a\tau+b}{c\tau+d} \;,
\label{tratau}
\ee
with $a$, $b$, $c$, $d$ integers and $ad-bc=1$. The group $\overline \Gamma$ is discrete, infinite and non-compact.
It has a presentation in terms of two generators $S$ and $T$:
\be
\tau\xrightarrow{S}-\frac{1}{\tau} \hspace{2cm} \tau\xrightarrow{T}\tau+1 \;,
\ee
satisfying:
\be
S^2=(ST)^3=\mathbb{1} \;.
\ee
The modular group is ubiquitous in string theory. It is the invariance group of a lattice $\Lambda$ defined in the complex plane $\mathbb{C}$. Two lattices $\Lambda$ and $\Lambda'$ with basis $(e_1,e_2)$ and $(e'_1,e'_2)$, such that ${\tt Im}(e_1/e_2)$ and ${\tt Im}(e'_1/e'_2)$ are both positive, coincide if and only if
\be
\left(\begin{array}{c}e'_1\\e'_2\end{array}\right)=
\left(\begin{array}{cc}a&b\\c&d\end{array}\right)
\left(\begin{array}{c}e_1\\e_2\end{array}\right) \;,
\ee
with $a$, $b$, $c$, $d$ integers and $ad-bc=1$. A frequently considered compactification of two extra dimensions gives rise to a torus, defined by the quotient $\mathbb{C}/\Lambda$ modulo rotations and scale transformations, which allow
to chose the basis of $\Lambda$ in the form $(\tau,1)$ $({\tt Im}(\tau)>0)$. It follows that two tori defined by $\tau$ and $\gamma\tau$ coincide, see fig.~\ref{Lattice}.
From this
viewpoint $\overline\Gamma$ can be thought as a gauge symmetry. With a gauge choice it is always possible
to restrict $\tau$ to a fundamental region, a representative of which is shown in fig \ref{FR}.
\subsection{Modular invariant supersymmetric theories}
We can define the action of $\overline\Gamma$ on a set of matter chiral multiplets $\phi^{(I)}$  by specifying a compact quotient of $\overline\Gamma$. A series of compact groups can be constructed by taking the quotient of
$\overline\Gamma$ by a principal congruence subgroup $\overline\Gamma(N)$ with elements obeying $a,d=1 \,({\tt mod}~ N)$, $b,c=0~({\tt mod}~ N)$, $N$ being a natural number called the {\em level}. $\overline\Gamma(N)$ are normal subgroup of $\overline\Gamma$ of finite index, so that the quotients $\Gamma_N=\overline\Gamma/\overline\Gamma(N)$ are finite
groups admitting finite-dimensional unitary representations. For the first few levels, they are isomorphic to permutation groups: 
$\Gamma_2=S_3$, $\Gamma_3=A_4$, $\Gamma_4=S_4$, $\Gamma_5=A_5$. 
We have $\partial(\gamma\tau)/\partial\tau=(c\tau+d)^{-2}$ and under the modular group the the matter fields $\phi^{(I)}$ transform as \cite{Ferrara:1989bc,Ferrara:1989qb}
\be
\phi^{(I)}\to (c\tau+d)^{k_I} \,\rho^{(I)}(\gamma)\phi^{(I)} \;.
\label{tramat}
\ee
The above transformation is completely defined by the {\em weight} $k_I$, the {\em level} $N$ and the unitary representation $\rho^{(I)}(\gamma)$ of $\Gamma_N$. We also recall that modular forms of {\em level} $N$ and {\em weight} $k$ are holomorphic functions $Y(\tau)$ of the modulus satisfying:
\be
Y(\gamma\tau)=(c\tau+d)^k Y(\tau)
\ee
for any $\gamma\in\overline\Gamma(N)$. They form a linear space ${\cal M}_{k}(\Gamma(N))$ of finite dimension $d_{k}(\Gamma(N))$ \cite{gunning}. Under the full modular group $\overline\Gamma$ 
a basis $Y(\tau)$ of ${\cal M}_{k}(\Gamma(N))$
transforms as $Y(\gamma\tau)=(c\tau+d)^k \rho(\gamma) Y(\tau)$, $\rho(\gamma)$ being a unitary, possibly reducible representation of $\Gamma_N$.

Turning off gauge interactions, the action ${\cal S}$ of an ${\cal N}=1$ global supersymmetric theory depending on the modulus $\tau$ and a set of supermultiplets $\phi$, comprising
matter fields $\phi^{(I)}$ of the same level $N$ and possibly different weights $k_{I}$, reads
\be
{\cal S}=\int d^4x \, d^2\theta \, d^2 \bar\theta \, K(\tau,\phi,\bar\tau,\bar\phi)+ \int d^4x \, d^2\theta \, w(\tau,\phi)+\int d^4x \, d^2\bar\theta\, \bar{w}(\bar\tau,\bar\phi) \;,
\ee
where $K$ and $w$ are the K\"ahler potential and the superpotential, respectively. Invariance under the transformations of eqs.~(\ref{tratau}) and~(\ref{tramat}) requires a modular invariant superpotential and a K\"ahler potential modular invariant
up to K\"ahler transformations
\bea
K(\tau,\phi,\bar\tau,\bar\phi)&\to&K(\tau,\phi,\bar\tau,\bar\phi)+f(\tau,\phi)+\bar{f}(\bar\tau,\bar\phi)\nn\\
w(\tau,\phi)&\to&w(\tau,\phi) \;.
\label{traKw}
\eea
Eq.~(\ref{traKw}) is easily satisfied by minimal forms of the K\"ahler potential, an example being
\be
K(\tau,\phi,\bar\tau,\bar\phi)=-h \log(-i\tau+i\bar\tau)+ \sum_I (-i\tau+i\bar\tau)^{k_I} |\phi^{(I)}|^2 \;,
\label{kalex}
\ee
where $h$ is a positive constant. On the contrary the requirement of modular invariance severely restricts the superpotential $w(\tau,\phi)$.  
Consider the expansion of $w(\tau,\phi)$ in power series of the supermultiplets $\phi^{(I)}$:
\be
w(\tau,\phi)=\sum_n Y_{I_1 \ldots I_n}(\tau)\, \phi^{(I_1)} \ldots \, \phi^{(I_n)} \;.
\label{psex}
\ee
For the $n$-th order term to be modular invariant, the functions $Y_{I_1 \ldots I_n}(\tau)$ should be holomorphic functions of $\tau$ transforming as
\be
Y_{I_1 \ldots I_n}(\gamma\tau)=(c\tau+d)^{k_Y(n)} \rho(\gamma) \, Y_{I_1 \ldots I_n}(\tau) \;,
\label{mfk}
\ee
with the weight $k_Y(n)$ and the unitary representation $\rho$ such that:
\begin{enumerate}
\item[1.]
The weight $k_Y(n)$ should compensate the overall weight of the product $\phi^{(I_1)}\ldots\, \phi^{(I_n)}$:
\be
k_Y(n)+k_{I_1}+ \ldots +k_{I_n}=0 \;.
\label{compensate}
\ee
\item[2.]
The product $\rho\times \rho^{{I_1}}\times \ldots \times \rho^{{I_n}}$ contains an invariant singlet.
\end{enumerate}
The holomorphic functions $Y_{I_1 \ldots I_n}(\tau)$ of eq.~(\ref{mfk}) are modular forms of level $N$ and weight $k=k_Y(n)$. This property sharply constrains the allowed Yukawa couplings, to the point of completely determining in some case the corresponding mass matrix as a function of $\tau$, up to an single overall constant. 

As an example, choose $N=3$ and consider 3 copies of lepton doublets $l$ and one Higgs supermultiplet $H_u$ transforming, respectively, as irreducible triplets of $\Gamma_3=A_4$
with weight -1 and as a singlet of $\Gamma_3$ with zero weight. Assuming neutrino masses described entirely by the Weinberg operator, the relevant superpotential reads:
\be
w_\nu=\frac{1}{2\Lambda}(l_iH)Y_{ij}(\tau) (l_jH) \;,
\label{W3}
\ee
where the holomorphic functions $Y_{ij}(\tau)$ should be modular forms of level 3, weight +2
transforming as one of the multiplet in the decomposition $(3\times 3)_{SYM}=1+1'+1''+3$.
The space ${\cal M}_{2}(\Gamma(3))$ is spanned by three linearly independent modular forms
$Y_i(\tau)$  $(i=1,2,3)$, transforming as a $3$ under $\Gamma_3$:
\bea
\label{craft}
Y_1(\tau)&=&\frac{i}{2 \pi}
   \left[
   \frac{\eta'\left(\frac{\tau }{3}\right)}{\eta \left(\frac{\tau}{3}\right)}
   +\frac{\eta'\left(\frac{\tau+1}{3}\right)}{\eta \left(\frac{\tau+1}{3}\right)}
   +\frac{\eta'\left(\frac{\tau +2}{3}\right)}{\eta\left(\frac{\tau +2}{3}\right)}
   -\frac{27 \eta'(3 \tau )}{\eta (3 \tau)}
   \right]\nn\\
Y_2(\tau)&=&\frac{-i}{\pi}
   \left[
   \frac{\eta'\left(\frac{\tau }{3}\right)}{\eta \left(\frac{\tau}{3}\right)}
+\omega^2 \, \frac{\eta'\left(\frac{\tau+1}{3}\right)}{\eta \left(\frac{\tau+1}{3}\right)}
   +\omega \, \frac{\eta'\left(\frac{\tau +2}{3}\right)}{\eta\left(\frac{\tau +2}{3}\right)}
   \right]\\
Y_3(\tau)&=&\frac{-i}{\pi}
   \left[
   \frac{\eta'\left(\frac{\tau }{3}\right)}{\eta \left(\frac{\tau}{3}\right)}
+\omega \, \frac{\eta'\left(\frac{\tau+1}{3}\right)}{\eta \left(\frac{\tau+1}{3}\right)}
   +\omega^2 \, \frac{\eta'\left(\frac{\tau +2}{3}\right)}{\eta\left(\frac{\tau +2}{3}\right)}
   \right] \;,\nn
\eea
where $\eta(\tau)$ is the Dedekind eta-function, defined in the upper complex plane:
\be
\eta(\tau)=q^{1/24}\prod_{n=1}^\infty \left(1-q^n \right) \hspace{2cm} q\equiv e^{i 2 \pi\tau} \;.
\ee
It follows that $w_\nu$ consists of a unique modular invariant combination and is fully determined up to an overall constant. In a suitable basis the neutrino mass matrix reads:
\be
m_\nu=m_0
\left(
\begin{array}{ccc}
2 Y_1(\tau)&-Y_3(\tau)&-Y_2(\tau)\\
-Y_3(\tau)&2Y_2(\tau)&-Y_1(\tau)\\
-Y_2(\tau)&-Y_1(\tau)&2Y_3(\tau)
\end{array}
\right) \;.
\label{examplemnu}
\ee
As long as supersymmetry is unbroken there are no corrections coming from higher dimensional holomorphic operators. The matrix $m_\nu$ in eq.~(\ref{examplemnu}) is exact and all the terms in the expansion in powers of $\tau$ are completely determined. 

Non-vanishing modular forms transforming under $\Gamma_N$ require even integer non-negative weights \cite{gunning}. Modular forms of vanishing weight are constant, that is $\tau$-independent.
Modular forms for the first few levels $N$ have been explicitly constructed and the combinations transforming
as irreducible representations of $\Gamma_N$ have been identified for the first few weights. The results for $\Gamma_2\approx S_3$\cite{Kobayashi:2018vbk},
$\Gamma_3\approx A_4$\cite{Feruglio:2017spp}, $\Gamma_4\approx S_4$\cite{Penedo:2018nmg}, $\Gamma_5\approx A_5$\cite{Novichkov:2018nkm,Ding:2019xna}, $\Gamma_7\approx \Sigma(168)$ \cite{Ding:2020msi}
are summarized in table \ref{tabmodform}. Modular forms of generic integer weights have been discussed in ref.~\cite{Liu:2019khw}, together with their application to neutrino mass models.
They have been shown to form representations of the homogeneous finite modular groups $\Gamma'_N$, double covering of $\Gamma_N$.
\begin{table}[h!] 
\centering
\begin{tabular}{|l|c|c|c|c|c|}
\hline
&$d_{k}(\Gamma(N))$&$k=2$&$k=4$&$k\ge6$\rule[-2ex]{0pt}{5ex}\\
\hline
$\Gamma_2\approx S_3$ &$k/2+1$&$2$&$1+2$&$\ldots $\rule[-2ex]{0pt}{5ex}\\
\hline
$\Gamma_3\approx A_4$&$k+1$&$3$&$1+1'+3$&$ \ldots $\rule[-2ex]{0pt}{5ex}\\
\hline
$\Gamma_4\approx S_4$&$2k+1$&$2+3'$&$1+2+3+3'$&$ \ldots $\rule[-2ex]{0pt}{5ex}\\
\hline
$\Gamma_5\approx A_5$&$5k+1$&$3+3'+5$&$1+3+3'+4+5+5$&$ \ldots $\rule[-2ex]{0pt}{5ex}\\
\hline
$\Gamma_7\approx \Sigma(168)$&$14k-2$&$3+7+8+8'$&$1+3+6+6'+7+7'+8+8'+8''$&$ \ldots $\rule[-2ex]{0pt}{5ex}\\
\hline
\end{tabular}
\caption{Dimension of ${\cal M}_{k}(\Gamma(N))$ and decomposition of multiplets of modular forms in representations
of the finite modular group $\Gamma_N$, for the first few levels and weights. Modular forms of higher weight can be
obtained from polynomials of modular forms of lower weight. Partial knowledge is available for modular forms of weight 2 for levels $8$ and $16$ \cite{Kobayashi:2018bff}.}
\label{tabmodform}
\end{table}

\subsection{Modular invariance and CP}
The action of CP on $\tau$ is uniquely determined, up to modular transformations \cite{Dent:2001cc,Dent:2001mn,Baur:2019kwi,Novichkov:2019sqv,Baur:2019iai}:
\be
\tau\xrightarrow{\text{CP}}-\tau^* \;.
\label{CPtransf}
\ee
Such a law corresponds to the outer automorphism of $\overline\Gamma$:
\be
S\xrightarrow{\text{CP}} S \hspace{2cm} T\xrightarrow{\text{CP}} T^{-1} .
\ee
By choosing a suitable basis for the generators $S$ and $T$, where both are described by
symmetric matrices in any representation of $\Gamma_N$, the action of CP on matter multiplets $\phi$
reduces to the canonical one:
\be
\phi\xrightarrow{\text{CP}} X_\text{CP} \,  \phi^* \;, \hspace{2cm} X_\text{CP}=\mathbb{1} \;.
\ee
In this basis the requirement of CP invariance amounts to restricting all the Lagrangian parameters
to real values. In such a theory CP invariance can only be spontaneously broken. The values of $\tau$ preserving CP
lie along the imaginary $\tau$ axis or along the border of the fundamental region shown in fig.~\ref{FR}, where $-\tau^*=\tau$, up to a modular transformation.

\subsection{Modular invariance and standard flavour symmetries}

It is worth to mention that in the low-energy theory arising from string theory compactification, the flavour group generally comprises both modular transformations and ordinary transformations, acting linearly on matter fields. The consistent combination
of the two types of transformations have been analyzed in \cite{Nilles:2020kgo,Nilles:2020nnc}.
The ordinary linear transformations belong to a group $G$, leave the modulus $\tau$ invariant and act on the fields 
$\phi^{(I)}$ through a unitary matrix $U^{(I)}(g)$:
\be
\tau\to \tau~~~~~~~~~~~~~~~~~~~~~~~
\phi^{(I)}\to \,U^{(I)}(g)\phi^{(I)} \;.
\label{traord}
\ee
The two sets of transformations (\ref{tratau},\ref{tramat}) and (\ref{traord}) should obey the consistency condition:
\be
\rho^{(I)}(\gamma)~U^{(I)}(g)~\rho^{(I)}(\gamma^{-1})=U^{(I)}(g')~~~,
\ee
for some element $g'\in G$. It follows that $G$ is a normal subgroup of the overall flavour group $G_{ecl}$, called eclectic by the authors, generated by both ordinary and modular transformations. At the same time the modular transformations define an automorphism of $G$ which, in the non-trivial cases, is of outer type. This construction allows for a unified
description of standard, non-linear and CP-like transformations. Not all groups $G$ can be embedded in such a framework,
which may open new possibilities in model building.

\subsection{Modular invariance and local supersymmetry}
This setup can be easily extended to the case of ${\cal N}=1$ local supersymmetry where K\"ahler potential and superpotential are not independent functions 
since the theory depends on the combination
\be
{\cal G}(\tau,\phi,\bar\tau,\bar\phi)=K(\tau,\phi,\bar\tau,\bar\phi)+\log w(\tau,\phi)+\log \bar{w}(\bar\tau,\bar\phi) \;.
\ee
The modular invariance of the theory can be realized in two ways \cite{Ferrara:1989bc}. Either $K(\tau,\phi,\bar\tau,\bar\phi)$ and $w(\tau,\phi)$ are separately modular invariant or
the transformation of $K(\tau,\phi,\bar\tau,\bar\phi)$ under the modular group is compensated by that of $w(\tau,\phi)$. An example of this second possibility is given
by the Kahler potential of eq.~(\ref{kalex}), with the superpotential $w(\tau,\phi)$ transforming as
\be
w(\tau,\phi)\to e^{i\alpha(\gamma)} (c\tau+d)^{-h} w(\tau,\phi) 
\ee
In the expansion (\ref{psex}) the Yukawa couplings $Y_{I_1 \ldots I_n}(\tau)$ should have weight $k_Y(n)$
such that $k_Y(n)+k_{I_1}+ \ldots +k_{I_n}=-h$ and the representation $\rho(\gamma)$ subject to the requirement 2
in eq.~(\ref{compensate}).
When we have $k_{I_1}+ \ldots +k_{I_n}=-h$, we get $k_Y(n)=0$ and the functions $Y_{I_1 \ldots I_n}(\tau)$ are $\tau$-independent constants.
This occurs for supermultiplets belonging to the untwisted sector in the orbifold compactification of the heterotic string.

\subsection{Models}
Models of lepton masses and mixing angles have been constructed for levels 2, 3, 4, 5, following
two different approaches, depending on whether the charged lepton mass matrix only
depends on $\tau$, as the neutrino one, or it depends on a separate set of flavons.
In either case the VEV of $\tau$ is usually treated as an additional parameter and scanned in order to
maximize the agreement with data. 
We show here an example for each possibilities. In both examples neutrino masses arise from the type I
seesaw mechanism and, after integrating out the right-handed neutrinos $N^c$, the low-energy superpotential reads:
\be
w=-{E^c}^T {\cal Y}_e H_d L-\dd\frac{1}{2\Lambda}(H_u L)^T \left({\cal Y}_\nu^T {\cal  C}^{-1} {\cal Y}_\nu\right) (H_u L) \;.
\ee
An example of the first possibility is the model of ref.~\cite{Novichkov:2018ovf,Novichkov:2019sqv} realized at level 4, with the particle content displayed in table
 \ref{tabmod4w}.
\begin{table}[h!] 
\centering
\begin{tabular}{|c|c|c|c|c|c|}
\hline
&$(E_1^c,E_2^c,E_3^c)$&$N^c$& $L$& $H_d$& $H_u$\rule[-2ex]{0pt}{5ex}\\
\hline
$\text{SU(2)}_L\times \text{U(1)}_Y$&$(1,+1)$ & $(1,0)$ & $(2,-1/2)$& $(2,-1/2)$& $(2,+1/2)$\rule[-2ex]{0pt}{5ex}\\
\hline
$\Gamma_4\approx S_4$&$(1',1,1')$& $3'$& $3$& $1$& $1$\rule[-2ex]{0pt}{5ex}\\
\hline
$k_I$&$(0,-2,-2)$& $0$& $-2$& $0$& $0$\rule[-2ex]{0pt}{5ex}\\
\hline
\end{tabular}
\caption{Chiral supermultiplets, transformation properties and weights of the model of ref.~\cite{Novichkov:2018ovf,Novichkov:2019sqv}.}
\label{tabmod4w}
\end{table}

\noindent
The matrices ${\cal Y}_e$,  ${\cal Y}_\nu$ and ${\cal C}$ are given by:
\be
{\cal Y}_e= \begin{pmatrix}
\a\, Y_3 & \a\, Y_5 & \a\, Y_4 \\
\b \left(Y_1Y_4 - Y_2Y_5\right) & \b \left(Y_1Y_3 - Y_2Y_4\right) & \b \left(Y_1Y_5 - Y_2Y_3\right) \\
\g \left(Y_1Y_4 + Y_2Y_5\right) & \g \left(Y_1Y_3 + Y_2Y_4\right) & \g \left(Y_1Y_5 + Y_2Y_3\right) \\
\end{pmatrix}\,,
\ee
\be
{\cal Y}_\nu= g \left[ 
\begin{pmatrix}
0 & -Y_1 & Y_2 \\
-Y_1 & Y_2 & 0 \\
Y_2 & 0 & -Y_1
\end{pmatrix} 
+ \frac{g'}{g} \begin{pmatrix}
2Y_3 & -Y_5 & -Y_4 \\
-Y_5 & 2Y_4 & -Y_3 \\
-Y_4 & -Y_3 & 2Y_5
\end{pmatrix}
\right], \hspace{2cm} {\cal C}= 
\left(
 \begin{array}{ccc}
1&0&0\\
0&0&1\\
0&1&0
 \end{array}
 \right)\; ,
\ee
where $Y_{1,2}$ and $Y_{3,4,5}$ are the five independent modular forms of weight 2 and level 4. They transform as $2$ and $3'$ under $\Gamma_4\approx S_4$, respectively.
Invariance under CP implies that $g/g'$ is real.  Charged lepton masses can be correctly reproduced by adjusting $\a$, $\b$ and $\g$. The remaining Lagrangian parameters
are an overall scale and $g/g'$. The VEV of $\tau$ is treated as an additional free parameter.

An example of the second possibility is the model of ref.~\cite{Criado:2018thu} realized at level 3, with the particle content displayed in table
 \ref{tabmod3w}.
\begin{table}[h!] 
\centering
\begin{tabular}{|c|c|c|c|c|c||c|}
\hline
&$(E_1^c,E_2^c,E_3^c)$&$N^c$& $L$& $H_d$& $H_u$&$\varphi$\rule[-2ex]{0pt}{5ex}\\
\hline
$\text{SU(2)}_L\times \text{U(1)}_Y$&$(1,+1)$ & $(1,0)$ & $(2,-1/2)$& $(2,-1/2)$& $(2,+1/2)$& $(1,0)$ \rule[-2ex]{0pt}{5ex}\\
\hline
$\Gamma_3\approx A_4$&$(1,1'',1')$& $3$& $3$& $1$& $1$& $3$ \rule[-2ex]{0pt}{5ex}\\
\hline
$k_I$&$-4$& $-1$& $+1$& $0$& $0$& $+3$ \rule[-2ex]{0pt}{5ex}\\
\hline
\end{tabular}
\caption{Chiral supermultiplets, transformation properties and weights of the model of ref.~\cite{Criado:2018thu}.}
\label{tabmod3w}
\end{table}

\noindent
The matrices ${\cal Y}_e$,  ${\cal Y}_\nu$ and ${\cal C}$ are given by:
\be
{\cal Y}_e=
\left(
 \begin{array}{ccc}
a\, \varphi_{1}&a\, \varphi_{3}&a\, \varphi_{2}\\
b\, \varphi_{2}& b\, \varphi_{1}&b\, \varphi_{3}\\
c\, \varphi_{3}&c\, \varphi_{2}&c\, \varphi_{1}
 \end{array}
 \right) \;, \qquad
{\cal Y}_\nu=y_0
 \left(
 \begin{array}{ccc}
1&0&0\\
0&0&1\\
0&1&0
 \end{array}
 \right) \;, \qquad
{\cal C}= 
\left(
 \begin{array}{ccc}
2 Y_1(\tau)&-Y_3(\tau)&-Y_2(\tau)\\
-Y_3(\tau)&2 Y_2(\tau)&-Y_1(\tau)\\
-Y_2(\tau)&- Y_1(\tau)& 2  Y_3(\tau) 
 \end{array}
 \right)  \;.
\label{su0}
\ee
Beyond the parameters $a$, $b$ and $c$, which control charged lepton masses, the low energy Lagrangian depends on a single parameter, the overall scale $y_0^2/\Lambda$. 
Additional parameters are provided by the VEVs of $\tau$ and of the flavon $\varphi$, assumed to be aligned along the $(1,0,{\tt Re}(\varphi_3))$ direction.
The results of the two models are collected in table \ref{tabMod}.
\begin{table}[!h]
\begin{flushleft}
\hspace*{0.25\textwidth}
\begin{tabular}{|c|c|c|c|c|c|}
\hline
$N$&$r\equiv|\Delta m^2_{sol}/\Delta m^2_{atm}|$&$\sin^2\theta_{12}$&$\sin^2\theta_{13}$&$\sin^2\theta_{23}$&$\delta/\pi$\rule[-2ex]{0pt}{5ex}\\
\hline
4&$0.0298$&$0.305$ &$0.0214$ &$0.486$ &$1.641$\rule[-2ex]{0pt}{5ex}\\
\hline
3&$0.0299$&$0.306$ &$0.0211$ &$0.459$ &$1.438$\rule[-2ex]{0pt}{5ex}\\
\hline
\end{tabular} \\
\vskip 0.5 cm
\hspace*{0.25\textwidth}
\begin{tabular}{|c|c|c|c|c|c|c|}
\hline
$N$&$\alpha_{21}/\pi$&$\alpha_{31}/\pi$&
$m_1 \,({\rm meV})$&$m_2 \,({\rm meV})$&$m_3 \,({\rm meV})$&$|m_{ee}| \,({\rm meV})$\rule[-2ex]{0pt}{5ex}\\
\hline
4&$0.346$&$1.254$&$12.1$&$14.8$&$51.4$&$12.0$\rule[-2ex]{0pt}{5ex}\\
\hline
3&$1.704$&$1.201$&$10.9$&$13.9$&$51.1$&$10.4$\rule[-2ex]{0pt}{5ex}\\
\hline
\end{tabular}
\end{flushleft}
\caption{Results of the model of ref.~\cite{Novichkov:2018ovf,Novichkov:2019sqv}, $N=4$, for $\tau=0.09922+i\, 1.0578$ and $g/g'=-0.02093$ and of the model of ref.~\cite{Criado:2018thu}, $N=3$, for $\tau=-0.2005+i\, 1.0578$ and $\varphi=(1,0,0.117)$.}
\label{tabMod}
\end{table}

In both models the mass ordering is normal. The atmospheric mixing angle is close to maximal, but predicted to lie in the first octant.
CP is broken spontaneously by the VEV of $\tau$ and both Dirac and Majorana phases are predicted. Also the absolute value of neutrino masses
and the combination relevant to $0\nu\beta\beta$ are predicted. Quite interestingly, the lightest neutrino has a mass close to $0.01$ eV, resulting in a relatively large $|m_{ee}|\approx 10$ meV for a normally ordered mass spectrum.

\begin{figure}[h!]
 \centering
 \includegraphics[width=0.7\textwidth]{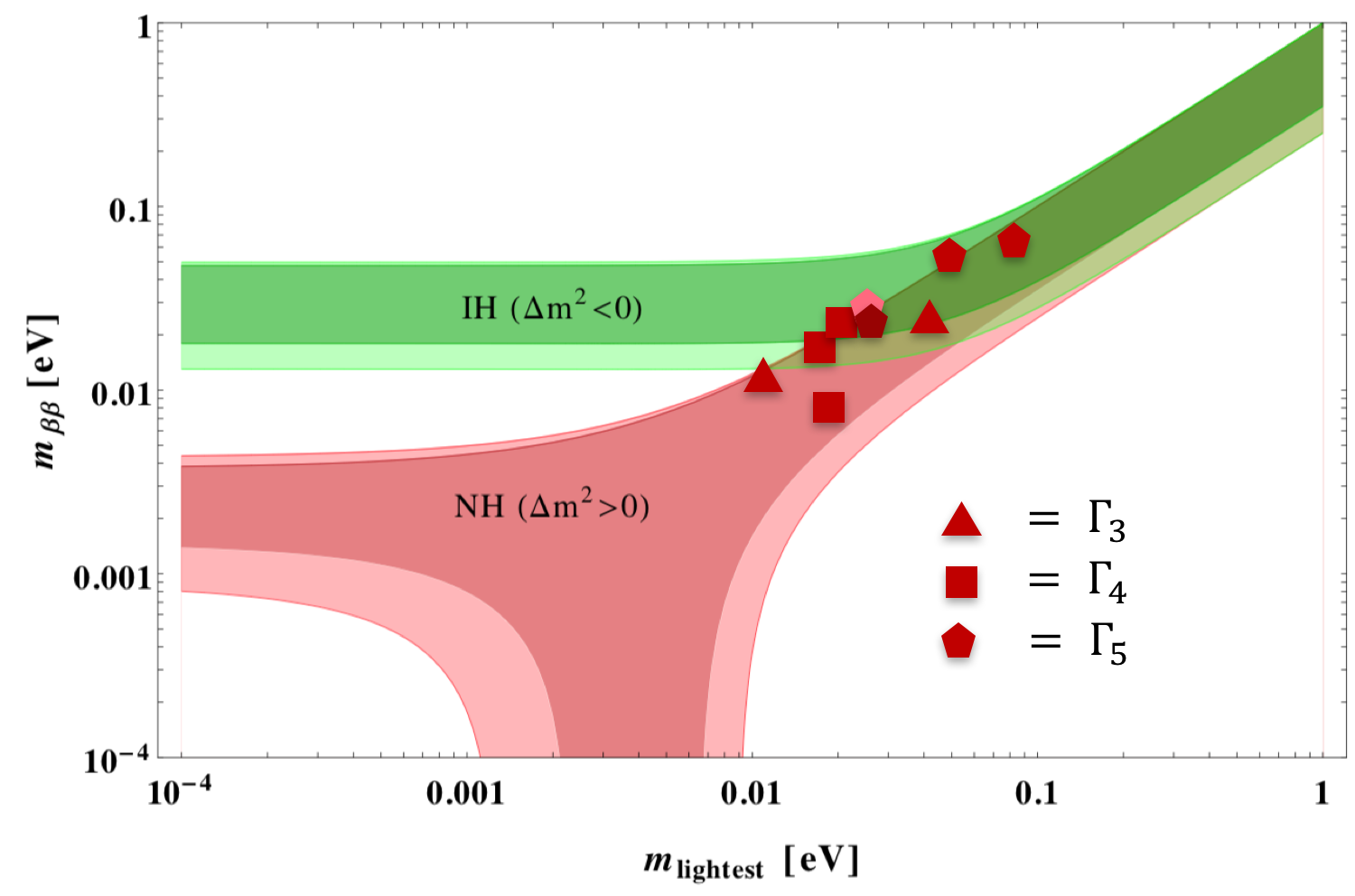}
 \caption{Regions allowed in the $(m_{\rm lightest},|m_{ee}|)$ plane for normal ordering (red) and inverted ordering (green) and predictions of modular invariant models
 at level 3, 4 and 5.}
\label{fig:mee}
\end{figure}

Several other models of lepton masses and mixing angles have been built at level 2 \cite{Kobayashi:2018vbk,Kobayashi:2018wkl}, level 3 \cite{Feruglio:2017spp,Criado:2018thu,Kobayashi:2018scp,Novichkov:2018yse,Nomura:2019yft,Ding:2019zxk,Gui-JunDing:2019wap}, level 4 \cite{Penedo:2018nmg,Novichkov:2018ovf,King:2019vhv,Criado:2019tzk,Liu:2020akv,Novichkov:2020eep} and level 5 \cite{Novichkov:2018nkm,Ding:2019xna,Criado:2019tzk}. The higher the level $N$, the more solutions are found in ${\cal H}$, corresponding to
physically distinct sets of predictions in good agreement with data. Most of the solutions predicting NO prefer a nearly degenerate spectrum with $m_1>10$ meV and $|m_{ee}|$ on the high side of allowed range. This is shown in figure \ref{fig:mee}.

A common feature of all the proposed models is the minimal form of the K\"ahler potential,
eq.~(\ref{kalex}). While this is the simplest choice, it is not the most general one compatible
with modular invariance. The symmetry of the K\"ahler potential $K$ of eq.~(\ref{kalex}) is bigger than the modular one. Indeed $K$ is invariant under transformation of $\text{SL}(2,\mathbb{R})$ and the modulus $\tau$ parametrizes the coset $\text{SL}(2,\mathbb{R})/\text{SO(2)}$. Such a continuous symmetry is broken by the superpotential down to
the modular group. In a bottom up approach there is no reason to exclude from the K\"ahler potential $K$ terms
that are invariant only under the discrete modular group. In particular a candidate modification of the K\"ahler potential (\ref{kalex}) is an additive contribution depending explicitly on both the matter supermultiplets and on the modular forms $Y(\tau)$ \cite{Feruglio:2017spp}. The power counting controlling the size of these contributions is
unknown, but examples in the string theory context suggest that, in the parameter region ${\tt Im}(\tau)=\ord{1}$
which is the one of interest to neutrino physics, they might be of similar importance as $K$ in eq.~(\ref{kalex}). 
Indeed these type of corrections have been analyzed in ref.~\cite{Chen:2019ewa},
showing that the new parameters appearing in the K\"ahler potential considerably reduce the predictability of the approach. At the moment the problem of better controlling the K\"ahler potential remains an open one.

An interesting question concerns the dynamical determination of the VEV of $\tau$. It has been conjectured \cite{Cvetic:1991qm,Kobayashi:2019uyt} that extrema of modular invariant scalar potentials of ${\cal N}=1$ supergravity theories lie on the imaginary $\tau$ axis or along
the border of the fundamental region ${\cal F}$ of figure~\ref{FR}. This is precisely the region where CP is unbroken if the theory is CP invariant. Interestingly,
in concrete models it suffices a small deviation from the border of ${\cal F}$ to allow for sizable CP violating effects. 
For instance in the model of ref.~\cite{Novichkov:2018ovf,Novichkov:2019sqv}, the value of $\tau$ that maximizes the agreement with data is $0.09922+i\, 1.0578$.
An attempt to dynamically determine the VEV of $\tau$ can be found in ref.~\cite{Kobayashi:2019xvz}, where modular invariance is realized in supergravity. At the minima of the scalar potential the energy density is negative, and some ad hoc mechanism should be invoked to
reproduce the correct cosmological constant. This is confirmed by the analysis of ref.~\cite{Gonzalo:2018guu} where no minima with positive energy density have been found.
Corrections from SUSY breaking have been shown to be negligible \cite{Criado:2018thu}, provided there is a sufficient gap between the sparticle masses and the messenger scale. 
The modulus-electron interactions can be directly tested
in neutrino oscillations, provided the modulus is extremely light \cite{Ding:2020yen}. In such a case scalar non-standard
neutrino interactions can affect lepton mass matrices and produce deviations in oscillation patterns in media with a sufficiently large electron
number density.

\subsection{Extension to quarks}

The possibility of extending modular invariance to the quark sector has also been investigated in \cite{Okada:2018yrn,Kobayashi:2018wkl,Okada:2019uoy,Okada:2020rjb} and, in a GUT context, in \cite{deAnda:2018ecu,Kobayashi:2019rzp}.
Description of the quark sector alone seems to require a large number of parameters, often larger than the number of observables. Having many parameters at disposal, it is not surprising that a unified description of leptons and quarks, adopting the same value of $\tau$ to simultaneously describe the two sectors, can be achieved. One of the major obstacle towards
the realization of a more economical model is the fact that each charged fermion mass requires
an independent parameter. In its present realization, modular invariance seems unable to
provide predictions concerning the charged fermion masses, which should be described by an 
ad-hoc set of parameters. To improve this aspect, two suggestions have been recently 
put forward.
If quark and charged lepton masses cannot be precisely predicted, at least their order
of magnitude can be captured by letting the modular weights play the role of Froggatt-Nielsen
charges \cite{Criado:2019tzk,King:2020qaj}. Assigning different weights to electroweak singlet fermions, we can achieve a natural relative
suppression of charged fermion masses, similarly to what happens in ordinary abelian symmetries. As a consequence, dimensionless free parameters are not reduced in number, but  
their values have the same order of magnitude.
A second observation is that
modular invariance can naturally enforce texture zeros, which are known to increase the predictive
power of flavour models. Along these lines, the authors of ref. \cite{Lu:2019vgm} have built several models
at level 3. They make use of odd weight modular forms and assign quarks to both singlet and doublet representations
of $\Gamma_3'\approx T'$, the double covering of $\Gamma_3$. In a specific model all 22 fermion mass/mixing observables are reproduced 
using 17 independent parameters and the best fit value of $\tau$ is intriguingly close to $-1/2+i \sqrt{3}/2$,
a fixed point under the action of $ST$.

For moderately large values of ${\tt Im} (\tau)$, modular forms have a nearly exponential dependence on $\tau$, which, at first sight, seems
ideal to describe the hierarchical mass spectrum we observe in quarks and in charged leptons.
This suggests that we might have not fully exploited all the possibilities offered by this approach.

\subsection{Outlook}
Modular invariance is an interesting candidate for a realistic flavour symmetry. Compared to the traditional linear realization of discrete symmetries, it allows to predict
not only mixing angles and phases but also neutrino masses. It requires less flavons: in minimal realizations no flavon beyond $\tau$ is needed. In the most favorable cases, as long as supersymmetry is exact,
the superpotential is completely determined by symmetry requirements, to any order in the $\tau$ power expansion,
up to an overall constant. In the exact supersymmetry limit the superpotential does not receive any perturbative or
nonperturbative corrections, a unique feature compared to the models based on linearly realized symmetries.
A lesson that we can learn from the proposed models is that a low level $N$ and modular forms of low weights
minimize the number of free parameters.
So far the approach allows no prediction for the charged lepton masses. The charged lepton sector might require a substantially different description, perhaps in terms
of additional moduli \cite{Ferrara:1989qb,deMedeirosVarzielas:2019cyj} or some conventional flavon.
The models proposed so far rely on a minimal form of the K\"ahler potential, which however is not justified in a bottom-up approach. Modular invariance allows for additional terms in the K\"ahler potential, and their impact in the
parameter region of interest to neutrinos has been shown to be important.

\vfill

\section{What have we learned?}
\label{sec:conclusions}
The discovery of neutrino oscillations has led to a major advance in our knowledge of the flavour sector. On the one side, there is still a considerable room for improvement of the data. The uncertainty on the absolute neutrino masses is very large, since only mass-squared differences have been measured.
CP-odd phases (in particular the Majorana ones, if present) will not really be known with good precision 
for a very long time. On the other end, the impressive experimental outcomes of the recent years have brought neutrino physics into a precision era, with several combinations of mass/mixing parameters known with a precision approaching the percent level.
Tracing those parameters back to some fundamental organizing principle is part of a very ambitious program, the solution of the flavour puzzle. In this wider context, we cannot avoid considering both leptons and quarks, most probably within some kind of unified framework emerging when physics is probed at a very high energy scale. Actually, the need of reconciling the very different features of the quark and lepton sectors might provide important clues to correctly address and solve the puzzle. Quark intergenerational hierarchy is much pronounced, especially in the up sector. Mixing angles are small, with the third generation very feebly coupled to the first two. On the contrary neutrino masses are of the same order of magnitude, with the possible exception of the lightest state, still compatible with being massless. The lepton mixing pattern is completely different from the quark one, the smallest mixing angle being similar in size to the Cabibbo angle. Nevertheless, the description of the lepton sector has borrowed many ideas and techniques originally developed in the context of the quark sector. 

An appealing approach that has pervaded the whole field for decades is the one based on flavour symmetries, supported by the success that symmetry considerations have collected during last century in the description of particle interactions. Flavour symmetries of the leptonic sector have been realized in a vast amount of ways, as shown by the extensive literature of the field. Maybe one of the most striking things that captures the attention is the fact that, despite all past efforts, a baseline model interpreting neutrino masses and mixings in the context of a flavour symmetry is still missing. Many early models have been discarded by gathering more and more precise data, but the range of remaining possibilities is still very large, even taking into account the constraints from the quark sector. This is closely related to the fact that in any realistic model of lepton masses relying on flavour symmetries and retaining some degree of predictability, the underlying symmetry is cleverly hidden and breaking effects are a decisive factor in constraining the relevant observables.

Actually, one of the few firm points is the fact that there cannot be exact flavour symmetries, neither for quarks nor for leptons alone. The observed masses and mixing angles break any initial flavour symmetry, except possibly for the total baryon and lepton numbers. Once excluded that exact flavour symmetries are allowed by data, we could wonder whether they can provide at least some reasonable first order approximation to the observed lepton mass/mixing pattern. It turns out that, under mild assumptions, symmetries compatible with this requirement are not very powerful. In the normal mass ordering case, the neutrino mass matrix is completely unconstrained and any neutrino masses and mixings are possible. The flavour symmetry is useless in the neutrino sector, where it leads to anarchy. Therefore, if the present hint for normal hierarchy were confirmed, we would conclude that 
symmetry breaking effects would play a leading role in a realistic non-trivial model of lepton masses.

Indeed, the common denominator of most predictive models is the breaking of the flavour symmetry induced by a set of spurions. The prototype of these models makes use of a spontaneously broken abelian continuous group. While abelian symmetries have played a pivotal role in the development of the field, they can lead to predictions matching the present experimental accuracy only in the presence of texture zeros, as each entry of the mass matrices is predicted order-of-magnitude wise, with intrinsic uncertainties of order one. Any successful model of neutrino masses and mixings based on flavour symmetries should rely on a sizable departure of the predictions from the symmetric limit, most often of a non-abelian group. 

If so well hidden, flavour symmetries can be difficult to identify from the data. Moreover in model building sizable breaking effects analyzed to the desired level of accuracy typically involve a non-negligible set of parameters, which weakens the aimed-for predictive power of the construction. In the absence of a symmetric limit reasonably close to observation, the whole symmetry approach seems undermined. So, why not to abandon it? We believe there are several counterexamples to this negative conclusion. Perhaps the most impressive one
is provided by the modular symmetry that, being non-linearly realized, does not allow any limit where the full modular group remains unbroken. The geometrical interpretation of this feature is particularly transparent. The modular transformations can be seen as gauge transformations describing all possible equivalent parametrizations of the same torus in terms of a modular parameter. For the modular group to be unbroken, we would need a torus not admitting distinct equivalent parametrizations, which is impossible by construction. Thus in modular invariant flavour models there is no notion of a symmetric limit, and this does not prevent predictability and precision, at least in principle. In the most favorable cases, the neutrino mass matrix is completely determined by symmetry requirements as a function of the modular parameter up to an overall constant. In the exact supersymmetry limit, the superpotential does not receive any perturbative or non-perturbative corrections, a unique feature compared to the models based on linearly realized symmetyries.

The requirement of being far from the symmetric limit does not forbid that, separately, the neutrino and the charged lepton sectors can be approximately invariant under independent symmetries, arising as subgroups of the full symmetry group. This occurs when spurions with different breaking properties are accidentally sequestered. Since the most general symmetry leaving the neutrino mass matrix invariant and its eigenvalues unconstrained is the Klein symmetry, the more economical realizations of such sequestering adopts discrete flavour groups. Due to unavoidable corrections, exact sequestering can hardly occur, and should be rather viewed as an ideal limit useful to identify approximate mixing patterns. It is remarkable that, even considering the smallest discrete groups allowing three-dimensional irreducible representations, semi-realistic mixing patterns such as the tribimaximal one can be easily obtained. Tribimaximal mixing is ruled out by data and, to identify realistic mixing patterns in the framework of exact sequestering, we should move to larger discrete groups. Otherwise, working with small discrete groups, we can relax sequestering by allowing sizable corrections or by reducing the residual symmetries. 

A weak point of this approach is that sequestering requires a specific vacuum alignment, that in turns is often realized at the price of a complicated scalar sector and of additional ad-hoc symmetries. Additional ingredients are needed in order to constrain masses and Majorana phases. The phases can be dealt with by exploiting flavour symmetries incorporating CP. The hierarchical nature of charged lepton masses can be accounted for with traditional suppression mechanisms, but order-of-magnitude uncertainties cannot be evaded. Also modular invariant models have limitations. The models proposed so far rely on a minimal form of the K\"ahler potential. 
Modular invariance alone allows for more general K\"ahler potentials, which introduces more parameters reducing the predictability of the approach. Such freedom and the related impact on predictability are common to all supersymmetric models independently from the specific flavour group, but they are particularly relevant in the modular case, where the superpotential can be almost uniquely determined and where realistic values of the modular parameter are non-perturbative. 

Even considering these limitations, flavour symmetries remain one of the few tools we have to address the flavour puzzle with the desired level of predictability and precision. In spite of the large number of relevant contributions to the field, that we have tried to highlight in this review, we believe there are still many  directions to be examined. We do not know if this approach will eventually succeed, but we feel certainly encouraged by the present results to proceed and further explore the new territory.

\begin{acknowledgments}
We are grateful to Claudia Hagedorn for useful discussions and comments on a part of this manuscript. The authors acknowledge partial support by INFN, the MIUR-PRIN project 2015P5SBHT ``Search for the Fundamental Laws and Constituents'' and by the European Union's Horizon 2020 research and innovation programme under the Marie Sklodowska-Curie grant agreements N$^\circ$ 674896 and 690575. 
\end{acknowledgments}

\bibliographystyle{apsrmp4-1}
\bibliography{biblio.bib}

\end{document}